\documentstyle[psfig]{ar}
\begin{document}

\def\ra{\rightarrow}
\def\RA{$\rightarrow$}
\def\Ra{\mbox{$\Rightarrow$}}
\def\PM{$\pm$}
\def\SIM{$\sim$}
\def\mum{$\mu$m}
\def\costh{$\cos\theta$}
\def\lumin{{\cal L}}
\def\pizero{\pi^0}
\def\piz{\pi^0}
\def\avg#1{\langle #1\rangle}
\def\abs#1{\vert #1\vert}
\def\eeto{e^+e^-\rightarrow}
\def\eetoqq{\mbox{$e^+e^-\rightarrow q\bar q$}}
\def\qqbar{\mbox{$q\bar q$}}
\def\BBbar{\mbox{$B\bar B$}}
\def\BBmix{$B^0$--$\bar B^0$ mixing}
\def\etal{{et al.}}
\def\etap{\eta^\prime}
\def\G{$\gamma$}
\def\T{$\tau$}
\def\PI{$\pi$}
\def\PIP{$\pi^+$}
\def\PIM{$\pi^-$}
\def\PIZ{$\pi^\circ$}
\def\RHO{$\rho$}
\def\MU{$\mu$}
\def\TAU{$\tau$}
\def\K{$K$}
\def\KP{$K^+$}
\def\KM{$K^-$}
\def\KS{$K_s$}
\def\KSH{$K_s^\circ$}
\def\KST{$K^*$}
\def\KSTP{$K^{*+}$}
\def\KSTZ{$K^{*\circ}$}
\def\PHI{$\phi$}
\def\PSIP{$\psi '$}
\def\B{$B$}
\def\D{$D$}
\def\BB{$\bar B$}
\def\BZ{$B^0$}
\def\BZB{$\bar B^0$}
\def\UPS{$\Upsilon$}

\def\displayscale#1#2#3#4#5{ 
   \begin{figure}[#1]
   \centerline{\psfig{figure=#3.ps,height=#4in,clip=}} 
   \caption{#5}
   \label{#2}
   \end{figure}}
\def\displaytwo#1#2#3#4#5{ 
   \begin{figure}[#1]
   \centerline{\psfig{figure=#2.ps,height=3.6in,clip=}}
   \caption{#4}\vskip 1.0truecm
   \label{#2}
   \centerline{\psfig{figure=#3.ps,height=3.6in,clip=}}
   \caption{#5}
   \label{#3}
   \end{figure}}

\def\displayabh#1#2#3#4#5#6{
   \begin{figure}[#1]
    \begin{center}
       \psfig{figure=#3.ps,height=#5in} 
       \psfig{figure=#4.ps,height=#5in} 
    \end{center}
    \caption{#6}
    \label{#2}
   \end{figure}}

\def\dbline{\noalign{\vskip 0.15truecm\hrule}\noalign{\vskip 2pt}\noalign{\hrule \vskip 0.15truecm}}
\def\sgline{\noalign{\hrule}}
\def\piz{\pi^0}
\newcommand{\xs}{\mbox{$X_S$}}
\newcommand{\Ebeam}{\mbox{$E_{\rm beam}$}}
\newcommand{\thethr}{\mbox{$\theta_{T}$}}
\newcommand{\DE}{\mbox{$\Delta E$}}
\newcommand{\mb}{\mbox{$m_B$}}
\newcommand{\xf}{\mbox{${\cal F}$}}
\newcommand{\hel}{\mbox{${\cal H}$}}
\newcommand{\A}{\mbox{$\cal{A}\;$}}
\newcommand{\gaga}{{\gamma\gamma}}
\newcommand{\mgg}{\mbox{$m_{\gaga}$}}
\newcommand{\meta}{\mbox{$m_\eta$}}
\newcommand{\momega}{\mbox{$m_\omega$}}
\newcommand{\omegappp}{\mbox{$\omega\ra\pi^+\pi^-\piz$}}
\newcommand{\etagamgam}{\mbox{$\eta\ra\gaga$}}
\newcommand{\etathreepi}{\mbox{$\eta\ra\pi^+\pi^-\pi^0$}}
\newcommand{\etagg}{\mbox{$\eta_{\gaga}$}}
\newcommand{\kshpp}{\mbox{$K_S\ra\pi^+\pi^-$}}
\newcommand{\etatogg}{\mbox{$\eta\ra\gaga$}}
\newcommand{\etathrpi}{\mbox{$\eta_{3\pi}$}}
\newcommand{\etacomb}{\mbox{$\eta_{comb}$}}
\newcommand{\costT}{\mbox{$\cos\theta_T$}}
\newcommand{\etaprepp}{\mbox{$\etapr\ra\eta\pi^+\pi^-$}}
\newcommand{\etaprrg}{\mbox{$\etapr\ra\rho^0\gamma$}}
\newcommand{\blankline}{& & & & & & \cr}
\newcommand{\kstz}{K^{*0}}
\newcommand{\kstp}{K^{*+}}
\newcommand{\calB}{\mbox{$\cal B$}}
\newcommand{\calL}{\mbox{$\cal L$}}

\newcommand{\Bkpi}{\mbox{$B^0\rightarrow K^+\pi^-$}}
\newcommand{\Bkzpi}{\mbox{$B^+\rightarrow K^0\pi^+$}}
\newcommand{\Bpipi}{\mbox{$B^0\rightarrow\pi^+\pi^-$}}
\newcommand{\kpi}{\mbox{$K^+\pi^-$}}
\newcommand{\pipi}{\mbox{$\pi^+\pi^-$}}
\newcommand{\kk}{\mbox{$K^+K^-$}}

\newcommand{\etapr}{{\eta^{\prime}}}
\newcommand{\etaprk}{\mbox{$\etapr K$}}
\newcommand{\etaprkp}{\mbox{$\etapr K^+$}}
\newcommand{\etaprkpd}{\mbox{$\etapr_{\eta\pi\pi}K^+$}}
\newcommand{\etaprkprg}{\mbox{$\etapr_{\rho\gamma}K^+$}}
\newcommand{\etaprkpfv}{\mbox{$\etapr_{5\pi}K^+$}}
\newcommand{\etaprkz}{\mbox{$\etapr K^0$}}
\newcommand{\etaprkzd}{\mbox{$\etapr_{\eta\pi\pi} K^0$}}
\newcommand{\etaprkzrg}{\mbox{$\etapr_{\rho\gamma} K^0$}}
\newcommand{\etaprpi}{\mbox{$\etapr\pi^+$}}
\newcommand{\etaprpid}{\mbox{$\etapr_{\eta\pi\pi}\pi^+$}}
\newcommand{\etaprpirg}{\mbox{$\etapr_{\rho\gamma}\pi^+$}}
\newcommand{\etaprpifv}{\mbox{$\etapr_{5\pi}\pi^+$}}
\newcommand{\etaprh}{\mbox{$\etapr h^+$}}
\newcommand{\etaprpiz}{\mbox{$\etapr\piz$}}
\newcommand{\etaprpizepp}{\mbox{$\etapr_{\eta\pi\pi}\piz$}}
\newcommand{\etaprpizrg}{\mbox{$\etapr_{\rho\gamma}\piz$}}
\newcommand{\etaprkstz}{\mbox{$\etapr K^{*0}$}}
\newcommand{\etaprkstzd}{\mbox{$\etapr_{\eta\pi\pi} K^{*0}$}}
\newcommand{\etaprkstp}{\mbox{$\etapr K^{*+}$}}
\newcommand{\etaprkstpd}{\mbox{$\etapr_{\eta\pi\pi} K^{*+}_{K^+\piz}$}}
\newcommand{\etaprkstpkz}{\mbox{$\etapr_{\eta\pi\pi} K^{*+}_{K^0\pi^+}$}}
\newcommand{\etaprrhoz}{\mbox{$\etapr\rho^0$}}
\newcommand{\etaprrhozd}{\mbox{$\etapr_{\eta\pi\pi}\rho^0$}}
\newcommand{\etaprrhop}{\mbox{$\etapr\rho^+$}}
\newcommand{\etaprrhopd}{\mbox{$\etapr_{\eta\pi\pi}\rho^+$}}
\newcommand{\etapreta}{\mbox{$\etapr\eta$}}
\newcommand{\etapretagg}{\mbox{$\etapr_{\eta\pi\pi}\eta_{\gaga}$}}
\newcommand{\etapretathrp}{\mbox{$\etapr_{\eta\pi\pi}\eta_{3\pi}$}}
\newcommand{\etapretarg}{\mbox{$\etapr_{\rho\gamma}\eta_{\gaga}$}}
\newcommand{\etapretargtp}{\mbox{$\etapr_{\rho\gamma}\eta_{3\pi}$}}
\newcommand{\etapretapr}{\mbox{$\etapr\etapr$}}
\newcommand{\etapretaprd}{\mbox{$\etapr_{\eta\pi\pi}\etapr_{\eta\pi\pi}$}}
\newcommand{\etapretaprrg}{\mbox{$\etapr_{\eta\pi\pi}\etapr_{\rho\gamma}$}}
\newcommand{\Betaprk}{\mbox{$B\ra\etapr K$}}
\newcommand{\Betaprkp}{\mbox{$B^+\ra\etapr K^+$}}
\newcommand{\Betaprkz}{\mbox{$B^0\ra\etapr K^0$}}
\newcommand{\Betaprks}{\mbox{$B^0\ra\etapr K_S^0$}}
\newcommand{\Betaprpi}{\mbox{$B^+\ra\etapr\pi^+$}}
\newcommand{\Betaprpiz}{\mbox{$B^0\ra\etapr\piz$}}
\newcommand{\Betaprkst}{\mbox{$B\ra\etapr K^*$}}
\newcommand{\Betaprkstz}{\mbox{$B^0\ra\etapr K^{*0}$}}
\newcommand{\Betaprkstp}{\mbox{$B^+\ra\etapr K^{*+}$}}
\newcommand{\Betaprrhoz}{\mbox{$B\ra\etapr\rho^0$}}
\newcommand{\Betaprrhop}{\mbox{$B^+\ra\etapr\rho^+$}}
\newcommand{\Betapreta}{\mbox{$B^0\ra\etapr\eta$}}
\newcommand{\Betapretapr}{\mbox{$B\ra\etapr\etapr$}}

\newcommand{\etak}{\mbox{$\eta K^+$}}
\newcommand{\etakgg}{\mbox{$\eta_{\gaga} K^+$}}
\newcommand{\etakthrp}{\mbox{$\eta_{3\pi} K^+$}}
\newcommand{\etapi}{\mbox{$\eta\pi^+$}}
\newcommand{\etapigg}{\mbox{$\eta_{\gaga}\pi^+$}}
\newcommand{\etapithrp}{\mbox{$\eta_{3\pi}\pi^+$}}
\newcommand{\etapiz}{\mbox{$\eta\piz$}}
\newcommand{\etapizgg}{\mbox{$\eta_{\gaga}\piz$}}
\newcommand{\etapizthrp}{\mbox{$\eta_{3\pi}\piz$}}
\newcommand{\etakz}{\mbox{$\eta K^0$}}
\newcommand{\etakzgg}{\mbox{$\eta_{\gaga} K^0$}}
\newcommand{\etakzthrp}{\mbox{$\eta_{3\pi} K^0$}}
\newcommand{\etaeta}{\mbox{$\eta\eta$}}
\newcommand{\etaetagg}{\mbox{$\eta_{\gaga}\eta_{\gaga}$}}
\newcommand{\etaetathrp}{\mbox{$\eta_{\gaga}\eta_{3\pi}$}}
\newcommand{\etaetasixp}{\mbox{$\eta_{3\pi}\eta_{3\pi}$}}
\newcommand{\etakstz}{\mbox{$\eta K^{*0}$}}
\newcommand{\etakstzgg}{\mbox{$\eta_{\gaga} K^{*0}$}}
\newcommand{\etakstzthrp}{\mbox{$\eta_{3\pi} K^{*0}$}}
\newcommand{\etakstp}{\mbox{$\eta K^{*+}$}}
\newcommand{\etakstpgg}{\mbox{$\eta_{\gaga} K^{*+}_{K^+\piz}$}}
\newcommand{\etakstpthrp}{\mbox{$\eta_{3\pi} K^{*+}_{K^+\piz}$}}
\newcommand{\etakstpggkz}{\mbox{$\eta_{\gaga} K^{*+}_{K^0\pi^+}$}}
\newcommand{\etakstpthrpkz}{\mbox{$\eta_{3\pi} K^{*+}_{K^0\pi^+}$}}
\newcommand{\etarhoz}{\mbox{$\eta \rho^0$}}
\newcommand{\etarhozgg}{\mbox{$\eta_{\gaga} \rho^0$}}
\newcommand{\etarhozthrp}{\mbox{$\eta_{3\pi} \rho^0$}}
\newcommand{\etarhop}{\mbox{$\eta \rho^+$}}
\newcommand{\etarhopgg}{\mbox{$\eta_{\gaga} \rho^+$}}
\newcommand{\etarhopthrp}{\mbox{$\eta_{3\pi} \rho^+$}}
\newcommand{\Betak}{\mbox{$B\ra\eta K$}}
\newcommand{\Betakp}{\mbox{$B^+\ra\eta K^+$}}
\newcommand{\Betapi}{\mbox{$B^+\ra\eta\pi^+$}}
\newcommand{\Betapiz}{\mbox{$B^0\ra\eta\piz$}}
\newcommand{\Betakz}{\mbox{$B^0\ra\eta K^0$}}
\newcommand{\Betaeta}{\mbox{$B^0\ra\eta\eta$}}
\newcommand{\Betakstz}{\mbox{$B^0\ra\eta K^{*0}$}}
\newcommand{\Betakstp}{\mbox{$B^+\ra\eta K^{*+}$}}
\newcommand{\Betarhoz}{\mbox{$B^0\ra\eta \rho^0$}}
\newcommand{\Betarhop}{\mbox{$B^+\ra\eta \rho^+$}}

\newcommand{\ksppd}{\mbox{$K^{0}\rightarrow K_S\rightarrow\pi^+\pi^-$}}
\newcommand{\kstzd}{\mbox{$K^{*0}\ra\K^+\pi^-$}}
\newcommand{\kstpd}{\mbox{$K^{*+}\ra\K^+\piz$}}
\newcommand{\kstpkz}{\mbox{$K^{*+}\ra\K^0\pi^+$}}

\newcommand{\Bomegapi}{\mbox{$B^+\rightarrow\omega\pi^+$}}
\newcommand{\Bomegapiz}{\mbox{$B^0\ra\omega\pi^0$}}
\newcommand{\Bomegak}{\mbox{$B\rightarrow\omega K$}}
\newcommand{\Bomegakp}{\mbox{$B^+\rightarrow\omega K^+$}}
\newcommand{\Bomegakz}{\mbox{$B^0\rightarrow\omega K^0$}}
\newcommand{\Bomegah}{\mbox{$B^+\rightarrow\omega h^+$}}
\newcommand{\Bomegakpi}{\mbox{$B^+\rightarrow\omega K^+/\pi^+$}}
\newcommand{\Bomegaetapr}{\mbox{$B^0\rightarrow\omega\etapr$}}
\newcommand{\Bomegaeta}{\mbox{$B^0\rightarrow\omega\eta$}}
\newcommand{\Bomegarhoz}{\mbox{$B^0\rightarrow\omega \rho^0$}}
\newcommand{\Bomegarhop}{\mbox{$B^+\rightarrow\omega \rho^+$}}
\newcommand{\Bomegakstz}{\mbox{$B^0\rightarrow\omega K^{*0}$}}
\newcommand{\Bomegakstp}{\mbox{$B^+\rightarrow\omega K^{*+}$}}
\newcommand{\Bomegaomega}{\mbox{$B^0\ra\omega\omega$}}

\newcommand{\omegak}{\mbox{$\omega K^+$}}
\newcommand{\omegakz}{\mbox{$\omega K^0$}}
\newcommand{\omegapi}{\mbox{$\omega\pi^+$}}
\newcommand{\omegapiz}{\mbox{$\omega\pi^0$}}
\newcommand{\omegah}{\mbox{$\omega h^+$}}
\newcommand{\omegaetapr}{\mbox{$\omega\etapr$}}
\newcommand{\omegaetaprd}{\mbox{$\omega\etapr_{\eta\pi\pi}$}}
\newcommand{\omegaetaprrg}{\mbox{$\omega\etapr_{\rho\gamma}$}}
\newcommand{\omegaeta}{\mbox{$\omega\eta$}}
\newcommand{\omegaetagg}{\mbox{$\omega\eta_{\gaga}$}}
\newcommand{\omegaetathrp}{\mbox{$\omega\eta_{3\pi}$}}
\newcommand{\omegakstz}{\mbox{$\omega K^{*0}$}}
\newcommand{\omegakstp}{\mbox{$\omega K^{*+}$}}
\newcommand{\omegakstpd}{\mbox{$\omega K^{*+}_{K^+\piz}$}}
\newcommand{\omegakstpkz}{\mbox{$\omega K^{*+}_{K^0\pi^+}$}}
\newcommand{\omegarhoz}{\mbox{$\omega \rho^0$}}
\newcommand{\omegarhop}{\mbox{$\omega \rho^+$}}
\newcommand{\omegaomega}{\mbox{$\omega\omega$}}

\newcommand{\Bphik}{\mbox{$B^+\ra\phi K^+$}}
\newcommand{\Bphikz}{\mbox{$B^0\ra\phi K^0$}}
\newcommand{\Bphiks}{\mbox{$B^0\ra\phi K_S^0$}}
\newcommand{\Bphipi}{\mbox{$B^+\ra\phi\pi^+$}}
\newcommand{\Bphipiz}{\mbox{$B^0\ra\phi\pi^0$}}
\newcommand{\Bphietapr}{\mbox{$B^0\ra\phi\etapr$}}
\newcommand{\Bphieta}{\mbox{$B^0\ra\phi\eta$}}
\newcommand{\Bphikstz}{\mbox{$B^0\ra\phi K^{*0}$}}
\newcommand{\Bphikstp}{\mbox{$B^+\ra\phi K^{*+}$}}
\newcommand{\Bphikstpd}{\mbox{$B^+\ra\phi K^{*+}(K^{*+}\ra K^+\piz$}}
\newcommand{\Bphikstpkz}{\mbox{$B^+\ra\phi K^{*+} (K^{*+}\ra K^0\pi^+$}}
\newcommand{\Bphikst}{\mbox{$B\ra\phi K^*$}}
\newcommand{\Bphirhoz}{\mbox{$B^0\ra\phi \rho^0$}}
\newcommand{\Bphirhop}{\mbox{$B^+\ra\phi \rho^+$}}
\newcommand{\Bphiomega}{\mbox{$B^0\ra\phi\omega$}}
\newcommand{\Bphiphi}{\mbox{$B^0\ra\phi\phi$}}

\newcommand{\phik}{\mbox{$\phi K^+$}}
\newcommand{\phikz}{\mbox{$\phi K^0$}}
\newcommand{\phipi}{\mbox{$\phi\pi^+$}}
\newcommand{\phipiz}{\mbox{$\phi\pi^0$}}
\newcommand{\phih}{\mbox{$\phi h^+$}}
\newcommand{\phietapr}{\mbox{$\phi\etapr$}}
\newcommand{\phietaprd}{\mbox{$\phi\etapr_{\eta\pi\pi}$}}
\newcommand{\phietaprrg}{\mbox{$\phi\etapr_{\rho\gamma}$}}
\newcommand{\phieta}{\mbox{$\phi\eta$}}
\newcommand{\phietagg}{\mbox{$\phi\eta_{\gaga}$}}
\newcommand{\phietathrp}{\mbox{$\phi\eta_{3\pi}$}}
\newcommand{\phikstz}{\mbox{$\phi K^{*0}$}}
\newcommand{\phikstzd}{\mbox{$\phi K^{*0}_{K^+\pi^-}$}}
\newcommand{\phikstzkz}{\mbox{$\phi K^{*0}_{K^0\piz}$}}
\newcommand{\phikstp}{\mbox{$\phi K^{*+}$}}
\newcommand{\phikstpd}{\mbox{$\phi K^{*+}_{K^+\piz}$}}
\newcommand{\phikstpkz}{\mbox{$\phi K^{*+}_{K^0\pi^+}$}}
\newcommand{\phikst}{\mbox{$\phi K^*$}}
\newcommand{\phirhoz}{\mbox{$\phi \rho^0$}}
\newcommand{\phirhop}{\mbox{$\phi \rho^+$}}
\newcommand{\phiomega}{\mbox{$\phi\omega$}}
\newcommand{\phiphi}{\mbox{$\phi\phi$}}

\newcommand{\etaprinc}{\mbox{$B\ra\etapr X_S$}}
\def\babar{{\sl B}$\scriptstyle\sl A${\sl B}$\scriptstyle\sl AR$}

\pagestyle{empty}
\large
\vbox{
      \hbox{\hspace{5.0in}               }
      \hbox{\hspace{5.0in}               }
      \hbox{\hspace{5.0in} COLO--HEP--395}
      \hbox{\hspace{5.0in} SLAC-PUB-7796}
      \hbox{\hspace{5.0in} HEPSY 98-1}
      \hbox{\hspace{5.0in} April 23, 1998 }
}
\vskip 1.0in
\centerline{\hskip 1.0in\LARGE PENGUIN DECAYS OF B MESONS}
\vskip 1.0in

\centerline{\hskip 1.0in Karen Lingel
\footnote{Work supported by Department of Energy contract
DE-AC03-76SF00515} }
\centerline{\hskip 1.0in Stanford Linear Accelerator Center}
\centerline{\hskip 1.0in Stanford University, Stanford, CA 94309}
\centerline{}
\centerline{\hskip 1.0in and}
\centerline{}
\centerline{\hskip 1.0in Tomasz Skwarnicki
\footnote{Work supported by National Science Foundation contract
PHY 9807034} }
\centerline{\hskip 1.0in Department of Physics, Syracuse University}
\centerline{\hskip 1.0in Syracuse, NY 13244}
\centerline{}
\centerline{\hskip 1.0in and}
\centerline{}
\centerline{\hskip 1.0in James G. Smith
\footnote{Work supported by Department of Energy contract
DE-FG03-95ER40894} }
\centerline{\hskip 1.0in Department of Physics, University of Colorado}
\centerline{\hskip 1.0in Boulder, CO 80309}
\vskip 1.7in
\centerline{\hskip 1.0in Submitted to {\it Annual Reviews of Nuclear and
Particle Science}}
\normalsize

\newpage
\pagestyle{headings}
\title{PENGUIN DECAYS OF B MESONS}
\markboth{\rm LINGEL \& SKWARNICKI \& SMITH}{\rm PENGUIN DECAYS}

\author{Karen Lingel {Stanford Linear Accelerator Center}\\
Tomasz Skwarnicki {Department of Physics, Syracuse University}\\
James G. Smith {Department of Physics, University of Colorado}}


\begin{abstract}
Penguin, or loop, decays of $B$ mesons induce effective flavor-changing 
neutral currents, which are forbidden at tree level in the Standard Model.
These decays give special insight into the CKM matrix and
 are sensitive to non-standard model effects.
In this review, we give a historical and theoretical introduction
to penguins and a description of the various types of penguin processes: 
electromagnetic, electroweak, and gluonic.   We review the
experimental searches for penguin decays, including the measurements
of the electromagnetic penguins $b \rightarrow s \gamma$ and
$B \rightarrow K^*\gamma$ and gluonic penguins $B \rightarrow K\pi$,
$B^+\rightarrow \omega K^+$ and $B\rightarrow \eta' K$, and their 
implications for the Standard Model and New Physics.  We conclude
by exploring the future prospects for penguin physics.

\end{abstract}
\maketitle

\section{INTRODUCTION}
\begin{figure}[htbp]
\centerline{\psfig{figure=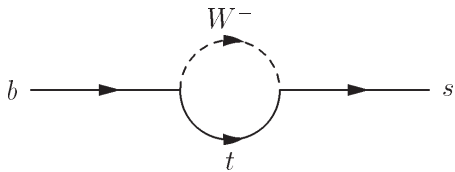,height=1.0in}}
\caption{$b\rightarrow s$ loop or ``penguin'' diagram.  Of course,
in order to conserve energy and momentum, an additional particle 
is understood to be emitted in the transition.}
\label{fig:simplepenguin}
\end{figure}

In the Standard Model, flavor-changing neutral currents (FCNC) are forbidden,
for example, there is no direct coupling between
the $b$ quark and the $s$ or $d$ quarks.  
Effective FCNC are induced by one-loop, or
``penguin", diagrams, where a quark emits and re-absorbs a
$W$ thus changing flavor twice, as in the
$b \ra t \ra s$ transition depicted in Fig.~\ref{fig:simplepenguin}.
Penguin decays have become increasingly appreciated
in recent years.  These loop diagrams with their interesting combinations
of CKM matrix elements give insight into the Standard Model.  In addition,
they are quite sensitive to new physics.

The weak couplings of the quarks are given by the Cabibbo-Kobayashi-Maskawa 
(CKM) \cite{ref:CKM} matrix of complex amplitudes:
\begin{equation}
V_{\rm CKM} = \left( \begin{array}{ccc}
V_{ud} & V_{us} & V_{ub} \\
V_{cd} & V_{cs} & V_{cb} \\
V_{td} & V_{ts} & V_{tb}
\end{array}\right)
\label{eq:CKM}
\end{equation}
For the Standard Model with three generations, the CKM matrix can be 
described completely by three Euler-type angles, and a complex phase.
In the Wolfenstein parameterization \cite{ref:wolfenstein}, 
the CKM matrix is approximated in terms of four real, independent, parameters,
$(\lambda,A,\rho,\eta),$ which makes clear the
hierarchical structure of the elements:
\begin{equation}
V_{\rm CKM} \approx \left( \begin{array}{ccc}
1 - {1\over 2}\lambda^2   & \lambda      & A\lambda^3(\rho-i\eta) \\
-\lambda                  & 1-{1\over 2}\lambda^2  & A\lambda^2 \\
A\lambda^3(1-\rho-i\eta)  & -A\lambda^2  & 1
		     \end{array}\right)
\label{eq:wolfenstein}
\end{equation}
Since $\lambda\equiv\sin\theta_C$, the well-known Cabibbo angle, is small 
($\lambda \approx 0.22$), this parameterization shows that the off-diagonal 
elements are small and the diagonal elements are close to 1.
The complex phase, which may be responsible for $CP$ violation
in the Standard Model, has been
assigned to the corner elements in this parameterization.

\subsection{History of Penguins}
The curious name penguin goes back to a game of darts in a Geneva pub in
the summer of 1977, involving theorists John Ellis, Mary K. Gaillard, Dimitri
Nanopoulos and Serge Rudaz (all then at CERN) and experimentalist
Melissa Franklin (then a Stanford student, now a Harvard professor).
Somehow the telling of a joke about penguins evolved to the resolution
that the loser of the dart game would use the word penguin in their next
paper.  It seems that Rudaz spelled Franklin at some point, beating Ellis
(otherwise we might now have a detector named penguin);
sure enough the seminal 1977 paper
on loop diagrams in $B$ decays \cite{ref:ellisB} refers to such diagrams as
penguins.  This paper contains a whimsical acknowledgment to Franklin
for ``useful discussions" \cite{dartgame}.


Prior to 1975, the loop diagram had been neglected.
Vainshtein, Zakharov, and Shifman \cite{ref:shifman}
discovered the importance of the penguin diagrams,
and suggested that penguins were responsible for the
enhancement of the $\Delta I = 1/2$ amplitude compared to
the $\Delta I = 3/2$ amplitude in
weak $K\rightarrow \pi\pi$ decays.  Penguins were considered in the
$B$ system by Ellis, {\it et al.} \cite{ref:ellisB},
and determined to be small compared to $b \rightarrow c$ amplitudes.
However,
Guberina, Peccei, and R\"uckl \cite{ref:guberina} later pointed out
that the penguin $b \ra s$ decays could have a rate as large as
tree-level $b \ra u$ decays.  From Eqn.~\ref{eq:wolfenstein},
we see that penguin decays involve $|V_{tb}V_{ts}| \propto \lambda^2$,
whereas $b \ra u$ transitions involve $|V_{ub}V_{ud}| \propto \lambda^3$.
In 1979, the role of penguins in $CP$ violation was pointed out by
Bander, Silverman and Soni \cite{ref:bander} who
showed that interference between penguin diagrams and tree-level
diagrams could give large $CP$ asymmetries in $B$ decays.
In 1982, Eilam \cite{ref:eilam}
added the gluonic penguin $b \rightarrow s g$ to the
inclusive penguin rate.  Later large QCD corrections
to the radiative penguin $b \rightarrow s \gamma$ \cite{ref:QCDcorr}
were calculated.
These corrections
increased the predicted $b \rightarrow s \gamma$ rate by a factor
of $\sim 3$,
enough to make experimentalists sanguine about
measuring the rate.  The inclusive gluonic penguin $b \rightarrow s g^*$
was later clarified to include the time-like $b\rightarrow sq\bar q$,
and space-like $b\bar q \rightarrow s\bar q$ as well as the light-like
$b \rightarrow s g$ \cite{ref:hou}, which increased the gluonic penguin
rate \cite{ref:lenz}.

In 1993, the CLEO Collaboration published the first evidence for
electromagnetic penguins (Sec.~\ref{sec:Kstgam})
in the channel $B \rightarrow K^*\gamma$.
In 1994 they also measured the inclusive $b \rightarrow s \gamma$
rate (Sec.~\ref{sec:bsgam})
which was in good agreement with the theoretical predictions.
In 1997, after many hints in several experiments, CLEO found first
evidence for gluonic penguin decays (Sec.~\ref{sec:glupeng}).

\subsection{The Importance Of Penguins}
Although $s\ra u$ 
loop diagrams are important in $K$ decays, those decays
are typically dominated by large non-perturbative effects.
A notable exception is $K^+ \rightarrow \pi^+\nu\bar\nu$ (charge
conjugate states are implied throughout this review).  This decay is 
expected to be dominated by electroweak penguins (Sec.~\ref{sec:EWintro})
and could eventually provide a measurement of $|V_{td}|$.
Penguin processes are also possible in
$c$ and $t$ decays, but
these particles have the CKM-favored decays $c \rightarrow s$
and $t \rightarrow b$  accessible to them.
Since the $b$ quark has no kinematically-allowed CKM-favored decay
(Eqn.~\ref{eq:wolfenstein}), the relative
importance of the penguin decay is greater.
The mass of the top quark, the main contributor to the loop,
is large, and the coupling of the $b$ quark to the $t$ quark, $|V_{tb}|$,
is very close to unity, both strengthening the effect of the penguin.
The $b \rightarrow s$ ($b\rightarrow d$)
penguin transition is sensitive to $|V_{ts}|$ ($|V_{td}|$) which
will be extraordinarily difficult to measure in top decay.  Information
from the penguin decay will complement information on $|V_{ts}|$ and
$|V_{td}|$ from $B_s$--$\bar B_s$ and \BBmix.

Since the Standard Model (SM) loops involve the heaviest known particles 
($t,W,Z$), rates for these processes are very sensitive to 
non-SM extensions with heavy charged Higgs or supersymmetric particles.
Therefore, measurements of loop processes constitute the most sensitive
low energy probes for such extensions to the Standard Model.

\section{THEORY}

In sections~\ref{sec:EMintro} through \ref{sec:otherhadintro} we 
give general descriptions of various kinds of penguins.  
Details of the effective Hamiltonian theory are given in
section \ref{sec:efftheory}.
Sections \ref{sec:CPintro} and \ref{sec:NP} discuss two
important topics in penguin decay: $CP$ Violation and 
New Physics.

\subsection{\it Electromagnetic Penguins}
\label{sec:EMintro}
\begin{figure}[htbp]
\centerline{\psfig{figure=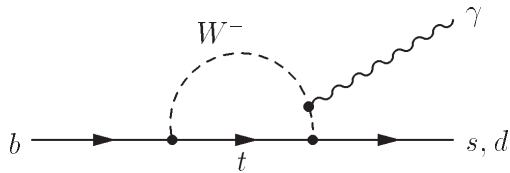,height=1.0in}}
\caption{Feynman diagram for the electromagnetic penguins
$b \ra s \gamma$ and $b \ra d \gamma$.  The photon can be emitted from
the $W$ (shown) or from any of the quarks.}
\label{fig:empenguin}
\end{figure}

In electromagnetic penguin decays such as $b \rightarrow s \gamma$,
a charged particle emits an external
real photon (Fig.~\ref{fig:empenguin}).
The hard photon emitted in these decays is an excellent
experimental signature.
The inclusive rate is dominated by short distance (perturbative)
interactions and can be reliably predicted.  QCD 
corrections enhance the rate and have been calculated precisely.  
Assuming unitarity of the CKM matrix to constrain $|V_{ts}|$ the Standard
Model predicts \cite{bsgammaTh}
${\cal B}(b\rightarrow s \gamma) = (3.5 \pm 0.3) \times 10^{-4}$.
Unfortunately, uncertainties in the hadronization process limit
the ability to predict individual exclusive rates
from first principles of the theory.  Phenomenological predictions
range from
1\% to 40\% \cite{kstargammaTh} for the ratio
$R_{K^*}\equiv{\cal B}(B \ra K^*\gamma)/{\cal B}(b \ra s \gamma)$.

The electromagnetic penguin decay $b \rightarrow d \gamma$ is further
suppressed by $|V_{td}|^2/|V_{ts}|^2$
and gives an alternative to \BBmix\ for extracting $|V_{td}|$.
Experimentally, inclusive $b \rightarrow d \gamma$ has
large backgrounds from the dominant $b \rightarrow s\gamma$
decays which must be rejected using good particle identification or kinematic
separation.

\subsection{Electroweak Penguins}
\label{sec:EWintro}
\begin{figure}[htbp]
\centerline{\psfig{figure=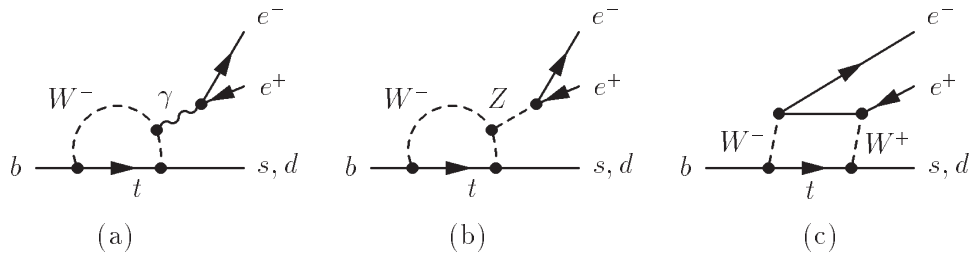,width=13.2cm}}
\caption{(a) Photon penguin (b) $Z^0$ penguin and (c)
box diagrams for the electroweak decay
$b \rightarrow (s,d)\ell^+\ell^-$.  
The diagrams for 
$b \rightarrow (s,d)\nu \bar\nu$ are similar, except that (a) does not
contribute.}
\label{fig:ewpenguin}
\end{figure}

The decay $b \rightarrow s\ell^+\ell^-$ can proceed via an electroweak
penguin diagram where an emitted virtual photon or $Z^0$ produces a
pair of leptons (Fig.~\ref{fig:ewpenguin}a,b).
This decay can also proceed via a box diagram (Fig.~\ref{fig:ewpenguin}c).
The SM prediction for the $b \rightarrow s\ell^+\ell^-$ decay rate is two 
orders of magnitude smaller than the $b \ra s \gamma$ rate \cite{Ali,refAH}.

The rate for $b \rightarrow s \nu \bar \nu$ is enhanced relative to
$b \rightarrow s \ell^+\ell^-$ primarily due to summing the
three neutrino flavors.  These decays are expected to be dominated
by the weak penguin, since neutrinos do not couple to photons.
The predicted rate is only a factor of 10 lower
than for $b \rightarrow s \gamma$ \cite{BurasWaw}.
Unfortunately, the neutrinos escape
detection, making this decay mode difficult to observe.

\subsection{\it Vertical Electroweak Penguins}
\label{sec:VEWintro}
\begin{figure}[htbp]
\centerline{\psfig{figure=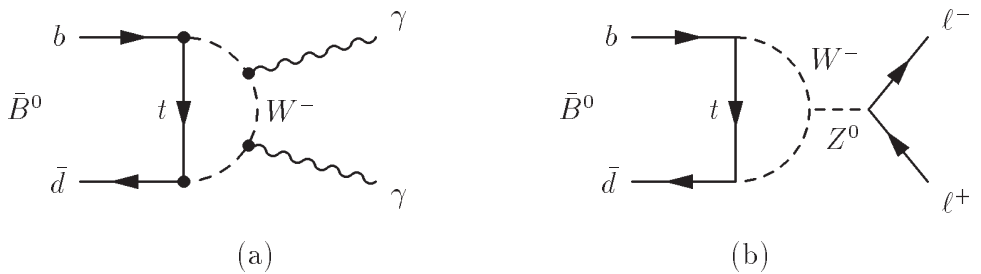,width=13.2cm}}
\caption{Vertical or annihilation penguins:  (a) $B \rightarrow
\gamma \gamma$ and (b) $B\rightarrow \ell^+\ell^-$.}
\label{fig:goofypeng}
\end{figure}

Another category of penguin is the so-called vertical or
annihilation penguin where the penguin loop connects the
two quarks in the $B$ meson (Fig.~\ref{fig:goofypeng}).  These rates
are expected to be highly suppressed in the Standard Model since they
involve a $b \rightarrow d$ transition and
are suppressed by $(f_B/m_B)^2 \approx 2\times 10^{-3}$,
where $f_B$ is the $B$-meson
decay constant which parameterizes the probability that the
two quarks in the $B$ meson will ``find each other'',
 and $m_B$ is the $B$ meson mass.   The
$B \rightarrow \gamma \gamma$ decay is suppressed 
relative to $b \rightarrow s \gamma$ by an additional
$\alpha_{QED}$.  The $B\rightarrow\ell^+\ell^-$ decays are
helicity-suppressed.
Because these decays are so suppressed in the Standard Model, they provide
a good opportunity to look for non-SM effects (Sec.~\ref{sec:NP}).

\subsection{\it Gluonic Penguins}
\label{sec:GLUintro}

\begin{figure}[htbp]
\centerline{\psfig{figure=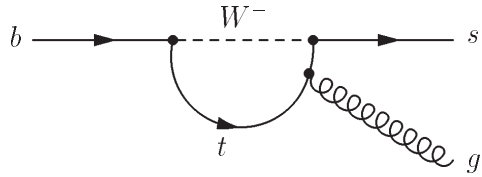,height=1.0in}}
\caption{Feynman diagram for the gluonic penguin
$b \rightarrow s g^*$.  The gluon can be emitted from any of the
quark lines and can be on-shell or off-shell.}
\label{fig:hadpenguin}
\end{figure}

An on- or off-shell gluon can also be emitted from the penguin loop.
(Fig.~\ref{fig:hadpenguin}).  While the on-shell $b \rightarrow s g$ rate
had been calculated to be ${\cal O}(0.1\%)$, the inclusive on- plus off-shell
$b \rightarrow s g^*$ rate includes contributions
from $b \rightarrow sq \bar q$ and $b \rightarrow sgg$ which increase the
inclusive rate to 0.5--1\% \cite{ref:hou,ref:lenz}.
The $b \rightarrow dg^*$ penguin rate is smaller by $|V_{td}/V_{ts}|^2$.
Unfortunately, there are several difficulties associated with gluonic penguins.
There is no good signature for the inclusive $b \rightarrow sg^*$ decay,
unlike the $b \rightarrow s \gamma$ case.  The branching fraction of individual
exclusive gluonic penguin channels is typically quite small and
hadronization effects are difficult to calculate.
In addition, many gluonic penguin final states are accessible via
other diagrams (Sec.~\ref{sec:otherhadintro}) so the gluonic penguin
is difficult to assess.
Thus the ``smoking-gun'' penguin processes such as $B^0 \rightarrow \phi K^0$
that have contributions only from gluonic
penguins are eagerly sought.

\subsection{\it Other Contributions to Hadronic Final States}
\label{sec:otherhadintro}
\begin{figure}[htbp]
\centerline{\psfig{figure=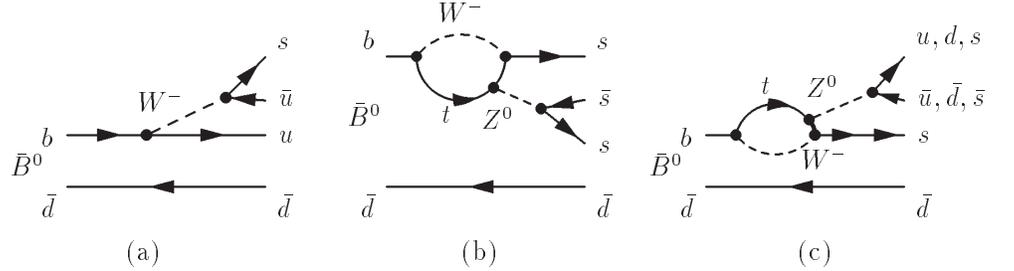,width=13.2cm}}
\caption{Examples of other diagrams which contribute to hadronic
final states: (a) tree-level Cabibbo-suppressed $b \rightarrow u\bar u s$,
(b) electroweak penguin (c) hairpin diagram.}
\label{fig:althad}
\end{figure}

While the gluonic penguin gives rise only to hadronic final states, several
other processes can contribute to the same final states.
One important contribution is from the tree-level 
$b\ra u$ decay.  For example, the $b \rightarrow u\bar u s$
transition (Fig.~\ref{fig:althad}a) and the 
$b \rightarrow s g^*$ penguin transition both contribute to
$B^0\rightarrow K^+\pi^-$.  However, the $b \rightarrow u\bar u s$ 
transition is Cabibbo-suppressed, so the penguin process
is expected to dominate.  On the other hand, in $B \rightarrow \pi^+\pi^-$
for example, the small 
$b \rightarrow d g^*$ contribution is expected to be dominated 
by the non-Cabibbo-suppressed tree-level $b \ra u \bar u d$ transition.  
In general, most decays to hadronic final states
with $\phi$ mesons or non-zero net strangeness are expected to be dominated by
gluonic penguins and hadronic final states with
zero net strangeness are expected to be dominated by 
tree-level $b \rightarrow u$.

Electroweak penguins also contribute to
hadronic final states.  Every gluonic penguin diagram
can be converted to an electroweak penguin by replacing the gluon with a
$Z^0$ or $\gamma$ (Fig.~\ref{fig:althad}b).  Electroweak penguins
with internal $Z^0$ or $\gamma$ emission are
suppressed relative to the corresponding strong gluonic penguin.
In the hairpin process (Fig.~\ref{fig:althad}c) the gluon,
$Z^0$, or $\gamma$ is emitted externally and subsequently forms a meson
(similar to the leptonic electroweak penguins in
Figs.~\ref{fig:ewpenguin}a and b).
External gluon emission is OZI-suppressed \cite{ref:OZI}:
the color-octet gluon has difficulties forming
a color-singlet meson!  These hairpin processes, such as 
$B\rightarrow \phi \pi$,
are expected to be dominated by electroweak penguins.
A possible exception involves decays such as \Betaprk, where it has been
suggested that a gluonic-hairpin diagram could be significant
(Sec.~\ref{sec:etaomega}).

The vertical electroweak penguin diagram,
similar to Fig.~\ref{fig:goofypeng}b,
with the lepton pair replaced by a di-quark pair, is highly suppressed
and is important only for decays such as $B^0\ra\phi\phi$, where no
other diagrams contribute.

In the annihilation diagram the $b$ and
$\bar u$ quarks in a $B^-$ meson annihilate to form a virtual $W^-$.
The annihilation diagram is suppressed by $|V_{ub}|$ and by 
$f_B/m_B$ and is expected to be 
mostly negligible.
In the exchange diagram, a $b \rightarrow u$ transition and
a $\bar d\rightarrow \bar u$ transition occur simultaneously
via the exchange of a
$W$ between the $b$ and $\bar d$ quarks in a $\bar B^0$ meson.
The exchange process is also suppressed by $|V_{ub}|$ and $f_B/m_B$, 
and is also 
expected to be negligible, except in decays such as $B^0\rightarrow K^+K^-$
where no favored diagrams contribute.

\subsection{Effective Theory of Penguin Decays}
\label{sec:efftheory}

\begin{figure}[tbhp]
\centerline{\psfig{figure=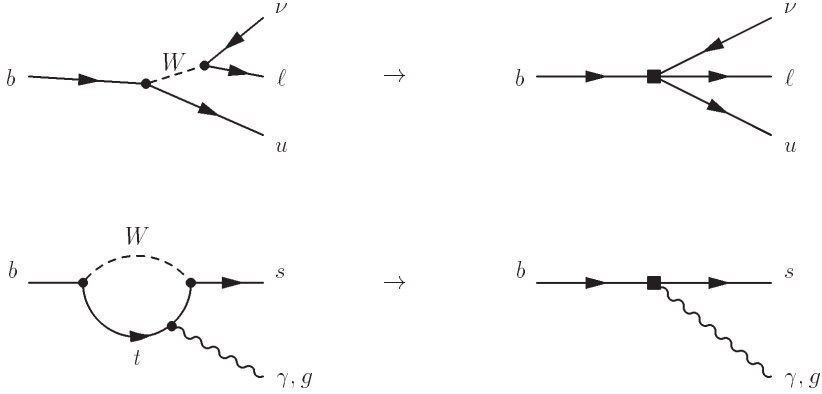,width=11.3cm}}
\caption{Derivation of effective low energy theory from high energy
         Quantum Field Theory. 
         Derivation of the Fermi theory of the $W$ exchange diagram
         for semileptonic $b\to u\,l\nu$ decay (\lq\lq $\beta-$decay'')
         is shown at the top.  Generalization
         for the loop processes $b\to s\,\gamma$ and $b\to s\,g$
         are shown at the bottom \cite{Falk}. \label{fig:Fermi}}
\end{figure}

At high energy scales, $\mu\sim M_W\sim80$ GeV, quark decays are 
governed by Feynman diagrams such as those depicted in the previous sections.
To obtain an effective low energy theory relevant 
for scales $\mu\sim m_b\sim5$ GeV, heavy degrees of freedom must be
integrated out to obtain an effective coupling for point-like 
interactions of initial and
final state particles \cite{EffectiveTheory}.
For semileptonic decays (e.g. the familiar $\beta-$decays in nuclear physics), 
this integration corresponds to derivation of the Fermi theory of 
point-like four-fermion interactions from Electroweak Quantum Field Theory.
The effective theory relevant for penguin decays is obtained 
by generalization of the Fermi theory, as depicted in Fig. \ref{fig:Fermi}.
The heavy degrees of freedom in loop decays are $W$, $Z^0$ and $t$.
After the integration they don't appear explicitly in the theory, but
their effects are hidden in the effective gauge coupling constants, 
running masses and, most importantly, 
in the so-called Wilson coefficients ($C_i$) 
describing the effective strength of the local operators ($Q_i$)
generated by electroweak and strong interactions.
The operators can be grouped into three categories:
$i=1,2$ --- current-current operators (Fig.~\ref{fig:althad}a);
$i=3,\dots,6$ --- gluonic penguin operators (related to diagrams
                  in Fig.~\ref{fig:hadpenguin});
$i=7,\dots,10$ --- electro-weak penguin operators 
                  (Figs.~\ref{fig:empenguin} and \ref{fig:ewpenguin}).
The effective Hamiltonian for $b\to s$ penguin decays has the following form:
$$ {\cal H}_{\rm eff} = - \frac{4\,G_F}{\sqrt{2}} \, V_{tb} V^*_{ts} \,
 \sum_{i=1}^{10} \, C_i(\mu) \, Q_i(\mu) $$
Technically, the calculations are performed at a high energy scale
$\mu\sim M_W$, and then evolved to a low energy scale $\mu\sim m_b$
using renormalization group equations. 
This evolution mixes the operators: 
$C_i(\mu)=\sum_j U_{ij}(\mu,M_W)C_j(M_W)$.
The renormalization guarantees that the $\mu$ dependence of $C_i$
is canceled by the $\mu$ dependence of $Q_i$, thus any
observable quantity should not depend on the renormalization scale $\mu$.
However, since the calculations are performed perturbatively,
the truncation of the perturbative series induces $\mu$ dependence of
the theoretical predictions, which often dominates the theoretical
uncertainty. Higher order terms must be included to minimize
the $\mu$ dependence.

Even though renormalization mixes the operators, 
specific processes are mostly sensitive to a small subset of
Wilson coefficients. For example: $b\to s\,\gamma$ to $C_7$,
$b\to s\,l^+l^-$ to $C_7$, $C_9$ and $C_{10}$.

Since extensions of the Standard Model contribute additional diagrams at
the high energy scale, they modify the values of the Wilson coefficients
in the effective low energy theory.

While Wilson coefficients represent short distance (i.e. high energy)
electro-weak and strong interactions, 
the  operator elements $<\!X| Q_i |B\!>$ 
are influenced by long
distance (i.e. low energy) strong interactions
(here $|B\!>$ represents the $B$ meson and  $|X\!>$ its decay mode). 
Therefore, unlike the Wilson coefficients,
the operator elements cannot be obtained perturbatively due
to the confining nature of strong interactions at large distances.

Fortunately, when $|X\!>$ represents an inclusive final state,
expansion in powers of $1/m_b$ shows that to leading order
$<\!X| Q_i |B\!>\approx<\!s| Q_i |b\!>$, where
$<\!s| Q_i |b\!>$ is an operator element for free quarks
which can be easily calculated. 
The first non-perturbative corrections are of second order \cite{mbexp}, 
${\cal O}(1/{m_b}^2)$, and are small, thanks to the heavy $b$ quark mass.
Experimentalists must sum over all possible hadronic final states
to determine an inclusive rate. For example, 
${\cal B}(B\to X_s\,\gamma)\approx {\cal B}(b\to s\,\gamma)$, where $X_s$ 
represents a collection of charmless hadrons with net strangeness -1.

When $|X\!>$ is an exclusive final state 
it is difficult to obtain $<\!X| Q_i |B\!>$
from first principles.
Numerical treatment of strong quantum fields (Lattice QCD)
has been useful so far only for the simplest cases
in which part of the final state is non-hadronic,
$|X\!>=| h\,L\!>$,
such as $B\to K^*\gamma$.
Since leptons and photons are not involved in long distance interactions,
the operator element factorizes into hadronic and 
non-hadronic currents, $<\!h\,L| Q_i |B\!> = 
<\!h| J_1 |B\!><\!L| J_2 |0\!>$.
The latter can be written explicitly.
Non-leptonic final states are the most difficult to calculate.
Heavy Quark Effective Theory, which is so useful for describing
ordinary $b\to c$ decays, is of little use for $b\to s,d$ decays, since
the final state quarks are light.  Phenomenological models 
used to predict rates for gluonic penguin decays make many
assumptions, the accuracy of which is often difficult to assess, and
the scope of the predictions is usually limited to two-body final states.
The approach which has often been employed \cite{aliGreub,dean,kps} 
is based on factorization \cite{Feynman}.
The rationale of factorization in hadronic decays
lies in the phenomenon of color-transparency
\cite{Bjorken}, in which one expects intuitively that a pair of fast-moving
quarks (in two-body decays $E_{h}\sim m_B/2$)
in a color-singlet state effectively decouples from 
soft gluons. Therefore, long distance 
final state interactions (FSI) can be neglected.
Short distance FSI mediated by hard gluon exchanges can
be included perturbatively. 
The latter are important for predictions of the strong phases
which make direct $CP$ violation possible.
With the factorization ansatz, the matrix elements
$<\! h_1 h_2 | Q_i |B\!>$ can be expressed as a product of two 
hadronic currents:
$<\! h_1 | J_1 |B\!><\!h_2 | J_2 |0\!>$. 
This involves both the matrix elements of
the singlet-singlet and octet-octet currents $J_i$.
Since the octet-octet matrix elements are not 
directly measured they are usually 
discarded. To compensate for this, the effective strengths of the 
singlet-singlet current matrix elements are renormalized by replacing the
inverse of the number of colors ($1/N_c$) by a phenomenological 
color parameter $\xi$.  While the assumption of factorization works well for 
tree-level $b\to c$ decays \cite{HonBrow,BSWetc}, 
its applicability to penguin and $b\to u$
decays needs to be verified. Also it is not clear
if a single parameter $\xi$ will suffice to 
incorporate non-perturbative effects in all
ten operators of the effective Hamiltonian.

\subsection{\it $CP$ Violation}
\label{sec:CPintro}
\begin{figure}
\centerline{\psfig{figure=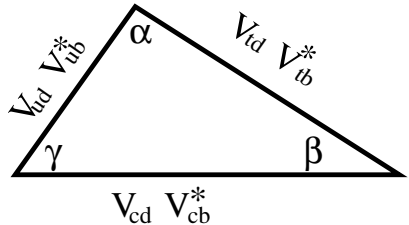,height=3cm}}
\caption{The Unitarity Triangle.}
\label{fig:UT}
\end{figure}

Unitarity of the CKM matrix (Eqn.~\ref{eq:CKM}) leads to several
constraints, the most interesting of which is the orthogonality 
of the first and third columns:
\begin{equation}
  V_{ud}V_{ub}^* + V_{cd}V_{cb}^* + V_{td}V_{tb}^* = 0
\label{eq:UT}
\end{equation}
Eqn.~\ref{eq:UT} defines a triangle in the complex plane (Fig.~\ref{fig:UT})
called the Unitarity Triangle.  The lengths of the sides of the Unitarity
Triangle are given by the magnitudes of the CKM matrix elements.
The angles are given by the phase of the CKM matrix elements:
$\beta$ is the phase of $V_{ub}$, $\gamma$ is the phase of $V_{td}$, and
$\alpha \equiv \pi - \beta - \gamma$.
Physicists wish to measure the sides of the triangle, and independently
measure the angles in order to check the consistency of the
Standard Model.

The $CPT$ theorem forbids partial rate asymmetries 
which are $CP$ violating to occur at
tree level.  The addition of the penguin diagram, with different
CKM matrix elements and thus different weak phases, allows
quantum interference between two amplitudes,
and therefore can produce direct $CP$ violation.
$CP$ violation caused by interference between the penguin and tree
processes can lead, for instance, to
rate asymmetries in $B^+$ {\it vs.}\  $B^-$  decays.
There are also schemes to measure $\gamma$
or $\alpha$
using relations between tree-level and penguin amplitudes and isospin
relations in $B \rightarrow K\pi$ and related decays
\cite{ref:dgr}.
Many of the most promising measurements of $CP$ violation in
the $B$ system rely on interference caused by \BBmix\
(indirect $CP$ violation).  For example, the penguin
decays \Bphiks\ and \Betaprks\ 
can be used to measure 
$\beta$ \cite{ref:lonpecc,ref:lonsoni}.  
The tree-dominated decay \Bpipi\ 
can be used to measure $\alpha$;
however, additional penguin diagrams
can cause direct $CP$ violation as well as indirect,
which may complicate extraction of $\alpha$.
This effect is known as penguin pollution.

\subsection{\it Non-Standard Model Possibilities}
\label{sec:NP}
Penguin diagrams are a very good place to look for new physics at low
energies \cite{ref:newphysrev}.
Loops are very sensitive to additional heavy particles,
for example, charged Higgs, SUSY particles, or fourth generation particles.
The inclusive $b \rightarrow s\gamma$ rate measured at CLEO has 
already placed constraints on charged Higgs models, and an anomalous $WW\gamma$
coupling, and other models (Sec.~\ref{sec:bsgam}).

Non-standard models with four generations of quarks can,
while respecting the experimental measurement of $b\rightarrow s \gamma$,
significantly decrease
the $b \rightarrow d\gamma$ electromagnetic penguin and
$b\rightarrow d$ gluonic penguin rates while noticeably increasing the
$b \rightarrow d$ electroweak and vertical penguin rates.
Models with $Z^0$-mediated flavor-changing neutral currents can enhance
$B^0 \rightarrow \ell^+\ell^-$ by two orders
of magnitude over the Standard Model value and $B_s^0 \rightarrow \ell^+\ell^-$
by one order of magnitude.  These models can also increase the
electroweak penguin-dominated decays
$B^+(B_s^0) \rightarrow \phi \pi^+(\phi\pi^0)$ by two (one) orders of magnitude
above the SM prediction
without violating the current limits on $b \rightarrow s \ell^+\ell^-$ or
$B^0 \rightarrow K_S^0 K_S^0$.   Multi-Higgs doublet models can
also enhance the rates of di-lepton processes such as 
$B^0 \rightarrow \ell^+\ell^-$, and in particular,
$b \rightarrow s \tau^+\tau^-$ can be enhanced without affecting
$b \rightarrow s \mu^+\mu^-$ or $b \rightarrow se^+e^-$.
In supersymmetric models, contributions from a charged Higgs in the penguin
loop can be cancelled by contributions from charginos and gluinos in the
loop, thus leaving $b \rightarrow s \gamma$ at the Standard Model value,
while increasing the $b \rightarrow s \ell^+\ell^-$ rate.
Similarly, if gluinos couple more readily to gluons than to
photons, the $b \rightarrow sg^*$
rate can be strongly enhanced without affecting $b \rightarrow s \gamma$
\cite{ref:gluinos}.  In particularly favorable scenarios,
SUSY ``penguino'' effects can dominate Standard Model penguin effects
in decays such as $B^0\rightarrow K^0\phi$ \cite{ref:aliev}.
Even if new physics conspires to give the same rates as Standard
Model predictions, there will likely be effects in $CP$ violation.
For instance London and Soni \cite{ref:lonsoni} point out that differences
between the value of $\sin2\beta$ measured from $B\ra\psi K_S^0$ 
and the value measured from 
penguins such as \Betaprks\ might indicate the presence of non-SM processes.
These are just a few of the many ways that new physics can be detected
in $B$ penguin decays.

Recently, there has been much discussion about new physics
enhancing the $b \rightarrow s g^*$ rate
\cite{ref:gluinos,ref:enhanced1,ref:enhanced2,ref:enhanced3},
thereby solving a couple
of ``mysteries'' in $B$ physics: the $B$ semileptonic
branching fraction is measured \cite{ref:sl_expt}
to be smaller than theoretical predictions \cite{ref:sl_theory};
and the number of charm particles
per $B$ decay is smaller than predicted \cite{ref:charmcount}.
Both of these mysteries can be solved by increasing the non-leptonic,
non-charm $B$ width, e.g., $b \rightarrow s g^*$ which is so far
not experimentally well-constrained.  A rate of ${\cal O}(10\%)$
seems sufficient.  However, limits on the $b \ra s g^*$ rate
are starting to rule out this explanation (Sec \ref{sec:gluinc}).

\section{EXPERIMENTAL OVERVIEW}

At present $b$ quark
decays are under investigation 
with high statistics data samples from three different
colliders: the Cornell Electron-positron Storage Ring (CESR)
producing $B\bar{B}$ pairs in decays of the 
$\Upsilon(4S)$ resonance just above the $e^+e^-\to B\bar{B}$ threshold;
the Large Electron-Positron collider (LEP) at the European Laboratory for
Particle Physics (CERN),
which produced $b\bar{b}$ pairs in $Z^0$ decays; and the Tevatron at 
Fermi National Accelerator Laboratory (FNAL)
producing $b\bar{b}$ pairs in $p\bar{p}$ collisions. 
Various production aspects of these machines are compared in Table 
\ref{tab:benvir}.

\begin{table}[bth]
\def\ufs{\Upsilon(4S)}
\caption{Various parameters characterizing experimental environments
          at three colliders used in analyses of $b$ quark data. 
          Explanation of the symbols: $\sigma-$cross section,
          ${\cal L}-$luminosity, $\beta-$velocity of $b$ quarks,
          $\beta\gamma c\tau-$mean decay path,
          $f-$fractions of $b$ hadron species produced.
}
\label{tab:benvir}
\small
\def\1#1{{#1}}
\def\2#1{{#1}}
\vspace{0.3cm}
\begin{tabular}{lrrr}
\hline\hline
 {} & {} & {} & {} \\    
Quantity      &  $\ufs$ (CESR)  & $Z^0$ (LEP) &   Tevatron \\
 {} & {} & {} & {} \\
\hline
 {} & {} & {} & {} \\
$\sigma(b\bar{b})$ (nb)  
  &   $1.1$  &   $9.2$  &  $\sim30000$ $(\sim6000)^\dag$ \\
$\sigma(b\bar{b})/\sigma(q\bar{q})$ 
           &  $\sim0.3$ & $\sim0.2$ & \2{$\sim0.001$} \\
${\cal L}^{peak}$ (10$^{31}$ cm$^{-2}$ s$^{-1}$ ) 
                              &  \1{$48.0$}  &   \2{$1.1$}  &   $2.5$  \\    
$\int{\cal L}dt$  (pb$^{-1}$) 
               analyzed   &  \1{$3100$}  &  $\sim120$ &  $\sim110$ \\
\phantom{$\int{\cal L}dt$  pb$^{-1}$}
   in~pipeline              &  $3600$  &      &       \\
$b\bar{b}$ pairs  analyzed ($10^6$)    
                    &  $3.3$   &  $0.9$  &  
$\sim3300$ $(\sim660)^{\dag}$ 
 \\
$\beta$ & \1{$\sim0.07$}   & $\sim1$ & $\sim1$ \\                             
$\beta\gamma c\tau$ ($\mu m$) &  \2{30}   & \1{2600} & \1{500} \\
fragmentation & & & \\
\qquad\quad background   &  \1{no}  & some  
  & \2{large} \\
$B$ energy 
       &  \1{$E_{\rm beam}$}  & $\sim0.7\,E_{\rm beam}$ 
  &    \\
spatial separation & & & \\
\qquad\quad of $b$ and $\bar{b}$ 
                   &  \2{no}  &  \1{yes}  &  \1{yes}   \\
$f_{B^+}\approx f_{B^0}$  &  $\sim0.5$  &  $\sim0.4$ & $\sim0.4$ \\
$f_{B^0_s}$      &  --  &  $\sim0.1$ & $\sim0.1$ \\
$f_{\Lambda_b}$  &  --  &  $\sim0.1$ & $\sim0.1$ \\
{} & {} & {} & {} \\
\hline 
{} & {} & {} & {} \\
main advantage & simple production, & vertexing,         & cross-section, \\
               & statistics         & one $b$ at a time  & vertexing  \\
{} & {} & {} & {} \\
\hline
\end{tabular} 
\newline
${\dag}$\quad {These numbers correspond to
the central region, $|y|<1$, with high transverse momentum of the
$b$ quark, $p_t>6$ GeV (the number in parentheses).}
\end{table}

CESR provides the highest 
luminosity and a simple production mechanism.
Since no fragmentation particles are produced, the energy of reconstructed
$B$'s can be constrained to the beam energy which provides for a powerful
reconstruction technique. Also, decay products
from the two $B$'s in the event populate the entire solid angle,
allowing background discrimination based on event shape.  However,
since the two $B$ mesons are
produced almost at rest, detection of a detached secondary $B$ decay
vertex is not possible on an event-by-event basis. 

At LEP, since the momentum of the $b$ hadrons is appreciable, their decay
products are easily separated into
back-to-back hemispheres. A large decay length gives rise to many
important analysis techniques. Even though the $b\bar{b}$ cross section
is much larger than at the $\Upsilon(4S)$ resonance, the high energy 
of the beam limits achievable luminosity. 
The small sample size is the limiting factor for the LEP experiments.

The main asset of the Tevatron experiments is
their huge production cross-section.
A large background cross-section is the main obstacle to overcome in
these experiments.
Specialized triggering is needed to limit data acquisition to
manageable rates.
So far, the results from the Tevatron
have been mostly limited to channels containing high $p_t$ muons. 
Again, vertexing is an important selection tool.

CESR houses only one experimental apparatus (CLEO II).
There are four experiments at LEP (ALEPH, DELPHI, OPAL, and L3),
and two at the Tevatron (CDF and D0).
All these collider detectors have a similar \lq\lq onion'' structure.
The beam collision point is surrounded by a thin vacuum pipe, followed
by subsequent layers of nearly cylindrical detectors.
Most of the experiments also have end-caps to maximize solid angle
coverage. 
The innermost layer is created by a silicon strip vertex detector
used to pinpoint production points of charged particles.
The vertex detector is followed by a larger gaseous charged-particle
tracker. Together they are 
used to determine particle momenta from curvature
in a solenoidal magnetic field (except for D0 which has no magnetic field).
In addition, most experiments measure
specific ionization ($dE/dx$) in the tracking devices to obtain
partial charged hadron identification.
To aid particle identification and triggering,
many detectors also have Time-of-Flight (ToF) scintillation counters 
surrounding the tracking system.
The DELPHI experiment has a Ring Imaging \v Cerenkov instead, providing
superior charged hadron identification.
In the next layer, an electromagnetic calorimeter measures 
electron and photon energies by integrating over electromagnetic showers 
developing in a dense medium.  Especially noteworthy are the CLEO II and L3 
scintillating-crystal calorimeters, which have superb energy resolution.
The final layer comprises a hadronic calorimeter which measures energies of
charged and neutral hadrons. 
The calorimeters also identify muons which penetrate to the
outermost layers. CLEO II has no hadron calorimeter; instead a thick layer
of iron is used to identify muons.
 
In our article, we also refer to the previous generation of experiments
at $\Upsilon(4S)$, CLEO I at CESR and ARGUS at DORIS (a CESR-like
$e^+e^-$ storage ring which operated at
DESY in Hamburg), and
in $p\bar p$ collisions, 
UA1 at Super Proton-antiproton Synchrotron (S$p\bar p$S) at CERN.
The amount of data collected by these older experiments was about two orders
of magnitude smaller than in the contemporary experiments.

Future $B$ experiments are discussed in section \ref{sec:future}.

\section{ELECTROMAGNETIC PENGUINS}

\subsection{$B\to X\gamma$ Exclusive decay modes}
\label{sec:Kstgam}

After a $b$ quark decays to $s\gamma$ via the penguin diagram 
(Fig. \ref{fig:empenguin}) the produced $s$ quark and the spectator $\bar q$ 
($\bar q=\bar u$ for $B^-$, and $\bar d$ for $B^0$) 
turn into hadrons. The final state usually contains one kaon
and at least one pion ($B\to K\gamma$ is forbidden by 
angular momentum conservation).
Hadronization may proceed via creation of an intermediate 
strange resonance: $K^*(892)$, $K_1(1270)$, $K_1(1400)$, etc.
The existence of penguin decays was first confirmed experimentally by 
the CLEO observation \cite{kstargammaPRL} of the exclusive decay 
$B\to K^*(892)\gamma$, with $K^*\to K\pi$.
The initial observation was based on 1.5 million
$e^+e^-\to \Upsilon(4S)\to B\bar{B}$ events.
Reconstruction of exclusive final states from $B$ mesons produced
at the $\Upsilon(4S)$ benefits from the beam energy constraint: $E_B=E_{beam}$. 
Thus, energies of the $B$ decay products must add up to the beam energy:
$\Delta E=(E_{K^*}+E_\gamma)-E_{beam}\approx 0$.
Also the $B$ meson mass resolution is improved by an order of magnitude
with the use of the beam constraint: 
$M_B=\sqrt{{E_{beam}}^2-(\vec{p}_{K^*}+\vec{p}_\gamma)^2}$.
These tight kinematic constraints are crucial in background suppression
and signal extraction. 
Eleven signal events were observed over a background of
two events, estimated from the $\Delta E$ and $M_B$ sidebands.

\begin{figure}[tbhp]
\hbox{
\psfig{figure=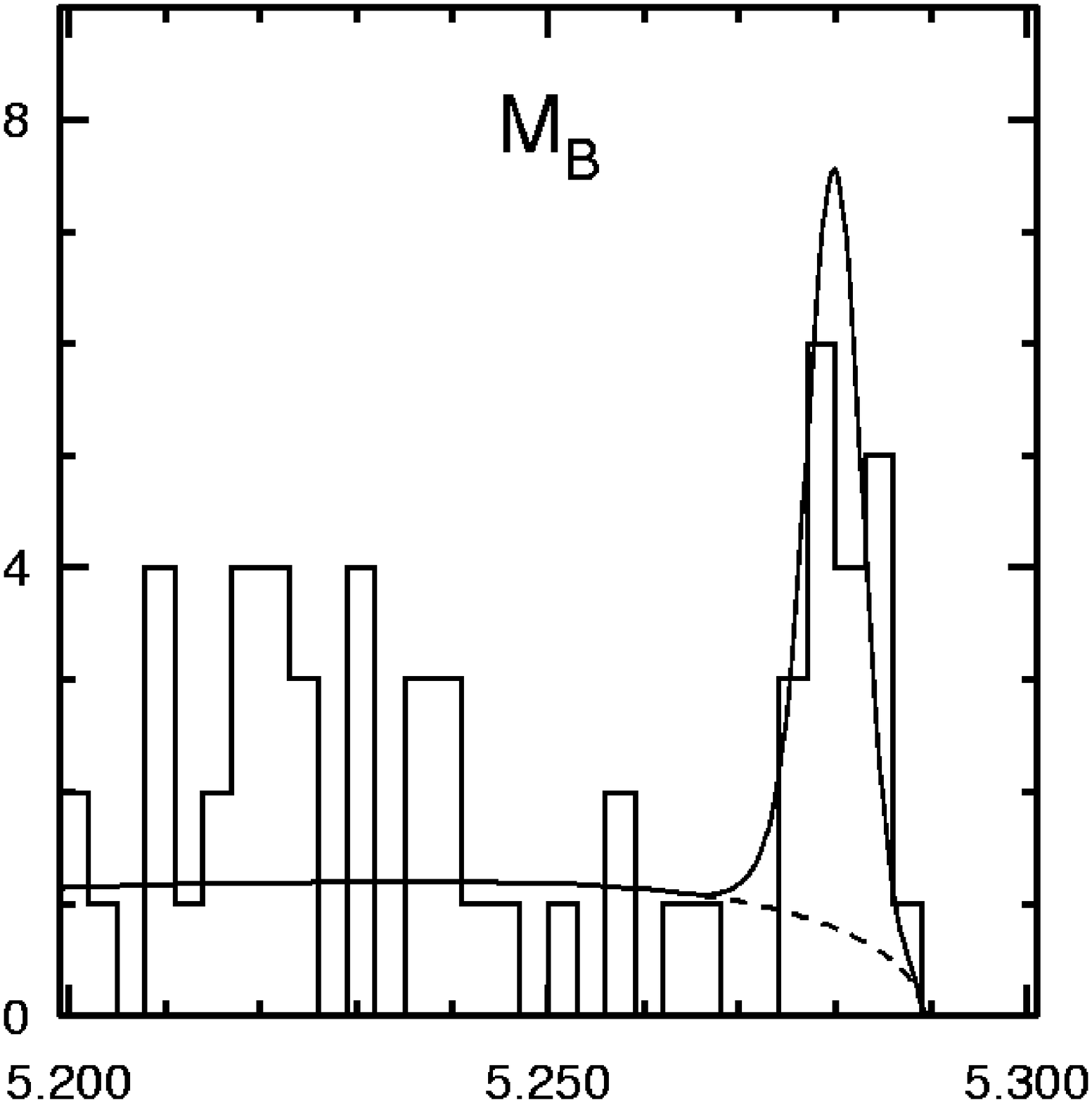,width=4cm}
\psfig{figure=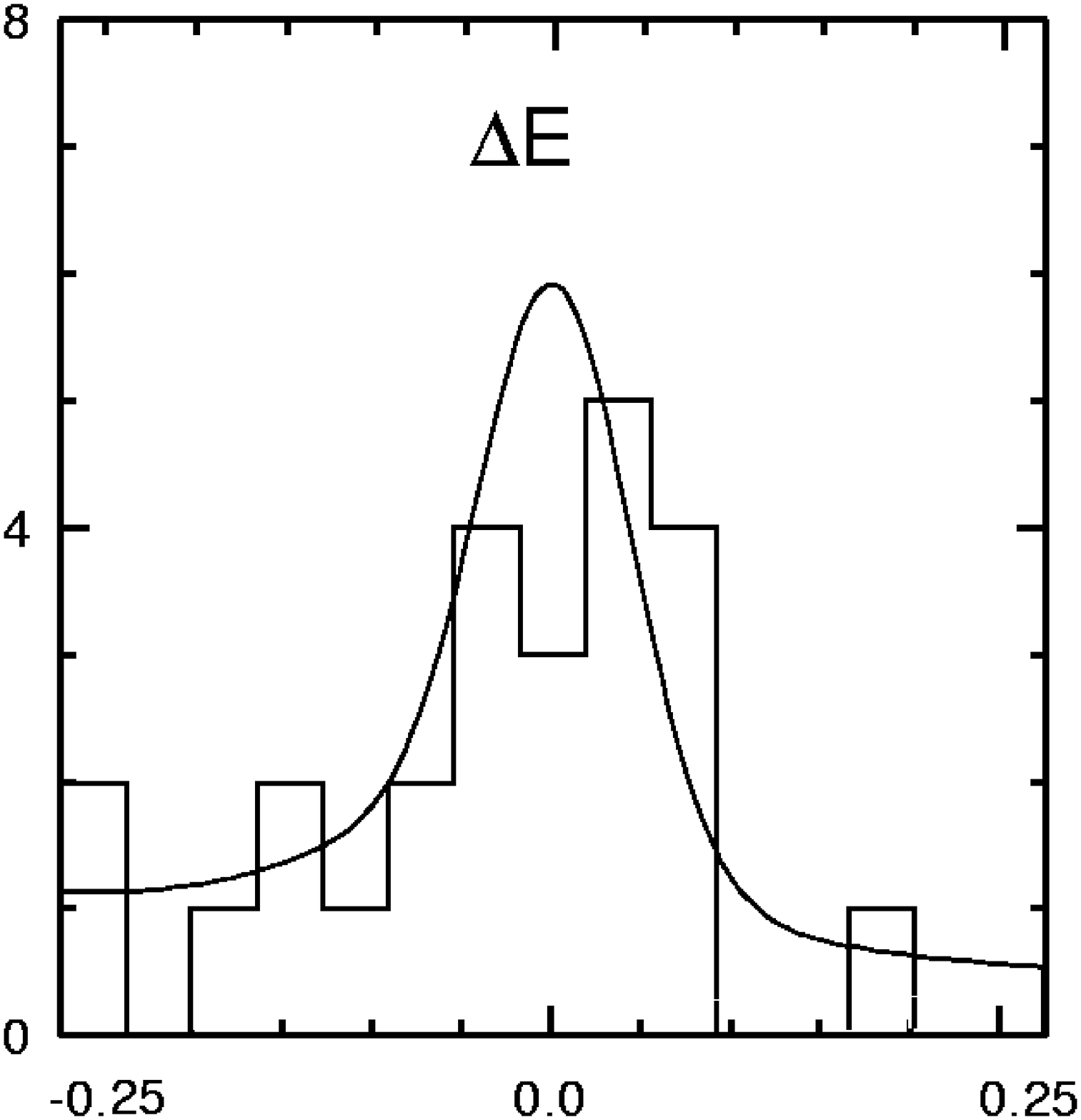,width=4cm}
\psfig{figure=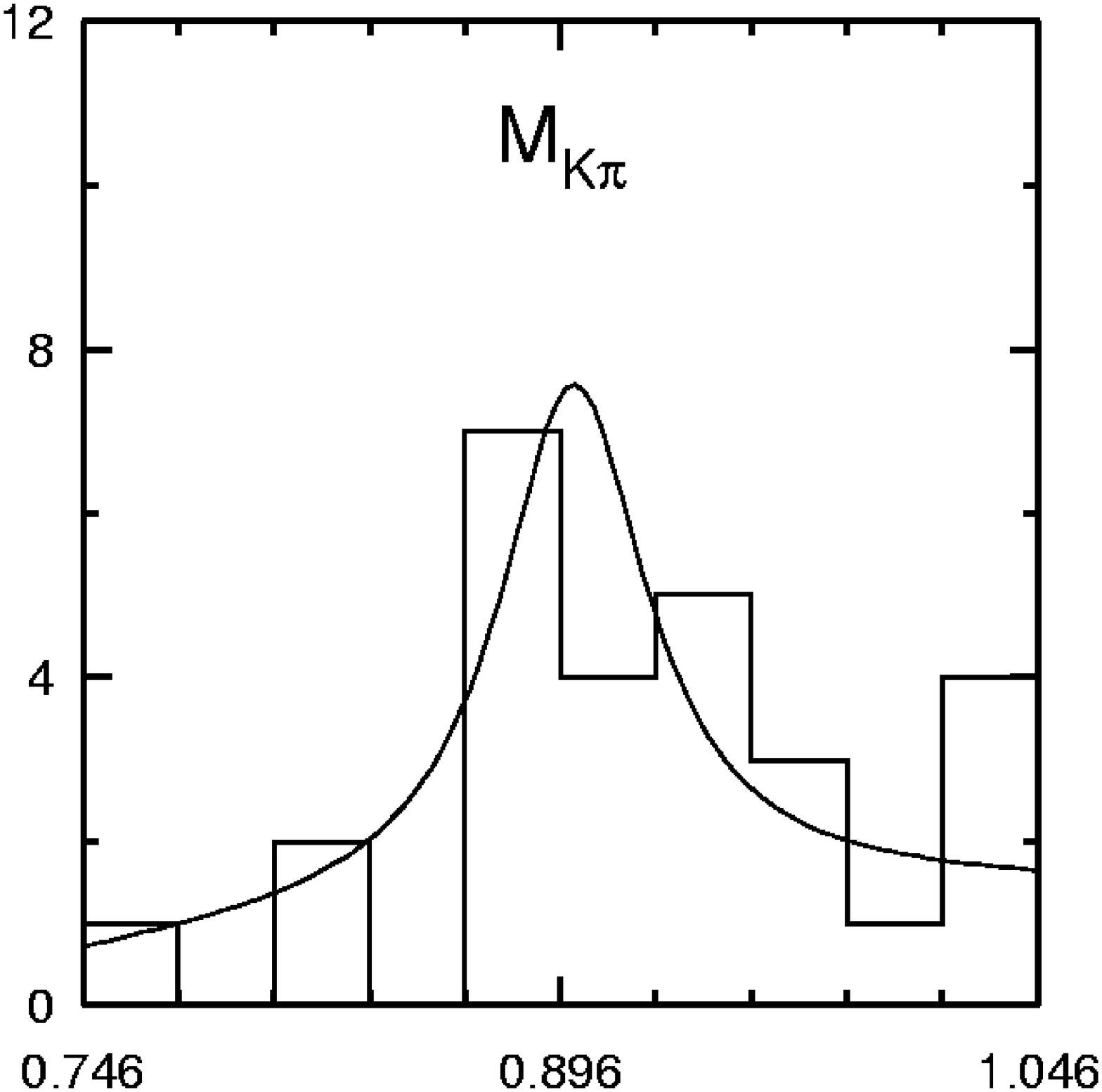,width=4cm}
}
\caption{
Projections of the maximum likelihood fit (solid lines) 
onto $M_B$, $\Delta E$, 
and $M_{K^*}$ for the $B^0\to(K^+\pi^-)\gamma$ data (histograms).
The horizontal scales are in GeV. The vertical scale gives 
number of events per bin. 
\label{fig:kstargamma}}
\end{figure}

Since the first observation, CLEO has presented 
an updated analysis based on larger statistics 
(2.6 million $B\bar{B}$ events in
$2.4$ $fb^{-1}$ of integrated luminosity)
and improved analysis techniques \cite{kstargammaWaw}.
Instead of cutting on various variables to define the signal region,
and then using sidebands to estimate the background, the improved analysis
used a maximum likelihood fit to determine signal and background yields.
In addition to $M_B$ and $\Delta E$, event shape variables and $M_{K^*}$
also were used in the fit. The event shape information was optimized to 
distinguish between the signal and the dominant background due to continuum 
production of lighter quarks ($e^+e^-\to q\bar q$, $q=d, u, s, c$).
This method improved the signal efficiency by a factor of two.
Even though the background also increased, 
the signal sensitivity increased by about 30\%\ beyond
the gain from the increased integrated luminosity.
Averaging over various charge modes ($K^+\pi^-$, $K^0_S\pi^0$,
$K^0_S\pi^-$, $K^-\pi^0$) CLEO obtained:
${\cal B}(B\to K^*\,\gamma)=(4.2\pm0.8\pm0.6)\times10^{-5}$.
Projections of the maximum likelihood fit onto $M_B$, $\Delta E$, 
and $M_{K^*}$ are shown in Fig.~\ref{fig:kstargamma} for the $K^+\pi^-$
channel which has the largest statistics.
%
%

The measured branching fraction is in the range predicted by the
Standard Model, $(1-15)\times 10^{-5}$ \cite{kstargammaTh}. 
It is also an order of
magnitude larger than would be expected if the penguin diagram were
not present \cite{ifnopenguin}.

LEP experiments looked for these decays in $e^+e^-\to Z^0\to b\bar{b}$ data
but were not able to observe the signal due to an insufficient number of
$b\bar{b}$ pairs. They also looked for $B_s\to\phi \gamma$ decays.
ALEPH set a 90\%\ C.L. upper limit of $29\times10^{-5}$ on the branching
fraction for these decays \cite{phigammaALEPH}.

\subsection{Inclusive measurements of $b\to s\gamma$ rate}
\label{sec:bsgam}

\subsubsection{Experimental results}

The measured rate for the exclusive mode $B\to K^*\,\gamma$ is in 
the ball-park of Standard Model predictions. 
Quantitative tests of the Standard Model with rates measured
for exclusive channels are severely handicapped by our inability
to calculate hadronization probabilities from the first principles
of the theory. Predictions of phenomenological models
for the $K^*$ fraction in $b\to s\,\gamma$ decays,
$R_{K^*}\equiv{\cal B}(B\to K^*\,\gamma)/{\cal B}(b\to s\,\gamma)$,
vary from 1--40\% \cite{kstargammaTh}.
One should notice improvement in 
recent lattice-QCD calculations in this area \cite{Flynn}.

Fortunately, when summed over all possible final states,
hadronization probabilities aren't relevant and the inclusively measured
rate should reflect the short distance interactions
which can be accurately predicted using the effective Hamiltonian
of the Standard Model. Since the first non-perturbative correction 
is expected to be of second order in the
$\Lambda_{QCD}/m_b$ expansion, it should be small, thanks
to the heavy $b$ quark mass.
Next-to-leading order perturbative calculations have been recently
completed for $b\to s\,\gamma$. 
Assuming unitarity of the CKM matrix to constrain $|V_{ts}|$
the Standard Model predicts \cite{bsgammaTh}:
${\cal B}(b\to s\,\gamma)=(3.5\pm0.3)\times10^{-4}$.

When reconstructing simple
exclusive final states such as $B\to K^*\,\gamma$, $K^*\to K\pi$,
backgrounds are usually low due to tight kinematic constraints, in this case
constraints to the $B$ and $K^*$ masses and the beam energy.
Inclusive measurements are more challenging and they are often background
limited.

The main background for CLEO is again from \eetoqq\ events.
This background can be subtracted reliably with data taken
below the $e^+e^-\to B\bar B$ threshold. However, statistical 
fluctuations in the background level
can easily swamp the signal unless the background is
efficiently suppressed.
Backgrounds from $B$ decays are less serious since $b\to s\,\gamma$   
decays are quasi-two-body and produce higher energy photons 
($E_\gamma\sim m_b/2$) than photons from typical $B$ decay modes.

\begin{figure}[tbhp]
\hbox{ 
\psfig{figure=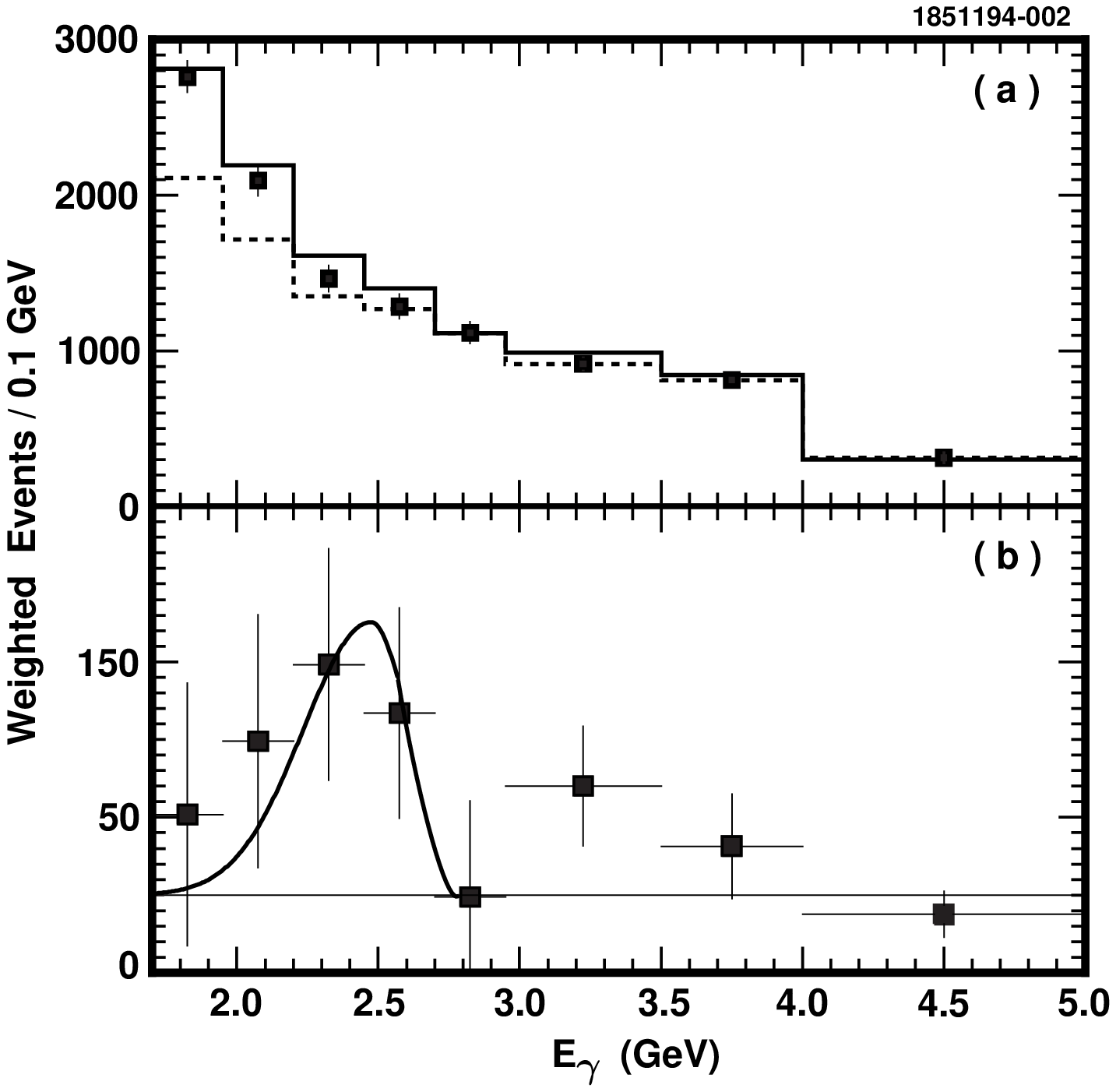,width=6.6cm}
\psfig{figure=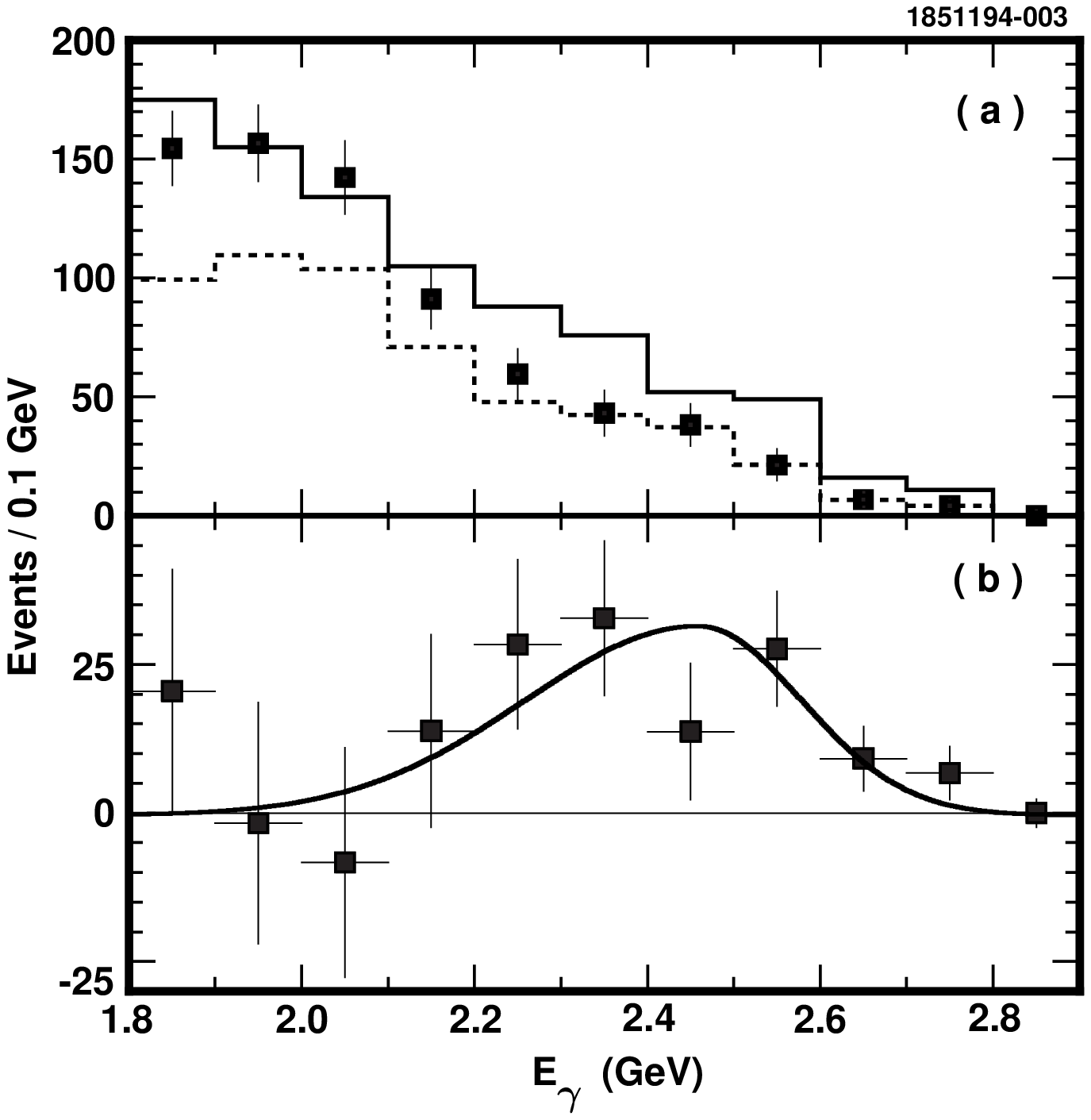,width=6.6cm}
}
\caption{Inclusive $E_\gamma$ spectra in the CLEO $b\to s\gamma$
         measurement obtained with the event-shape analysis (left)
         and with inclusive $B$ reconstruction (right).
         (a) $\Upsilon(4S)$ data (solid histogram), scaled 
             below-threshold data (dashed histogram)
             plus estimated $\Upsilon(4S)$ backgrounds (points with error bars).
         (b) Background-subtracted data (points) and Monte Carlo
             prediction based on Ref.~\cite{bsgE} 
             for the shape of the $b\to s\,\gamma$ signal
             (solid curve).
         Note that the range of $E_\gamma$ is very different for the
         left and right plots.
\label{fig:CLEObsg}}
%
\hbox{ 
\psfig{figure=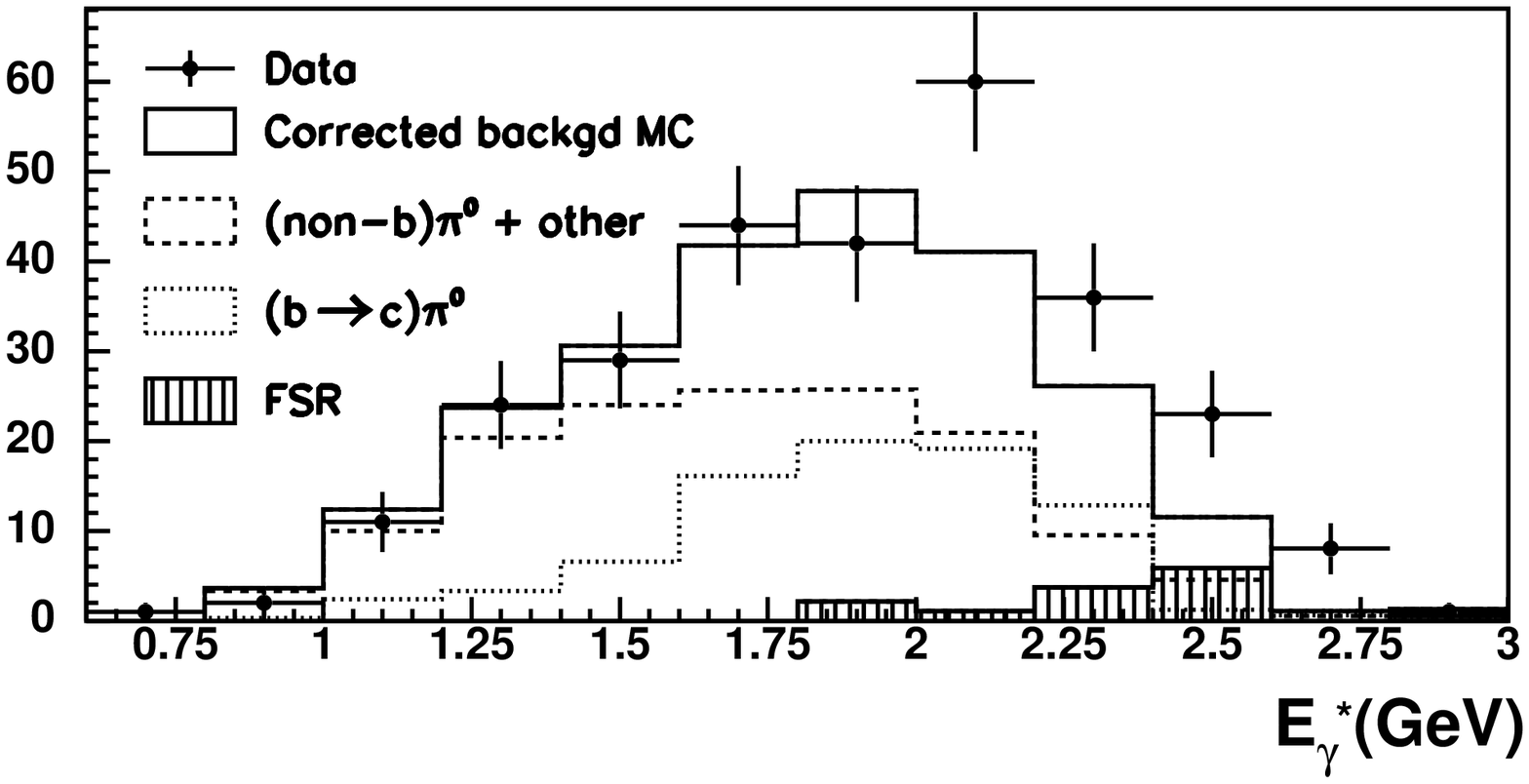,width=6.6cm}
\psfig{figure=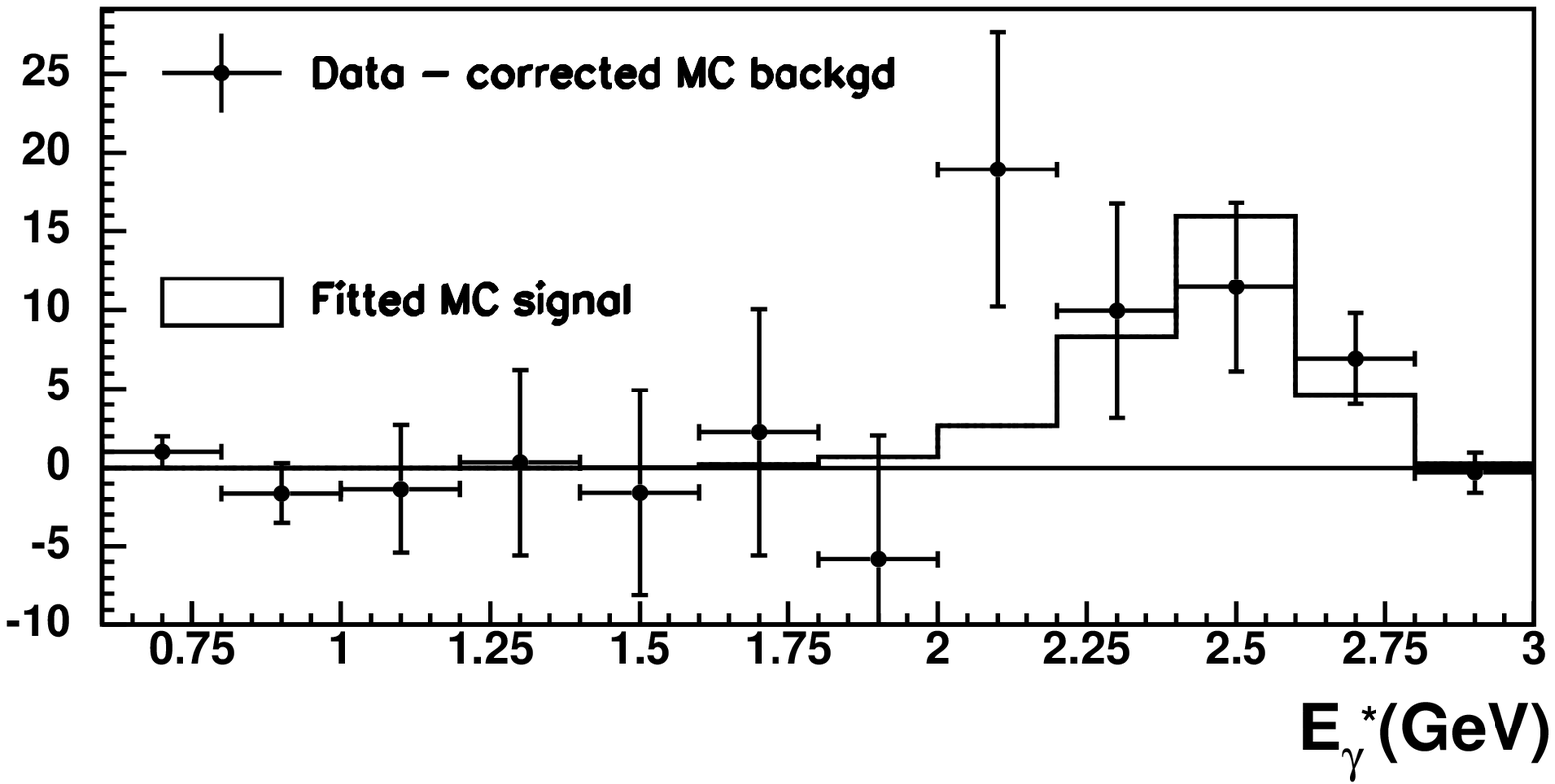,width=6.6cm}
}
\caption{Inclusive $E_\gamma^*$ spectrum in the ALEPH $b\to s\gamma$
         measurement. On the left: data (points), total estimated 
         background (solid histogram), $\pi^0$ background from
         $b\to c$ decays (dotted histogram), final state radiation
         background (shaded histogram), and all other backgrounds 
         (dashed histogram). The latter 
         comes mainly from $\pi^0$ decays from non-$b$ sources
         and from $\eta$ decays.
         On the right: background-subtracted
         data (points) and Monte Carlo
         prediction \cite{bsgE}
         for the shape of the $b\to s\,\gamma$ signal
         (solid histogram).
\label{fig:aleph}}
\quad
\end{figure}

CLEO used two complementary
approaches to suppress the continuum background \cite{bsgammaCLEO}.
In one approach only the photon among $b\to s\,\gamma$ decay products was
explicitly reconstructed.
Shape differences between $B\bar B$ events
and \qqbar\ background were used for background suppression.  $B\bar B$
events are nearly spherical since the $B$ mesons are nearly at rest at
the $\Upsilon(4S)$, while \qqbar\ events have a distinct two-jet appearance.
For the best sensitivity all shape variables were combined withe the use of 
a neural-net technique. 
The signal amplitude was extracted from a one-parameter fit to the
neural net output variable, with the signal shape and
the $B\bar{B}$ backgrounds taken from Monte Carlo
simulation, and the continuum background subtracted using the 
below-threshold data. 
In the second approach, all products of the $b\to s\,\gamma$ decay
were reconstructed as in exclusive reconstruction. Thus, the constraints
to the $B$ mass and beam energy could be used. 
The final state recoiling against the photon, denoted $X_s$,
was required to contain a kaon 
candidate (a charged track consistent with $K^{\pm}$ by $dE/dx$ and ToF,
or a $K^0_s\to\pi^+\pi^-$ candidate) and 1--4 pions (including at most
one $\pi^0$).   
In Fig.~\ref{fig:CLEObsg}, we show
the photon energy spectra measured by CLEO with these two methods in
a sample of $2.2\times10^6$ $B\bar{B}$ events.
The first method has rather large continuum background but also
high signal efficiency ($32\%$). 
The second method is very good in suppressing continuum background, but the
signal efficiency is much smaller ($9\%$).  The
sensitivity of these two approaches is nearly equal, and the measurements of
signal amplitudes are only slightly correlated.
After combining these two methods, CLEO
measured ${\cal B}(b\to s\,\gamma)=(2.32\pm0.57\pm0.35)\times10^{-4}$,
in agreement with the Standard Model calculations.

\begin{figure}[htbp]
\centerline{\psfig{figure=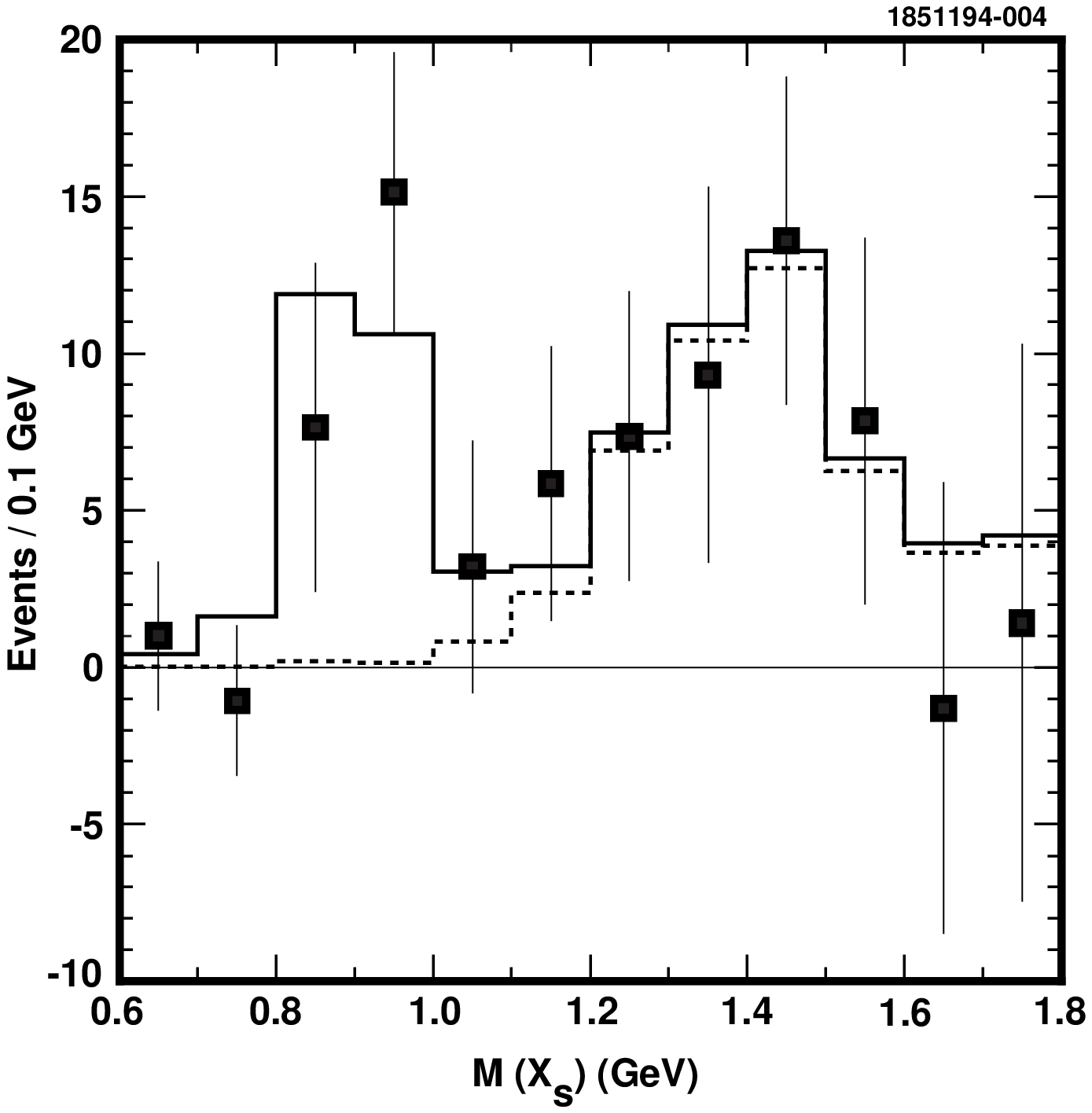,height=6cm}}
\caption{The $X_s$ mass distribution in $B\to X_s\gamma$ decays 
         from the CLEO inclusive $B$ reconstruction analysis.
         The solid curve is a fit to the expected distribution
         from a spectator model. The dashed curve shows
         the non-$K^*(892)$ component of the fit. 
\label{fig:mXsCLEO}}
\end{figure}

The $X_s$ mass distribution from the CLEO inclusive $B$-reconstruction
analysis is shown in Fig.~\ref{fig:mXsCLEO}.
A clear $K^*(892)$ peak is observed followed by a dip 
which is expected since the next excited kaon resonance is
$K_1(1270)$. A broad enhancement at and above the $K_1(1270)$ is observed; 
this is also expected since many resonant states exist in this region.
The present experimental statistics are insufficient to establish a 
positive signal for any of the resonances beyond $K^*(892)$ taken separately.

Combining the inclusive and the exclusive measurements, CLEO determines
$R_{K^*}=(18.1\pm6.8)\%$ in agreement with 
some phenomenological estimates \cite{kstargammaTh}.
In particular the calculations which take the most from QCD,
QCD sum rules \cite{QCDsumR}
and 
recent lattice-QCD calculations \cite{Flynn},
are in good agreement with the data.

Inclusive $b\to s\gamma$ decays also have been observed recently
in $Z^0\to b\bar b$ data by ALEPH \cite{bsgammaALEPH}. 
The ALEPH analysis, based on 0.8 million $b\bar{b}$ pairs, searches 
for these decays by selecting events with a single high energy 
photon and a jet in the opposite hemisphere 
which is $b-$like from its observed detached vertex.  
The $B$ hadron candidate in the photon hemisphere 
was reconstructed by combining the photon with other particles in the 
same hemisphere until the ensemble matches the $B$ meson mass within the 
experimental resolution.  The particles were selected by their probability 
of belonging to a $B$ meson which was assigned according to their momenta 
and impact parameters at the interaction point.   Up to eight particles 
were allowed in addition to the photon, including charged tracks,
$\pi^0$ mesons, and $K^0_L$ mesons detected in the hadron calorimeter.  
Once the $B$ candidate was
reconstructed, the photon was boosted to its rest frame where the 
energy  $E_\gamma^*$ is nearly monochromatic for signal photons 
($E_\gamma^*\sim m_b/2$). Further background suppression was achieved by 
imposing requirements on the boosted sphericity and photon angle 
in the rest frame of the $B$ meson candidate.

To measure the background the sample was divided into eight subsamples,
seven of which contained little $b\to s\gamma$ signal.
The variables used for this division were: lateral shower size in the
electromagnetic calorimeter to distinguish prompt photons from merged
$\pi^0$ mesons; vertex detachment in the opposite hemisphere to distinguish
$b\bar b$ events from $q\bar q$ backgrounds;
and energy of the $B$ candidate to discriminate against final state radiation
background.
The shapes of the contributing backgrounds (split into four major
components) and of the $b\to s\gamma$ 
signal are computed from Monte Carlo while
their normalizations are fixed by a fit to the $E_\gamma^*$ distributions 
in the eight subsamples. Fig.~\ref{fig:aleph}a 
shows the $E_\gamma^*$ distribution in 
the signal sensitive subsample for both the data and the background 
and Fig.~\ref{fig:aleph}b 
shows the data after background subtraction. 
The total reconstruction efficiency ($\sim13\%$) is similar to the
efficiency obtained in the CLEO inclusive $B$ reconstruction analysis.
In spite of 2.75 times fewer $b\bar b$ events, ALEPH is 
able to observe a significant inclusive signal.
This should be attributed to a better suppression of the
light quark backgrounds (the dominant background at the $\Upsilon(4S)$)
by the detached vertex cuts. 
The background composition is illustrated in Fig.~\ref{fig:aleph}a.  
The branching fraction measured by ALEPH,
${\cal B}(b\to s\,\gamma)=(3.11\pm0.80\pm0.72)\times10^{-4}$,
is consistent with the CLEO measurement and the Standard Model
predictions.

The other LEP experiments were not able to detect a $b\to s\,\gamma$ signal
and set upper limits consistent with the CLEO and ALEPH measurements:
\newline
DELPHI \cite{delphirarepub}  $<5.4\times10^{-4}$, 
L3 \cite{bsgammaL3} $<12\times10^{-4}$ (90\%\ C.L.)

Combining the CLEO and the ALEPH results, we obtain:
$$  {\cal B}(b\to s\,\gamma)=(2.54\pm0.57)\times10^{-4} $$

\subsubsection{Theoretical implications}

Dividing the measured value of ${\cal B}(b\to s\,\gamma)$ 
by the Standard Model predictions, we obtain:
$$\left|\frac{V_{ts}^*}{V_{cb}} V_{tb}\right| 
=0.85\pm0.10\hbox{(experiment)}\pm0.04\hbox{(theory)}\ ,$$
consistent with the unitarity constraint \cite{Ali}:
$$\left|\frac{V_{ts}^*}{V_{cb}} V_{tb}\right|\approx|V_{cs}|=1.01\pm0.18$$
Using the measured values to eliminate 
$V_{tb}=0.99\pm0.15$ \cite{Vtb} and $V_{cb}=0.040\pm0.002$ \cite{Vcb}, 
we obtain:
$$|V_{ts}|=0.034\pm0.007$$

\begin{figure}[htbp]
\psfig{figure=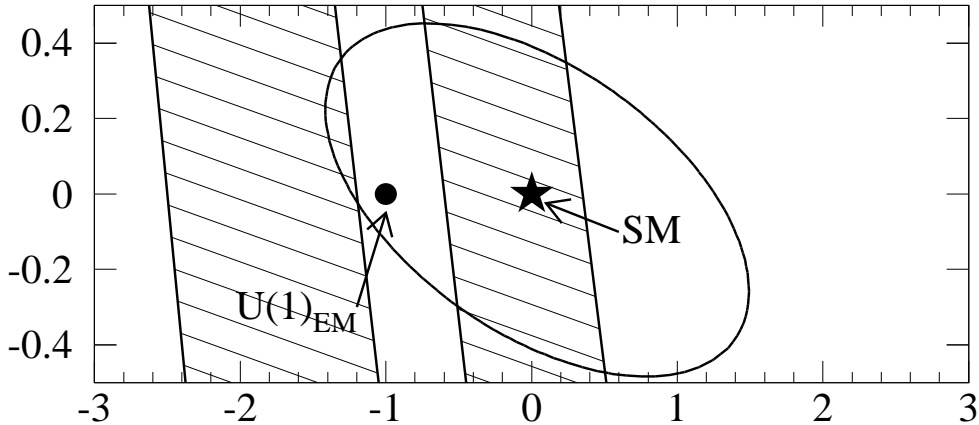,width=13.2cm}
\caption{
 Limits on the anomalous $WW\gamma$ coupling parameters $\lambda$ and
 $\Delta\kappa$. The hatched regions are consistent at 95\%\ confidence
 level with the
 $b\to s\,\gamma$ rate measured by CLEO. The leading-log calculations,
 together with their uncertainties were used to obtain these bands
 \cite{bsgammaCLEO}.
 The yield of $p\bar p\to W\gamma\,X$ measured by D0 limits the
 parameters to the interior of the ellipse (95\%~C.L.) \cite{D0anomalous}.
 The Standard Model (broken U(1)$\times$SU(2)) and pure U(1) theory
 are displayed. 
 \label{fig:anomalous}}
\end{figure}

The agreement between the measured and the Standard Model rates
(including the CKM matrix unitarity) 
leaves little room for non-standard contributions.
For example, limits on anomalous $WW\gamma$ couplings can be 
obtained \cite{anomalous-theory}.
Such anomalous couplings can arise from internal structure of
the gauge bosons or loop corrections involving new particles. 
They are parameterized by $\Delta\kappa$ 
and $\lambda$, which are both zero in the Standard Model.
Non-zero values of these parameters would generate anomalous 
magnetic dipole and electric quadrupole moments of the $W$:
$\mu_W=\frac{e}{2\,M_W} (2+\Delta\kappa+\lambda)$,
$Q_W^e= - \frac{e}{2\,M_W^2} (1+\Delta\kappa-\lambda)$.
They would either increase or decrease the $b\to s\,\gamma$ rate.
The region of the $\Delta\kappa-\lambda$ space consistent with the
CLEO measurement is shown in Fig.~\ref{fig:anomalous}.
The region allowed by a $p\bar p\to W\gamma\,X$ measurement by the D0
experiment \cite{D0anomalous} is also shown.
The two types of measurements are complementary.
The $b\to s\,\gamma$ measurement alone excludes the $U(1)$ theory in
which neutral bosons of electromagnetic and weak interactions
do not mix.

\begin{figure}[htbp]
\centerline{\psfig{figure=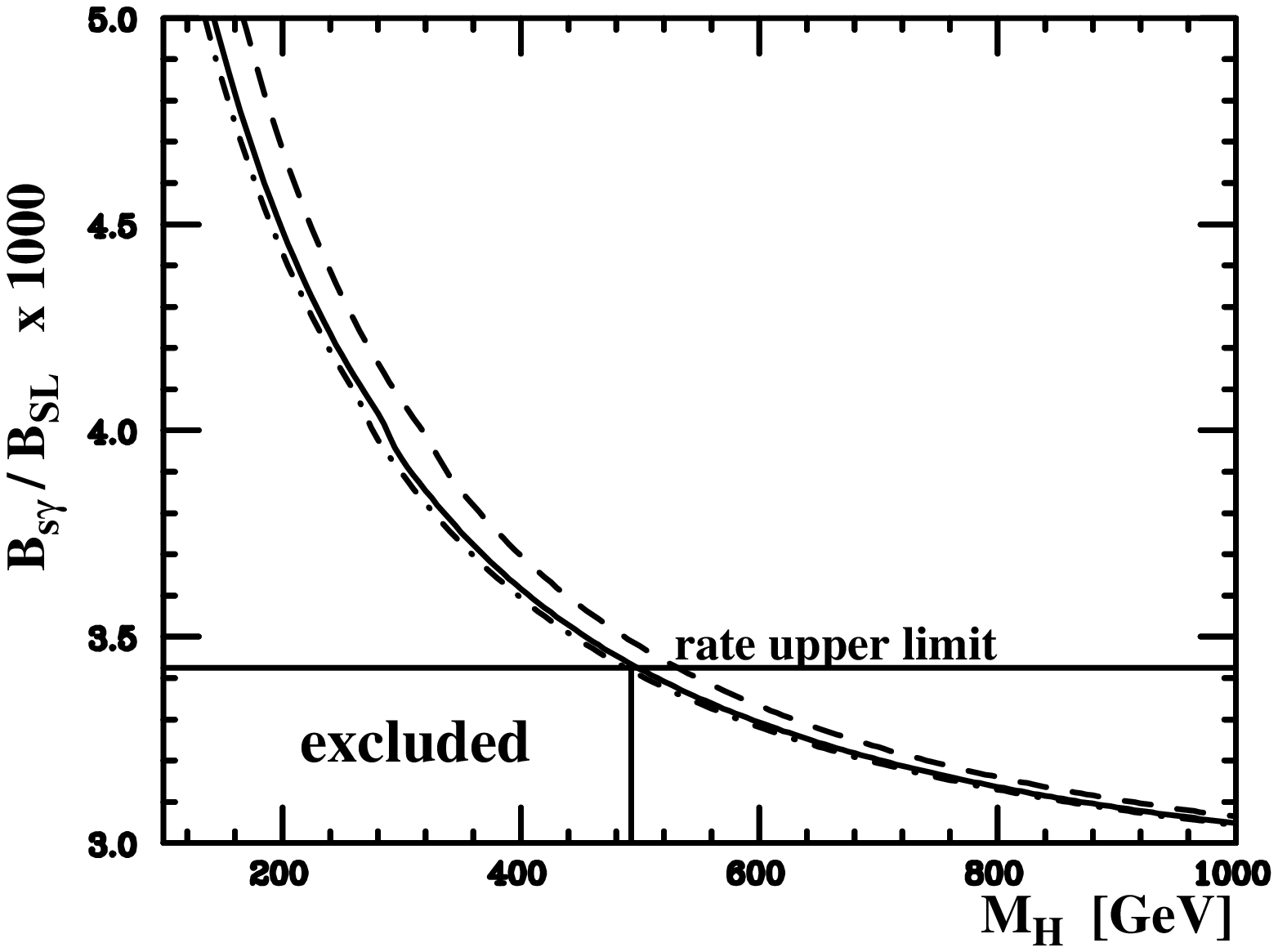,width=10cm}}
\caption{
 Exclusion region in the plane 
 ${\cal B}(b\to s\gamma)/{\cal B}(b\to c l\nu)$ versus $M_H$
 for $\tan\beta=1$ (dashed line), $\tan\beta=2$ (solid line),
 and $\tan\beta=5$ (dot-dashed line) as calculated by
 M.~Ciuchini \etal\ \cite{THDM}
 including next-to-leading QCD corrections and 
 combining theoretical uncertainties linearly.
 The upper limit on ${\cal B}(b\to s\gamma)/{\cal B}(b\to c l\nu)$ is
 indicated (horizontal line), together with a corresponding
 lower limit on $M_H$  for $\tan\beta=2$ (vertical line).
 \label{fig:THDM}}
\end{figure}

A charged Higgs boson can be exchanged instead of the charged $W$ in the
penguin loop. Models with two Higgs doublets are divided into two
categories: Model I -  both up-type and down-type 
quarks get their masses from the same
Higgs doublet; Model II - up-type quarks get masses from one Higgs doublet,
whereas down-type quarks get masses from the other Higgs doublet.
The free parameters of these models are the mass of the exchanged Higgs and
$\tan\beta$, which is the ratio of vacuum expectation values for the
two Higgs doublets. In Model I the $b\to s\,\gamma$ rate is decreased
relative to the Standard Model prediction. Since the CLEO measurement
is somewhat below the Standard Model expectation, the data are consistent
with Model I and a small Higgs mass. In Model II, the Higgs contribution 
always adds constructively to the Standard Model rate. 
Recently, next-to-leading order QCD corrections have been calculated
for this model \cite{THDM}. 
Using both the CLEO and ALEPH measurements, we obtain
${\cal B}(b\to s\,\gamma)/{\cal B}(b\to c l\nu)<3.4\times10^{-3}$
(95\%\ C.L.) which translates to a lower limit
on the charged Higgs mass with slight $\tan\beta$ dependence:
$M_H > 490$ GeV for $\tan\beta=2$,
as illustrated in Fig.~\ref{fig:THDM}.
The limit is almost $\tan\beta$ independent for $\tan\beta\ge2$.

Minimal supersymmetric extensions of the Standard Model (MSSM)
include Model II charged Higgs doublets.
However, the above lower limit on $M_H$ does not directly apply 
to the supersymmetric model since a chargino-stop loop may destructively
interfere with the charged Higgs and W-top contributions \cite{bsgMSSM}.
Nevertheless, the data impose interesting constraints on the
minimal supersymmetry. Naively speaking, either charged Higgs, chargino
and stop are all heavy or all light.

\begin{figure}[htbp]
\centerline{\hbox{\psfig{figure=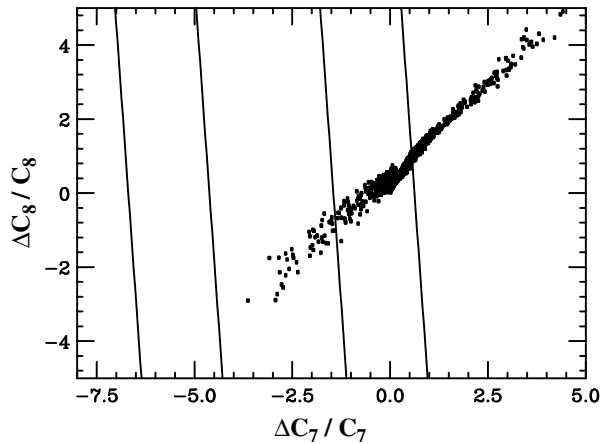,width=8cm}}}
\caption{Bounds on the contributions from new physics to $C_7$ and $C_8$.
         The region allowed by the CLEO data corresponds to the area
         inside the two bands. Various minimal supergravity models consistent
         with the direct searches for supersymmetric particles
         at LEP and the Tevatron are represented by points \cite{HewettWells}.
\label{fig:Hewett}}
\end{figure}

The limits on new physics imposed by the measurements can be
expressed in a model independent way by bounds imposed on 
Wilson coefficients. The $b\to s\,\gamma$ decay rate is sensitive mostly to 
the value of the $C_7$ coefficient, with slight $C_8$ dependence as 
illustrated in Fig.~\ref{fig:Hewett} \cite{HewettWells}. 
This figure also illustrates possible MSSM models, with additional
theoretical symmetries motivated by supergravity. All of these models
are made consistent with all direct searches for supersymmetric particles
at LEP and the Tevatron. They are also out of reach of LEP-II.
We see that the $b\to s\gamma$ results severely constrain the minimal 
supergravity models.

For constraints on other extensions of the Standard Model imposed
by the measured $b\to s\,\gamma$ rate see e.g. Ref.\cite{HewettSB}.

\subsection{$b\to d\gamma$ and $V_{td}$}

Detection of  $b\to d\,\gamma$ is difficult because the rates are
suppressed by $|V_{td}|^2/|V_{ts}|^2$:
${\cal B}(B\to d\,\gamma)=(0.017-0.074)\times {\cal B}(B\to s\,\gamma)$
\cite{AliAsa}.
Rejection of the dominant background from 
$b\to s\,\gamma$ decays requires good
particle identification, except for the simplest exclusive final states
in which kinematic cuts alone are very effective.
CLEO searched for  $B\to (\rho,\omega)\,\gamma$ 
decays \cite{kstargammaWaw}.
No evidence for the signal was found due to lack of sufficient experimental
statistics ($2.6\times 10^6$ $B\bar{B}$ pairs).
The following upper limits were set (90\%\ C.L.):
${\cal B}(B^0\to \rho^0\,\gamma)<3.9\times10^{-5} $,
${\cal B}(B^0\to \omega\,\gamma)<1.3\times10^{-5} $, and
${\cal B}(B^-\to \rho^-\,\gamma)<1.1\times10^{-5} $.
The ratio
${\cal B}(B\to (\rho,\omega)\,\gamma)/{\cal B}(B\to K^*\,\gamma)$
can be used to determine 
$|V_{td}|^2/|V_{ts}|^2$
after corrections for phase space and $SU(3)-$flavor symmetry-breaking
effects. Unfortunately the latter are somewhat model dependent.
Long distance interactions may further 
complicate the analysis \cite{AbS,Ali}.
From the present experimental limits CLEO obtains:
$|V_{td}|^2/|V_{ts}|^2 <0.45$--0.56,
where the range indicates the uncertainty in the theoretical factors.

\subsection{$b\to s l^+l^-$}

In the Standard Model, the  $b \to s l^+ l^-$  decay rate is expected 
to be nearly two orders of magnitude lower than the rate for
$b \to s \,\gamma$ decays \cite{Ali,refAH}.
Nevertheless, 
the $b \to s l^+ l^-$ process has received considerable
attention since it offers a deeper insight into 
the effective Hamiltonian
describing FCNC processes in $B$ decays \cite{Ali}.
While $b \to s \,\gamma$ is only sensitive to the
absolute value of the $C_7$  
Wilson coefficient 
in the effective Hamiltonian,
$b \to s l^+ l^-$ is also sensitive
to the sign of $C_7$ and
to the $C_9$ and $C_{10}$ coefficients, 
where the relative contributions vary with $l^+l^-$ mass.
These three coefficients are related to the three 
different processes contributing to $b \to s\, l^+l^-$: 
$b\to s\gamma^*\to sl^+l^-$, $b\to sZ^*\to sl^+l^-$, 
and the box diagram (see Fig.~\ref{fig:ewpenguin}).
Processes beyond the Standard Model 
can alter both the magnitude and the sign
of the Wilson coefficients.

\subsubsection{Searches in exclusive modes}

The simplest allowed final states are 
$B\to K \,l^+l^-$, and $B\to K^* \,l^+l^-$. 
Each of them is expected to constitute
$\sim10\%$  of the total $b\to s \,l^+l^-$ rate.
The most sensitive searches for these decays were 
performed by the CDF and CLEO experiments.

\begin{figure}[p]
\psfig{figure=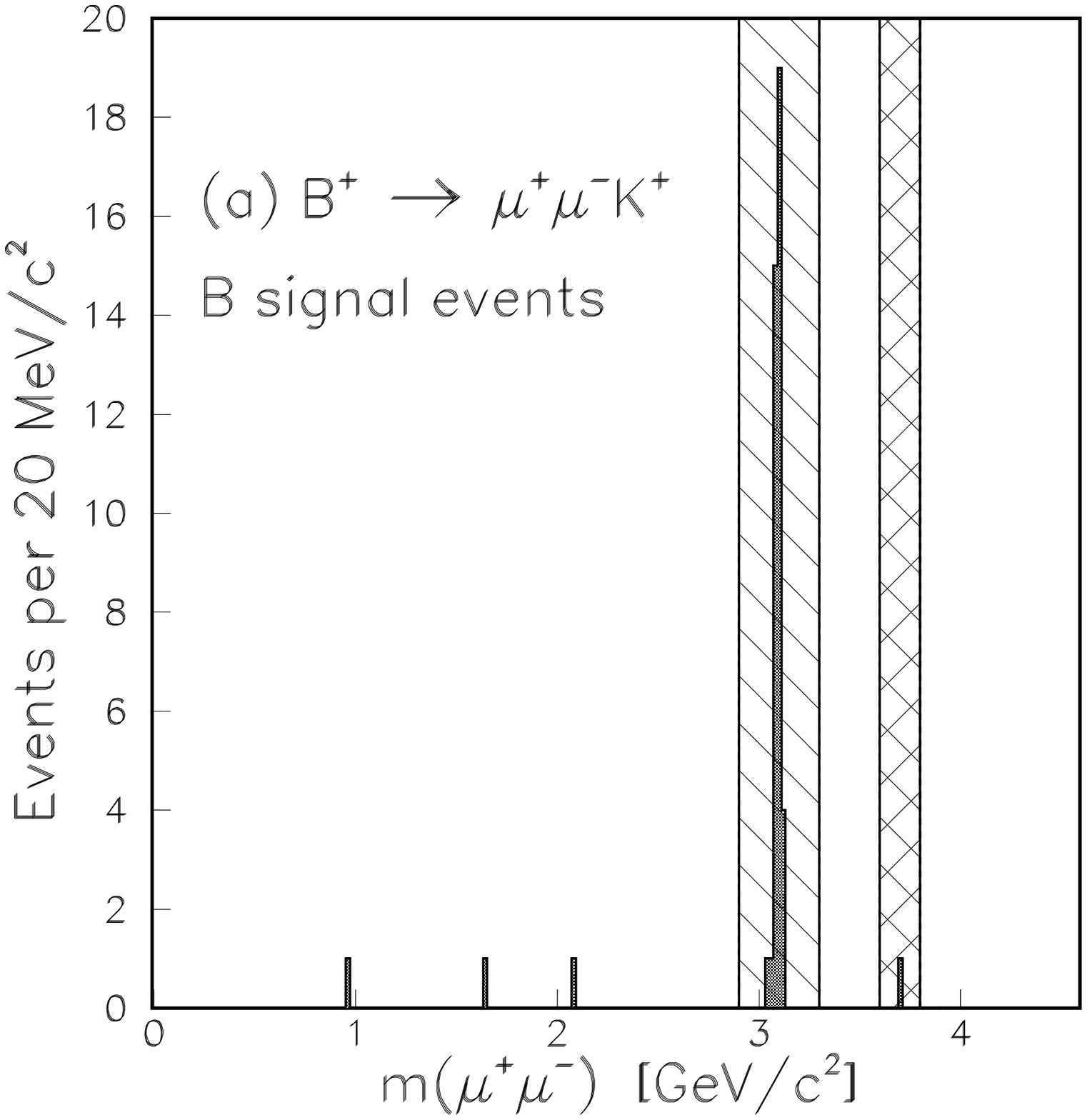,width=6.6cm}
\psfig{figure=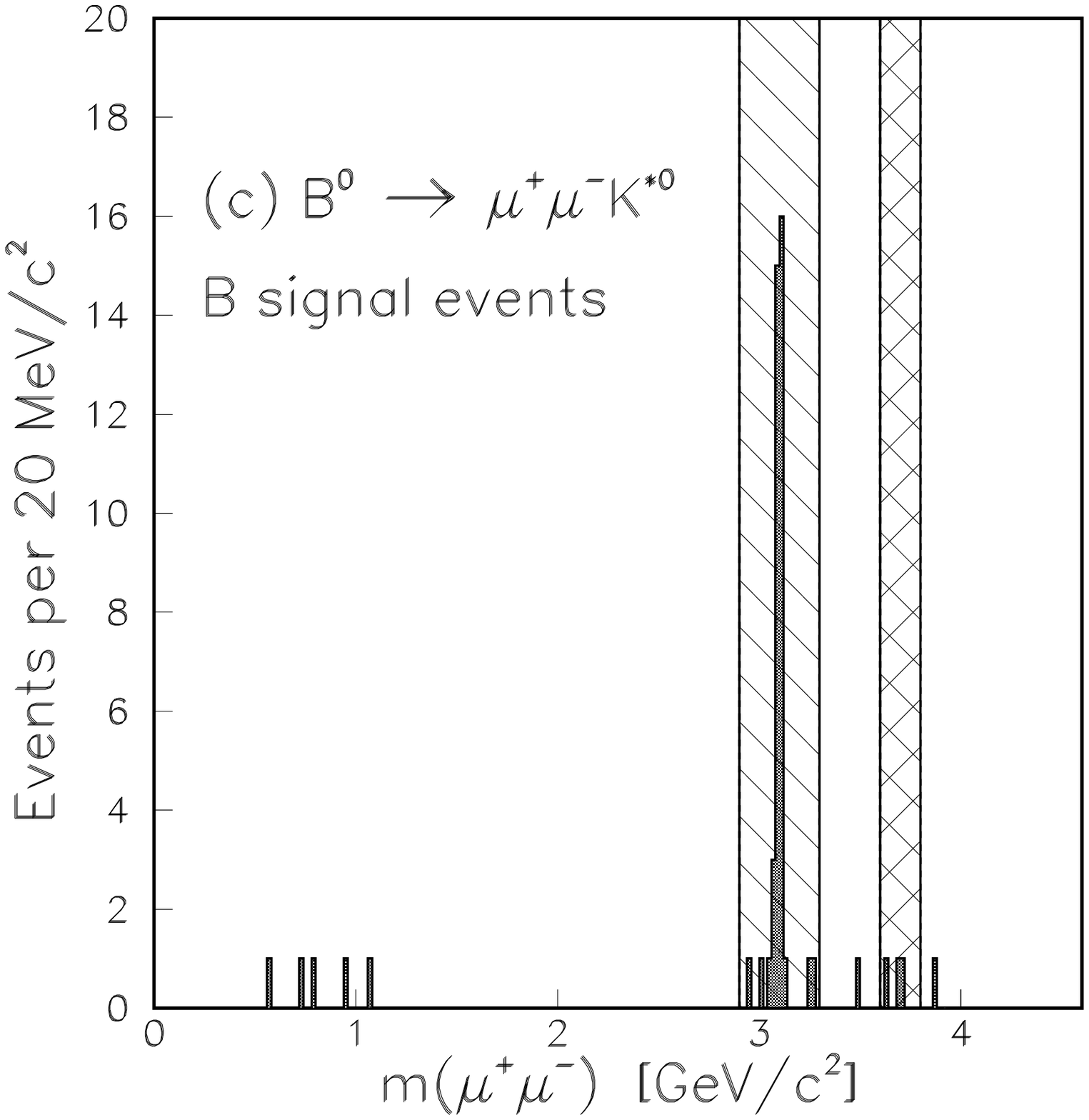,width=6.6cm}
\caption{Di-muon mass distributions in the CDF 
         search for $b\to s\,\mu^+\mu^-$ via {\bf exclusive}
         final states $B^+\to K^+\,\mu^+\mu^-$ (left) and
         $B^0\to K^{*0}\,\mu^+\mu^-$ (right).
\label{fig:CDFsll}}
%
\begin{center}
\psfig{figure=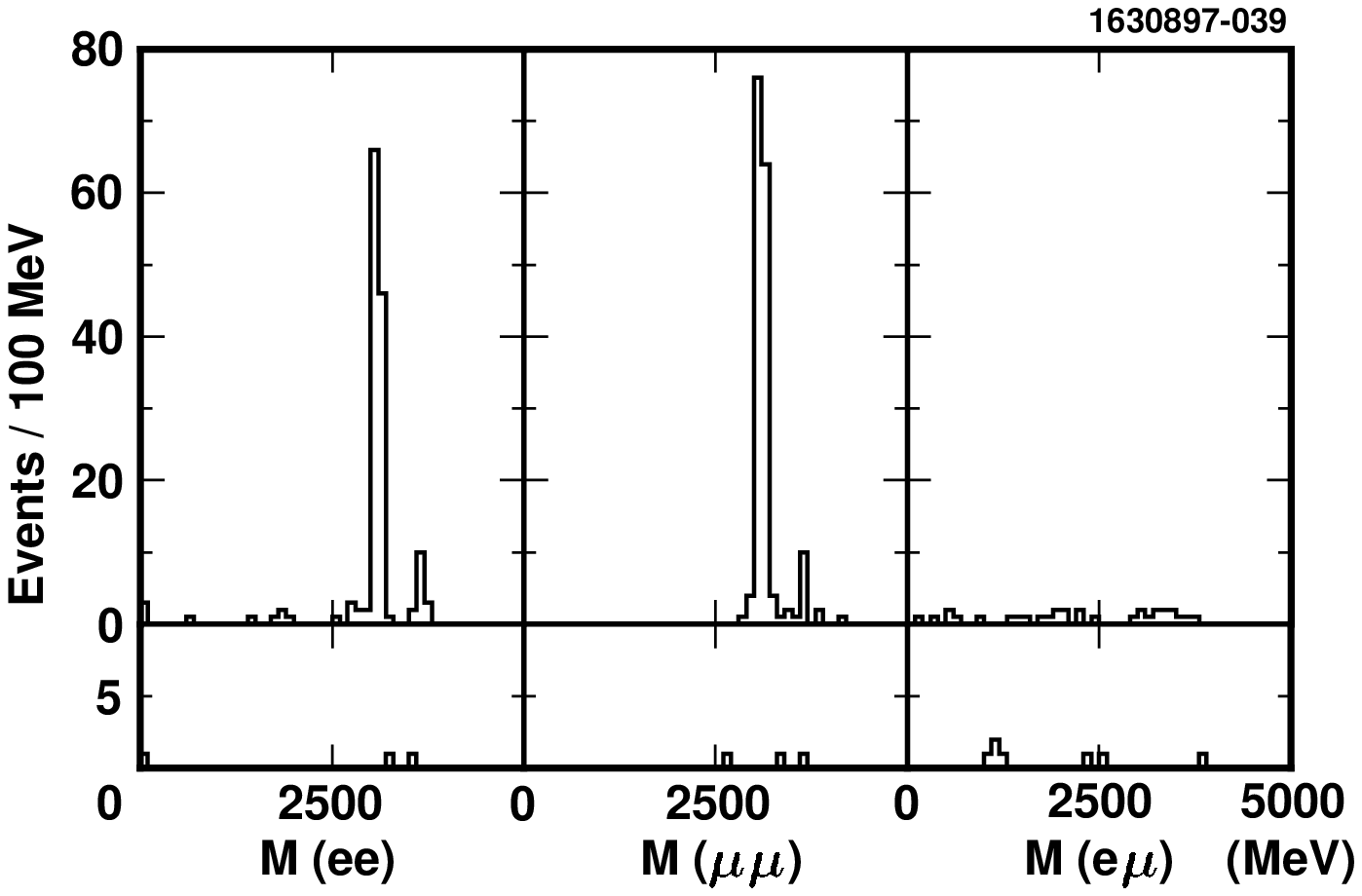,width=13.2cm}
\end{center}
\caption{Di-lepton mass distribution in the CLEO {\bf inclusive}
         search for $b\to s\,l^+l^-$. 
        On-resonance (top) and below-threshold (bottom) data are shown.
\label{fig:CLEOsll}}
\quad
\end{figure}

The CDF search \cite{sllCDF} 
is based on 17.8 pb$^{-1}$ of data 
($\sim5\times10^8$ $b\bar b$ pairs for $|\eta|<1$)
and a di-muon trigger. The backgrounds are suppressed
by transverse momentum cuts ($P_t(\mu_1)>2$, $P_t(\mu_2)>2.5$ GeV,
$P_t(K^{(*)})>2$ GeV, $P_t(B)>6$ GeV), a detached vertex
cut ($c\tau(B)>100 \mu m$), an isolation requirement and
a $B$ mass cut. The resulting di-muon mass distributions
are shown in Fig.~\ref{fig:CDFsll}.
Signals from the decays $B\to K^{(*)}\psi^{(')}$ can be seen. 
Since the branching fractions for these decays had been measured previously 
by other experiments, CDF used these
signals for normalization.
Reconstruction efficiencies are roughly
$0.13\%$ for the $K$, and $0.07\%$ for the $K^*$ modes.
A few events observed outside the $\psi$ (hatched) and
$\psi'$ (cross-hatched) bands are consistent with the background
estimates. They find the 90\%\ C.L. upper limits 
${\cal B}(B^-\to K^-\,\mu^+\mu^-)<1.0\times10^{-5}$ and
${\cal B}(B^0\to K^{*0}\,\mu^+\mu^-)<2.5\times10^{-5}$.

The CLEO II experiment searched for these decays
in a sample of $b\bar b$ pairs 
two orders of magnitude smaller 
($\sim2.2\times10^6$ $B\bar{B}$) than in the CDF analysis,
though with efficiencies larger also by two orders of 
magnitude ($\sim15\%$ for $K$ and $\sim5\%$ for $K^*$)
and suitably low backgrounds.
Thus, by coincidence, the sensitivity of the
CDF and CLEO II experiments were very similar.
In addition to the limits in the di-muon mode,
${\cal B}(B^-\to K^-\,\mu^+\mu^-)<0.9\times10^{-5}$ and
${\cal B}(B^0\to K^{*0}\,\mu^+\mu^-)<3.1\times10^{-5}$,
CLEO also set limits using di-electrons:
${\cal B}(B^-\to K^- \,e^+e^-)<1.2\times10^{-5}$ and
${\cal B}(B^0\to K^{*0} \,e^+e^-)<1.6\times10^{-5}$,
 
The experimental limits are an order of magnitude 
away from the Standard Model predictions.

\subsubsection{Inclusive searches}

The new CLEO analysis \cite{sllCLEO} 
looks for inclusive $b\to s \,l^+l^-$ decays using the inclusive $B$
reconstruction technique previously described
for $b\to s\,\gamma$ decays.
The obtained di-lepton mass spectra
are shown in Fig.~\ref{fig:CLEOsll}.
Again clear signals for $B\to X_s\psi$ and
$B\to X_s\psi'$ are observed. Events outside the
$\psi$ and $\psi'$ bands are consistent with
$B\bar{B}$ background estimates (the \qqbar\ background
is small). With a sample of $3.3\times10^6$ $B\bar{B}$
pairs and reconstruction efficiencies around $5\%$,
CLEO sets 90\%\ C.L.\ upper limits,
${\cal B}(b\to s \,e^+e^-)<5.7\times10^{-5}$ and
${\cal B}(b\to s \,\mu^+\mu^-)<5.8\times10^{-5}$
(combined: ${\cal B}(b\to s \,l^+l^-)<4.2\times10^{-5}$),
The SM predictions \cite{refAH},
${\cal B}(b\to s \,e^+e^-)=(0.8\pm0.2)\times10^{-5}$ and
${\cal B}(b\to s \,\mu^+\mu^-)=(0.6\pm0.1)\times10^{-5}$,
are again an order of magnitude below the experimental limits.

The upper limit on inclusive $b\to s\,\mu^+\mu^-$
previously presented by the UA1 experiment 
at $Sp\bar pS$ collider \cite{UA1}
has been shown recently to be
based on overestimated sensitivity \cite{sllCLEO,D0}.
A similar analysis recently completed by the D0 experiment at the 
Tevatron resulted in a less stringent limit, $<32\times10^{-5}$  \cite{D0},
than achieved by CLEO.

\subsection{$b\to s \nu\bar\nu$}

\begin{figure}[htbp]
\begin{center}
\psfig{figure=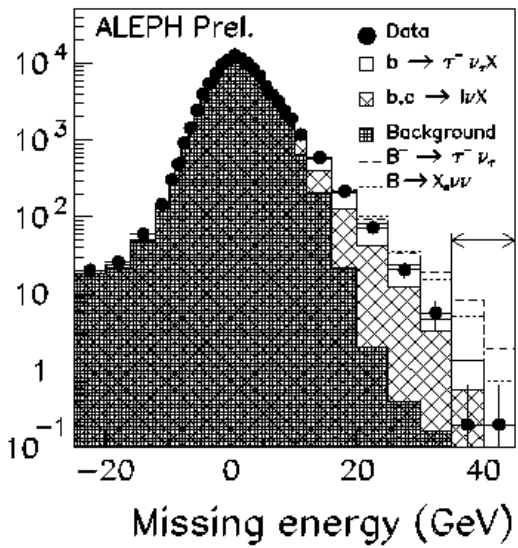,width=7cm}
\end{center}
\caption{Missing energy in a hemisphere for the selected $b\bar{b}$
         events by ALEPH (points). Shaded histograms show 
         the estimated background distribution. The expected 
         $b\to s\,\nu\bar\nu$ signal shape is indicated by a
         dotted line. The two highest bins are used to set
         the upper limit.}
\label{fig:nunu}
\end{figure}

The rate for  $b\to s\nu\bar \nu$
is enhanced compared to the
$b\to s \,l^+l^-$ decays primarily
by summing over three neutrino flavors 
($b\to s\tau^+\tau^-$ has a small expected rate and
will be difficult to detect experimentally).
The predicted rate is only a factor of ten lower than
for $b\to s\,\gamma$ \cite{BurasWaw}: $(3.8 \pm 0.8)\times 10^{-5}$.  
In principle, these decays are the cleanest theoretically 
among all penguin decays.
Therefore, a measurement of the inclusive rate for this
process would be of considerable interest.
Unfortunately, the neutrinos escape detection
making it difficult for experimentalists to control
the backgrounds. 
So far, only LEP experiments have been able to
probe these decays by requiring very large 
missing energy in a hemisphere \cite{nunuALEPH,nunuThE}. 
Semileptonic backgrounds are reduced by eliminating events
with an identified lepton in the signal hemisphere.
A detached vertex in the opposite hemisphere 
suppresses non-$b\bar b$ backgrounds.
The missing energy distribution in a $b-$hemisphere
obtained by ALEPH \cite{nunuALEPH}
in a sample of $\sim0.5\times 10^6$ $b\bar{b}$ pairs
is shown in Fig.~\ref{fig:nunu}.
A signal would be seen as an excess of events at large missing energy.
The lack of such an excess is used by ALEPH to obtain a 90\%\ C.L.  limit,
${\cal B}(b\to s\,\nu\bar\nu)<7.7\times10^{-4}$.

The exclusive mode $B\to K^*\,\nu\bar\nu$ is expected to 
constitute about 30\%\ of the total rate \cite{am}.
DELPHI obtains 90\%\ C.L upper limits \cite{delphirarepub}:
${\cal B}(B^0\to K^{*0}\,\nu\bar\nu)<1.0\times10^{-3}$ and
${\cal B}(B_s\to \phi\,\nu\bar\nu)<5.4\times10^{-3}$.

The inclusive limit set by ALEPH is an order of magnitude
away from the expected rate.
Unfortunately, no more data are expected at the $Z^0$ peak
at LEP. 
Perhaps $\Upsilon(4S)$ experiments will be able to
develop analysis techniques which can probe these decays
with future high statistics data samples.
It is hard to imagine that experiments at hadronic
colliders will ever have any sensitivity to these decays.

\subsection{$B\rightarrow \gamma\gamma$}

The vertical penguin $B^0 (B^0_s) \rightarrow \gamma\gamma$
(Fig.~\ref{fig:goofypeng}a)
decay rate is expected to be of order
$10^{-8}\ (10^{-7})$ \cite{ref:llpred,ref:bsgamgam}.
QCD corrections have been found to enhance the
$B_s^0 \rightarrow \gamma\gamma$ rate by $\sim 50\%$ \cite{ref:bsgamgam}.
The L3 detector at LEP has a high resolution
calorimeter, which is well suited to
this analysis, with a photon resolution better than 2\% for energies
greater than 1 GeV \cite{ref:L3gamgam}.  Using their $B$-enriched data sample,
L3 looks for pairs of high momentum photons with invariant mass 
near the $B$ mass.
The data are fit simultaneously for $B^0$ and $B_s^0$ signals
with an exponential background.
(No candidates are found in a $\pm 2\sigma$ mass window.)
The L3 90\% C.L. upper limits are given in
Table~\ref{tab:goofy}. Unfortunately, these limits are several
orders of magnitude away from constraining new physics.

\begin{center}
\begin{table}
\caption{Results for vertical penguins.
We give the 90\% confidence level upper limit on the branching
fractions (UL $\cal B$) and the theoretical predictions.}
\vspace{0.3cm}
\begin{tabular}{l l r c | l l r c}
\hline\hline
$B^0$ decay    &        & UL ${\cal B}$  &             & $B_s$ decay   &      & UL ${\cal B}$ &           \\
mode           & Exp.   &    ($10^{-6}$) & Theory      & mode          & Exp. & ($10^{-6}$)   & Theory    \\ \hline
$\gamma\gamma$ & L3     & 38             & $10^{-8} $  &$\gamma\gamma$ & L3   & 148           & $10^{-7 }$ \\
$e^+e^-$       & CLEOII & 5.9            & $10^{-15}$  &$e^+e^-$       & L3   & 54            & $10^{-14}$ \\
$\mu^+\mu^-$   & CDF    & 0.68           & $10^{-10}$  &$\mu^+\mu^-$   & CDF  & 2             & $10^{-9 }$ \\
$\tau^+\tau^-$ &        &                & $10^{-8 }$  &$\tau^+\tau^-$ &      &               & $10^{-7 }$ \\
\hline
\end{tabular}
\label{tab:goofy}
\end{table}
\end{center}

\subsection{$B \rightarrow \ell^+\ell^-$}
The Standard Model predictions \cite{ref:llpred}
for $B \rightarrow \ell^+\ell^-$ are given in Table~\ref{tab:goofy}.
The search for $B^0 \rightarrow e^+e^-$ and
$B^0 \rightarrow \mu^+\mu^-$ at CLEO \cite{ref:CLEOleplep} is similar to
their other rare $B$ decay searches.  The small background is suppressed
using mild event shape requirements.  No signal events are observed
in either channel giving limits on each channel of $5.9\times 10^{-6}$.

Since the CDF experiment has
excellent muon identification and a muon trigger, 
$B \rightarrow \mu^+\mu^-$
decays could be recorded with good efficiency \cite{ref:CDFmumu}.
$B^0_s$ mesons are also produced at CDF.
Background rejection is achieved using a detached vertex and
an isolation requirement.  In the CDF Run IA and IB data sample
($\approx 100~{\rm pb}^{-1}$) they find one candidate which
falls into both the $B^0$ and $B^0_s$ mass windows.  This event
is consistent with background, but is assumed to be signal
for calculating upper limits.  Results are included in
Table~\ref{tab:goofy}.

Searches for $B \rightarrow \ell^+\ell^-$ have also been done
by L3 \cite{ref:L3leplep}.  The small background is suppressed
by requiring large $B$ energy.  They observe no significant signals
and fit the data simultaneously
for $B^0$ and $B_s^0$ to extract upper limits on the branching
fractions.  The L3 limits for $B^0
\rightarrow e^+e^-\ (\mu^+\mu^-)$ are $1.4(1.0)\times 10^{-5}$,
less stringent than the CLEO and CDF limits.
The L3 limit for $B^0_s
\rightarrow e^+e^-$ is $ 5.4 \times 10^{-5}$, the only
limit for this mode.  Their limit for
$B_s^0 \rightarrow \mu^+\mu^-$,  $3.8\times 10^{-5}$,
is less restrictive than the CDF limit.

\section{GLUONIC PENGUINS}\label{sec:glupeng}

\begin{figure}[tb]
\centerline{\psfig{figure=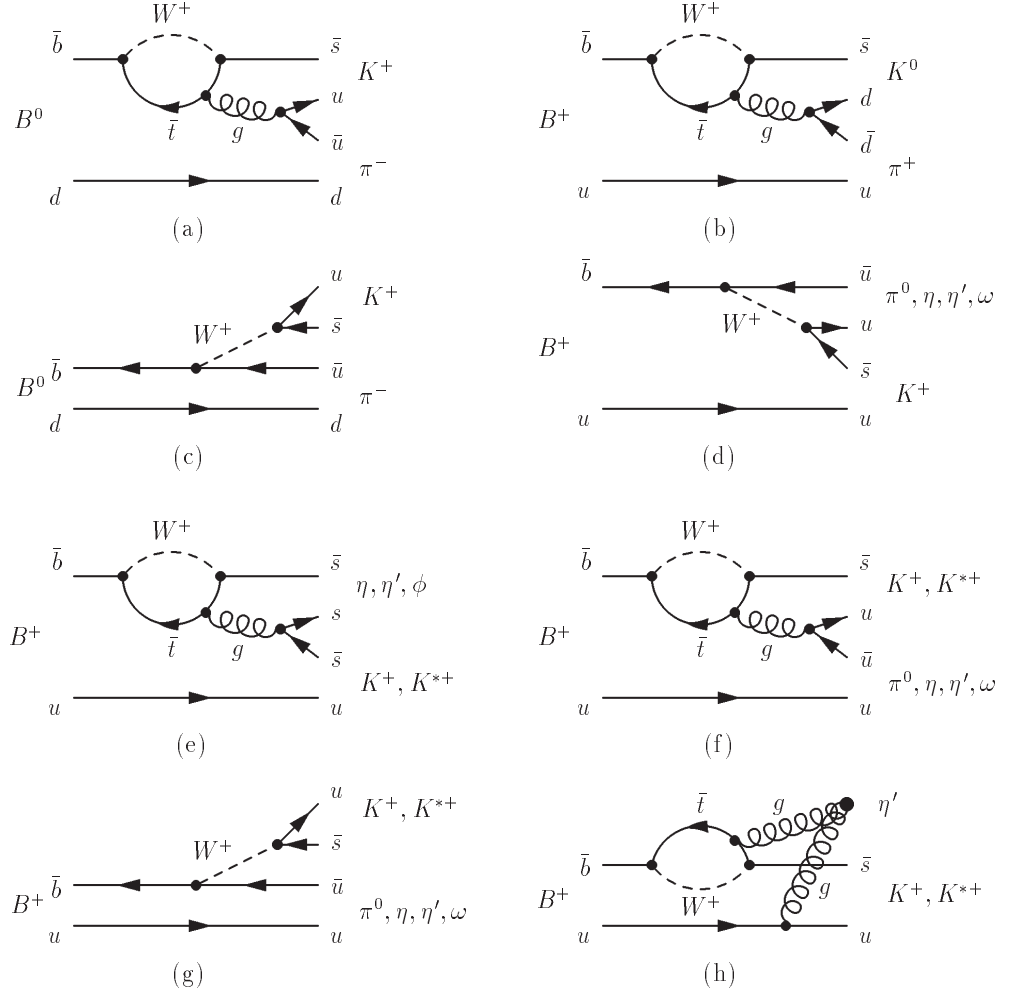,width=13.2cm}}\bigskip
\caption{Feynman diagrams for some of the penguin and tree processes 
which are expected to be dominant for the modes described in this paper.}
\label{fig:glufeyn}
\end{figure}

Some representative Feynman diagrams relevant for charmless hadronic $B$ decays
are shown in Fig. \ref{fig:glufeyn}.  In addition to the penguin diagrams
of primary interest in this review, we indicate examples of $b\ra u$
tree diagrams in Fig. \ref{fig:glufeyn}c, d, and g.  Such diagrams
are suppressed for decays with a strange particle in the final state but
are dominant for many decays with no strange particle.  Since penguins
play a significant role in such decays without strange particles and the
two processes can be difficult to separate experimentally, we will
include results for both processes in our discussion.  However we do not
include results for decays involving the dominant $b\ra c$ decay
mechanism.

The first indication of signals in charmless hadronic $B$ decays came in a 1993 
CLEO publication \cite{CLEObkpi} in which a quite significant signal was found
for $B\ra h^+h^-$, where $h$ is either $K$ or $\pi$.  Statistics were not 
sufficient to obtain significant signals for \Bkpi\ or \Bpipi\ separately.
Subsequently CLEO
updated these results \cite{bigrare}, still without an observation of either 
mode individually, and provided limits for many related modes.

\begin{figure}[htbp]
\label{fig:alephglu}
\begin{center}
\psfig{figure=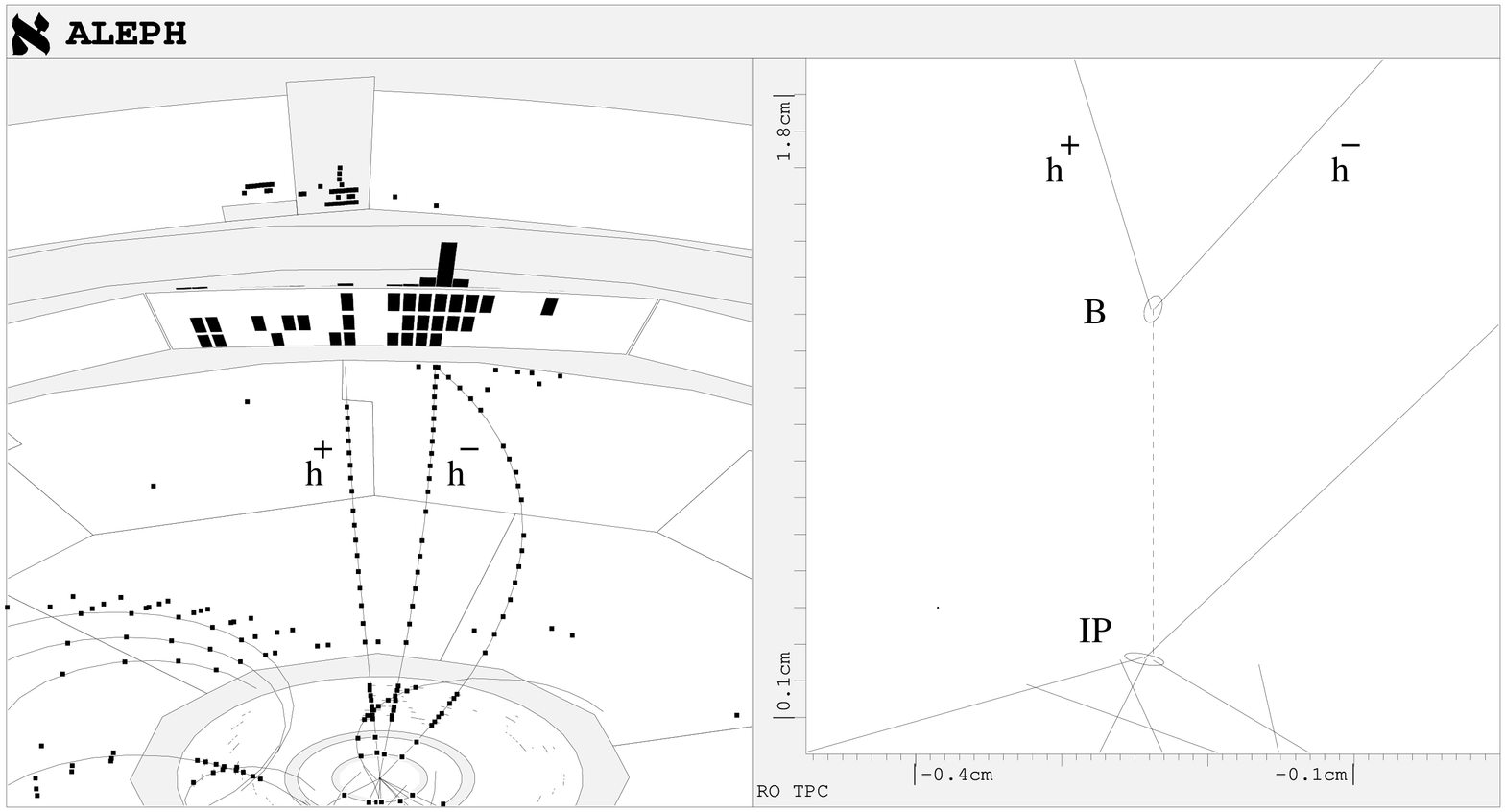,height=2.9in,angle=270}\bigskip
\caption{Event from the ALEPH experiment showing a \Bkpi\ decay
candidate as reconstructed with use of the ALEPH silicon vertex detector.}
\end{center}
\bigskip
\begin{center}
\psfig{figure=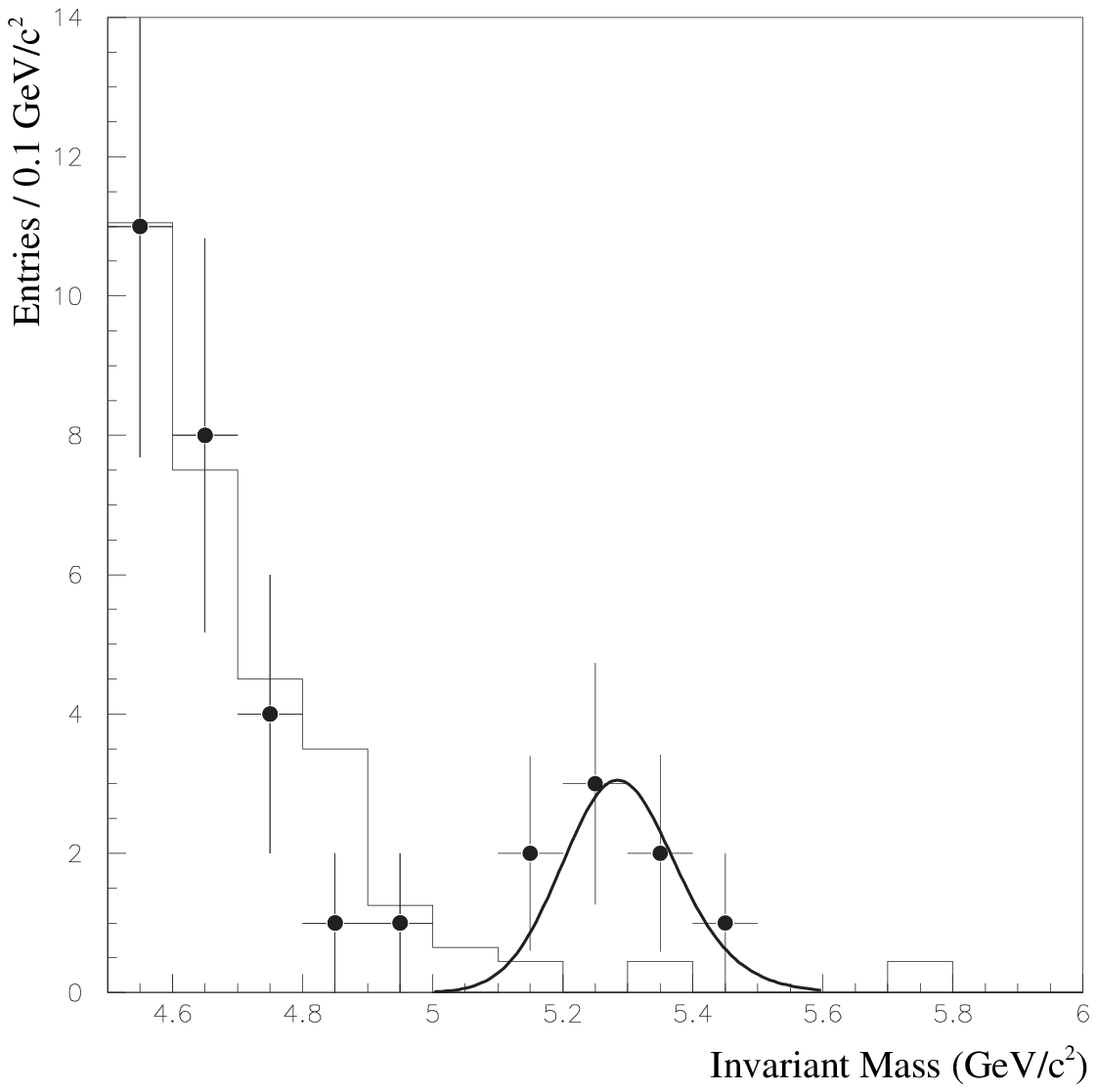,height=3.5in}\bigskip
\caption{Invariant mass distribution obtained by DELPHI for 
ten charmless hadronic $B$ decay candidates, background events at lower mass, 
and the Monte Carlo expectation for the background.}
\label{fig:delphirare}
\end{center}
\end{figure}

Several LEP experiments have used the excellent
vertex resolution provided by their silicon vertex detectors to obtain
virtually background-free evidence for charmless hadronic $B$ decays.
Examples of such events are shown in Fig. \ref{fig:alephglu} for the ALEPH
experiment \cite{alephrarepub}. The LEP experiments also have some
ambiguity between the decays \Bkpi, \Bpipi, and
$B_s^0\ra K^+K^-$.  Fig. \ref{fig:delphirare} shows
the mass distribution for ten candidate charmless hadronic $B$ decays
from the DELPHI experiment \cite{delphirarepub}.  The final states include
\pipi, \kpi, \kk, $\rho^0\pi^+$, $K^{*0}\pi^+$, $K^+\rho^0$, and $K^+a_1^-$.

\subsection{Exclusive decays of $B$ mesons}

In this section we discuss the recent observation of several new
decay modes by CLEO and many upper limits which are beginning to
approach theoretical expectations.

\subsubsection{Experimental method}

In addition to the 3.3 million \BBbar\ pairs from $\Upsilon(4S)$ running
(see Table \ref{tab:benvir}), a sample of events
obtained at a center-of-mass energy below \BBbar\ threshold is
used for studies of backgrounds from continuum \qqbar\ production.
Resonance states are reconstructed from charged tracks and photons
with the decay channels: \etaprepp, \etaprrg, $K^0$ via $K_S\ra\pi^+ \pi^-$,
$\rho^0\ra\pi^+\pi^-$, $\rho^+\ra\pi^+\pi^0$, $\pi^0 \ra \gamma\gamma$, 
\etagamgam, \etathreepi, \omegappp, $\phi\ra K^+K^-$, $K^{*0}\ra K^+\pi^-$,
$K^{*+}\ra K^+\pi^0$, and $K^{*+} \ra K^0\pi^+$.

The primary means of identification of $B$ meson candidates is through their
measured mass and energy.  The dominant background process for all
decays considered here is continuum \eetoqq\ production.  Signal events
have a total energy consistent with the beam energy (5290 MeV), while
most background events have smaller energy.  As discussed in 
Sec. \ref{sec:Kstgam}, CLEO uses the variable $\DE=E_{\rm cand}-E_{\rm beam}$, 
which peaks at or near zero for signal events.  Signal events also have 
mass $M$ consistent with $m_B$ (5280 MeV), only 10 MeV below
the maximum value, while most background events have much smaller mass.
For states decaying to the resonances mentioned above, the resonance 
mass is also important in discriminating against background processes.
In the case of vector-pseudoscalar decays and the \etaprrg\ channel, 
the helicity angle distribution is also used.
For modes with a high-momentum charged track or a charged track
paired with a $\pi^0$, $dE/dx$ information is also used to identify the
charged track as a pion or kaon.

The large \qqbar\ background can be reduced by about an order of magnitude
with the use of event shape information.  Since $B$ mesons are 
produced nearly at rest, \BBbar\ events tend to be spherical,
whereas \qqbar\ events tend to be quite collimated (``jetty").

In order to keep the efficiency high while still effectively rejecting
\qqbar\ background, a maximum likelihood (ML) fit has been performed for 
all of the recent CLEO analyses.  The inputs to
the fit are the quantities discussed in the previous two paragraphs,
while the outputs are the number of signal and background events.  
This procedure provides an efficiency of 20-50\% for most modes, at least a
factor of two larger than previous CLEO and
ARGUS analyses, with a comparable effective background.  

\subsubsection{$B\ra K\pi$ and related decays}
\label{sec:kpi}

The simplest modes, both experimentally and theoretically are the
decays to two-body final states without resonances.  This includes
$K\pi$ and $\pi\pi$ final states with both charged and neutral pions.
The CLEO results \cite{kpipub} include observation of the decay \Bkpi, strong
evidence for the decay \Bkzpi, and limits for the other five final
states.  The signals are summarized in Table \ref{sigtab}; the
limits for the penguin modes are summarized in Table \ref{pengtab}
while those for the modes dominated by $b\ra u$ processes are included in 
Table \ref{treetab}.

\begin{figure}[hbtp]
\begin{center}
\psfig{figure=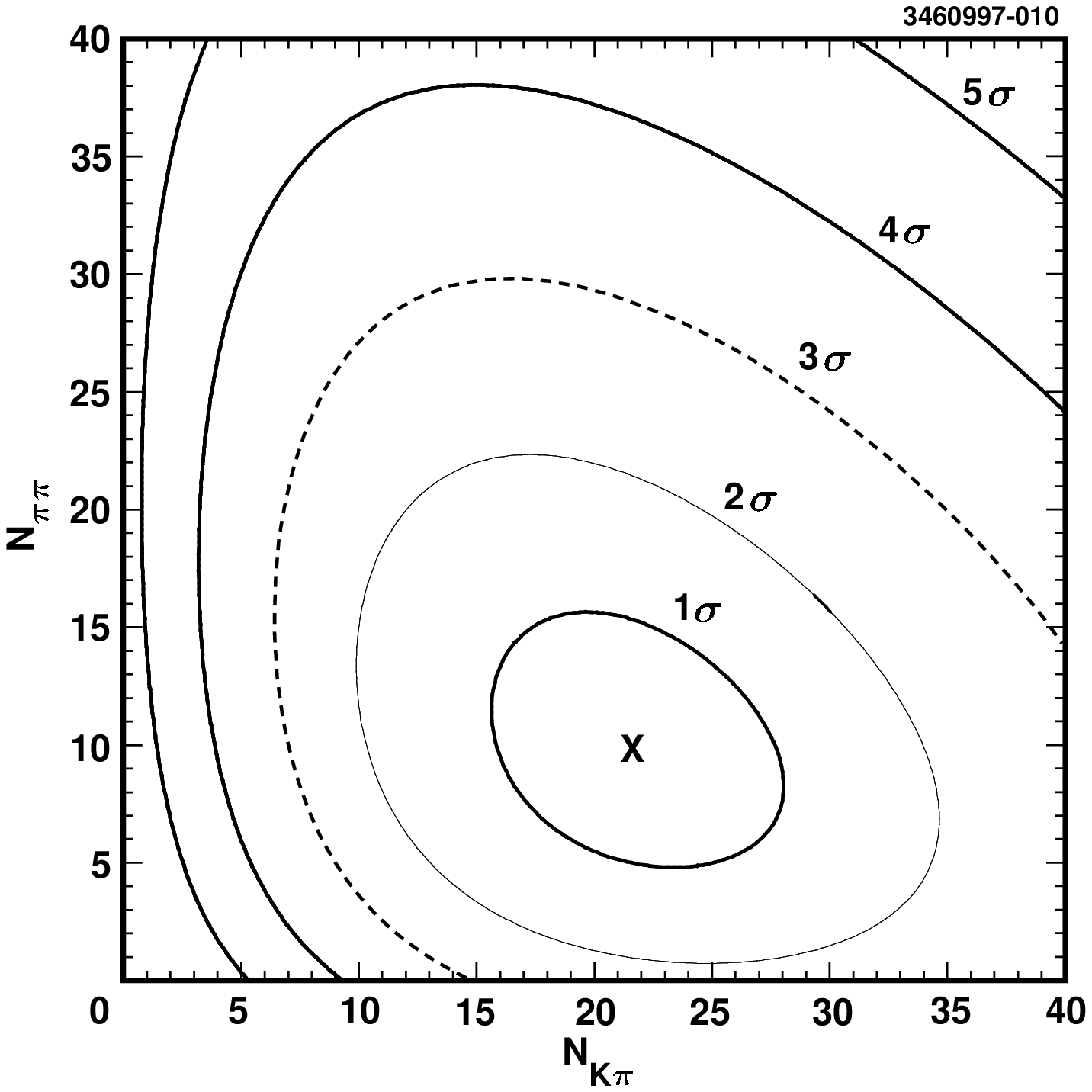,height=1.7in}
\psfig{figure=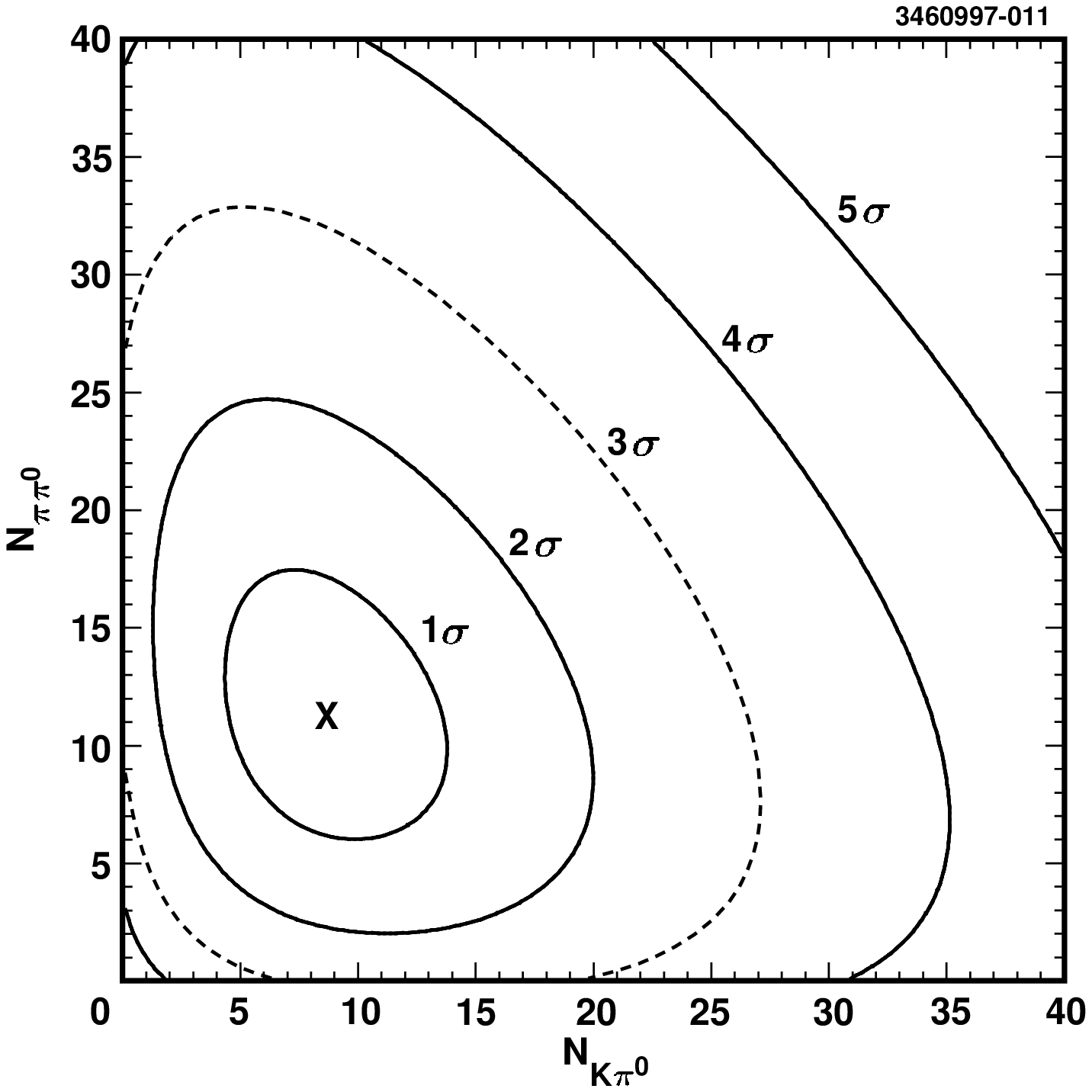,height=1.7in}
\psfig{figure=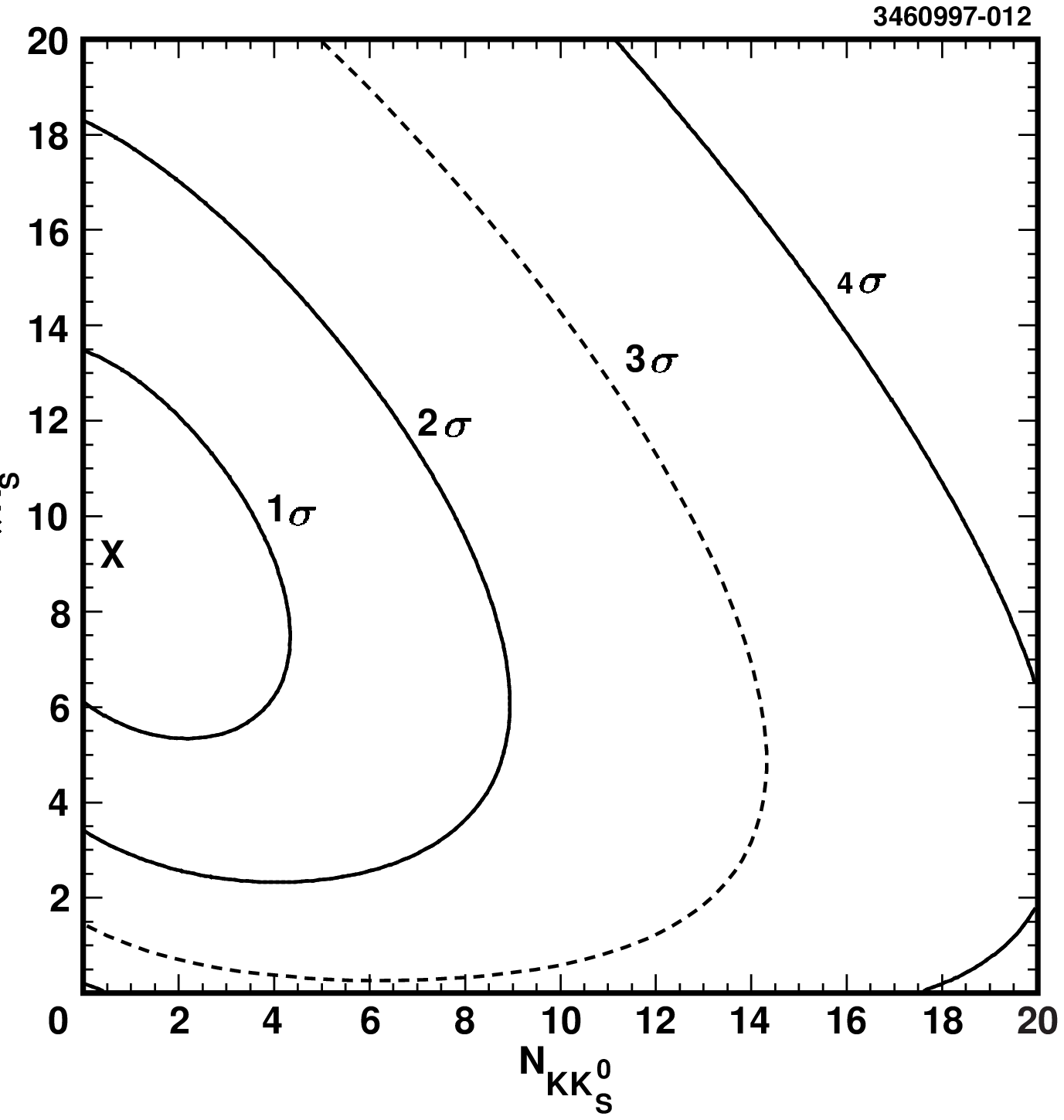,height=1.7in}
\end{center}
\caption{Likelihood contours for:
(a) $B^0\ra K^+\pi^-$ and $B^0\ra \pi^+\pi^-$; (b) 
$B^+\ra K^+\pi^0$ and $B^+\ra \pi^+\pi^0$; (c) 
$B^+\ra \bar{K}^0K^+$ and $B^+\ra K^0\pi^+$.}
\label{fig:contourkpi}
\begin{center}
\psfig{figure=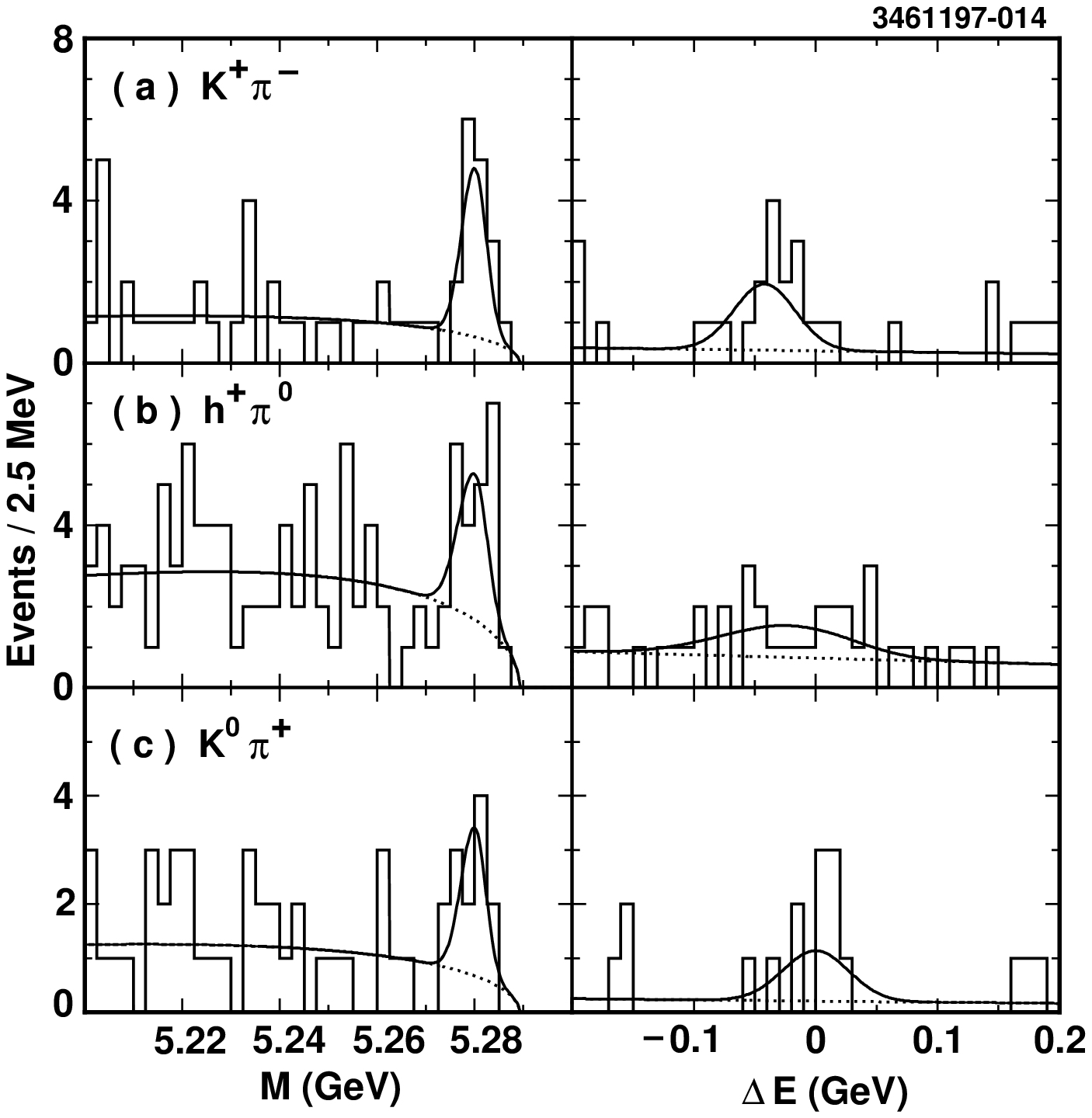,height=3.5in}
\end{center}
\caption{$M$ and $\Delta E$ plots for (a) $B^0\ra K^+\pi^-$,
(b) $B^+\ra h^+\pi^0$, and (c) $B^+\ra K^0\pi^+$.
The scaled projection of the total likelihood fit (solid curve)
and the continuum background component (dotted curve) are overlaid.}
\label{fig:kpiproj}
\end{figure}

Fig.~\ref{fig:contourkpi} shows
the likelihood contours for the three cases with
significance greater than three standard deviations.  While the
significance for the $K^+\piz$ and $\pi^+\piz$ final states are both
below $3\sigma$, there is strong evidence for their sum ($h\piz$).
This is similar to the
case of the $h^+h^-$ final state several years ago \cite{CLEObkpi}.
Fig.~\ref{fig:kpiproj} shows projections of the fit onto the $M$ and \DE\ axes; 
for all projection plots, cuts have been made on other ML variables
to better reflect the background near the signal region.

\begin{table}[htbp]
\caption{Results for the five decay modes which have signals with
significance larger than 3$\sigma$.  The second
efficiency column includes $B_r$, the product branching fraction of
resonance secondary decays.}
\vspace{0.4cm}
\def\notext{ & & \cr}
\begin{center}
\begin{tabular}{lcccccc}
\dbline
          & Signal   &         &                &                    & 
      &Theory\cr
Decay mode&events& Signif. & $\epsilon$(\%) &$\epsilon\calB_r$(\%)&\calB($10^{-5})$&\calB($10^{-5})$\cr

\sgline
\Bkpi     &$21.6^{+6.8}_{-6.0}$&5.6$\sigma^\dag$&44&44&$1.5^{+0.5}_{-0.4}\pm0.1$ & 0.1--2.4 \cr
\Bkzpi    & $9.2^{+4.3}_{-3.8}$&3.2$\sigma^\dag$&35&12&$2.3^{+1.1}_{-1.0}\pm0.4$ &  0.5--2.0\cr
\Betaprkp &$33.1^{+8.1}_{-6.8}$&7.5$\sigma$     &28&15&$6.5^{+1.5}_{-1.4}\pm0.9$&0.7--4.1\cr
\Betaprkz &$ 7.1^{+4.1}_{-3.0}$&3.8$\sigma$     &25& 4&$4.7^{+2.7}_{-2.0}\pm0.9$&0.9--3.3\cr
\Bomegakp &$12.2^{+5.5}_{-4.5}$&3.9$\sigma$     &28&25&$1.5^{+0.7}_{-0.6}\pm0.2$&0.1--0.7\cr
\sgline
\end{tabular}
\end{center}
$\dag$ ~~The significance does not include systematic uncertainties.
\label{sigtab}
\end{table}

\def\notext{\omit&\omit&\omit&\omit&\omit&\omit&\omit&\omit\\}
\begin {table}[p]
\begin{center}
\caption{Results from measurements of penguin-candidate $B$ decay modes.  We 
give the 90\% confidence level upper limit on the branching fractions (UL 
$\cal B$), and the range of theoretical predictions 
\cite{oldglurefs,newglurefs,baryonrefs}.
$K^*$, $K_1$ and $K_2^*$ are shorthand designations for $K^*(892)$, 
$K_1(1400)$ and $K^*_2(1430)$, respectively.  Modes indicated by ``**" 
have been observed (see Table \ref{sigtab}).  Limits below
the upper end of the theoretical range are indicated in bold face.
Experiment key: AR (ARGUS \cite{arguspeng}), C1 (CLEO I
\cite{cleo1peng}), C2 (CLEO II \cite{bigrare,kpipub,etaprpub,omegapub}), 
D (DELPHI \cite{delphirarepub}).}
\begin {tabular}{l c r c|l c r c}
\dbline
$B^0$ final &  & UL $\cal B$ &Theory &$B^+$ final&  & UL $\cal B$&Theory\\
state& Expt. & $(10^{-6})$ &($10^{-6}$)&state & Expt.& $(10^{-6})$&($10^{-6}$)\\
\sgline
$K^+\pi^-$      &C2& ** & 1--26   &$K^+\pi^0$      &C2&  16&  3--15\\
$K^0\pi^0$      &C2&  41&  2--10  &$K^0\pi^+$      &C2& ** &  5--36\\
$K^+K^-$        &C2&   4&         &$K^+K^0$        &C2&  21&0.6--3\\
$K^0K^0$        &C2&  17&0.6--1.4 & & & \\
\notext\notext
$K^{*+}\pi^-$   &C2&  72& 1--19   &$K^{*+}\pi^0$   &C2&  99&0.5--9\\
$K^{*0}\pi^0$   &C2&  28&  1--5   &$K^{*0}\pi^+$   &C2&  41& 4--12 \\
$K^{*0}K^0$     &  &    &0.3--1   &$K^{*+}K^0$     &  &    &0.0005--0.04\\
$K^{*+}K^-$     &  &    &         &$K^{*0}K^+$     &  &    &0.3--1 \\
$K^{*0}K^{*0}$  &  &    &0.3--1   &$K^{*+}K^{*0}$  &  &    &0.3--1 \\
\notext\notext
$K^+\rho^-$     &C2&  35&0.2--2   &$K^+\rho^0$     &C2&  19&0.1--2\\
$K^0\rho^0$     &C2&  39&0.04--0.8&$K^0\rho^+$     &C2&  48&0.01--0.4\\
\notext\notext
$K^{*0}\rho^0$  &AR& 460& 0.5--6  &$K^{*+}\rho^0$  &AR& 900&0.06--8\\
$K^{*+}\rho^-$  &  &    & 0.3--18 &$K^{*0}\rho^+$  &  &    & 5--13\\
\notext\notext
$K^0\eta$       &C2&  33&0.07--3  &$K^+\eta$       &C2&  14& 0.2--5\\
$K^{*0}\eta$    &C2&  30&0.03--9  &$K^{*+}\eta$    &C2&  30& 0.2--6\\
$K^0\etap$      &C2& ** & 9--41   &$K^+\etap$      &C2& ** & 7--65\\
$K^{*0}\etap$   &C2&  39&0.05--8  &$K^{*+}\etap$   &C2& 130&0.03--1.5\\
\notext\notext
$K^0\omega$     &C2&  57&0.02--10 &$K^+\omega$     &C2& ** &0.2--13\\
$K^{*0}\omega$  &C2&  23&0.6--12  &$K^{*+}\omega$  &C2&  87&0.4--15\\
\notext\notext
$K^0f_0$        &C1& 360&         &$K^+f_0$        &C1&  80& \\
$K^{*0}f_0$     &C1& 170&         & & & \\
\notext\notext
$K^{+}_1\pi^-$  &AR&1100&         &$K^{0}_1\pi^+$  &AR&2600&  \\
$K^{0}_1\rho^0$ &AR&3000&         &$K^{+}_1\rho^0$ &AR& 780&  \\
$K^{*+}_2\pi^-$ &AR&2600&         &$K^{*0}_2\pi^+$ &AR& 680&  \\
$K^{*0}_2\rho^0$&AR&1100&         &$K^{*+}_2\rho^0$&AR&1500& \\
\notext\notext
$K^+a_1^-$      & D& 230&         & & & \\
\notext\notext
$K^0\phi$       &C2&  31&0.3--18  &$K^+\phi$       &C2&{\bf5}&0.3--14\\
$K^{*0}\phi$    &C2&{\bf21}&0.2--31&$K^{*+}\phi$   &C2&  41&0.2--31\\
$K^{0}_1\phi$   &AR&5000&  21     &$K^{+}_1\phi$   &AR&1100& 21\\
$K^{*0}_2\phi$  &AR&1400&  0.7    &$K^{*+}_2\phi$  &AR&3400& 0.7\\
\notext\notext
$\pi^0\phi$     &C2&   5&0.002--0.2&$\pi^+\phi$    &C2&   5&0.005--0.4 \\
$\rho^0\phi$    &C2&  13&0.002--0.3&$\rho^+\phi$   &C2&  16&0.004--0.5 \\
\notext\notext
$\eta\phi$      &C2&   9&0.001--0.1& & & \\
$\etapr\phi$    &C2&  31&0.001--0.1& & & \\
$\omega\phi$    &C2&  21&0.002--0.3& & & \\
$\phi\phi$      &C2&  12&         & & & \\
\notext\notext
                &  &    &         &$\bar\Lambda p$ &C1&  60&       \\
\sgline
\end {tabular}
\label{pengtab}
\end{center}
\end {table}

\def\notext{\omit&\omit&\omit&\omit&\omit&\omit&\omit&\omit\\}
\begin {table}[p]
\begin{center}
\caption{Results from measurements of other charmless $B$ decays.  We 
give the 90\% confidence level upper limit on the branching fractions (UL 
$\cal B$), and the range of theoretical predictions 
\cite{oldglurefs,newglurefs,baryonrefs}.
Limits below the upper end of the theoretical range are indicated in bold face.
Experiment key: AL (ALEPH \cite{alephrarepub}), AR (ARGUS \cite{argustree}), 
C1 (CLEO I \cite{cleo1tree}), 
C2 (CLEO II \cite{bigrare,kpipub,etaprpub,omegapub}).}
\vspace{0.3cm}
\begin {tabular}{l c r c|l c r c}
\dbline
$B^0$ decay &  & UL $\cal B$ &Theory &$B^+$ decay&  & UL $\cal B$&Theory\\
mode& Expt. & $(10^{-6})$ &($10^{-6}$)&mode & Expt.& $(10^{-6})$&($10^{-6}$)\\
\sgline
$\pi^+\pi^-$      &C2&{\bf15}&7--18&$\pi^+\pi^0$   &C2&  20& 3--20\\
$\pi^0\pi^0$      &C2&   9&0.1--1.3& & & & \\
\notext\notext
$\rho^\pm\pi^\mp$ &C2&  88& 26--52 &$\rho^+\pi^0$  &C2&  77&11--27\\
$\rho^0\pi^0$ 	  &C2&  24& 0.9--2 &$\rho^0\pi^+$  &AL&  32&0.4--8\\
\notext\notext
$\rho^+\rho^-$    &AR&2200& 13--34 &$\rho^+\rho^0$ &AL& 120& 6--23\\
$\rho^0\rho^0$    &AL&  40& 0.5--3 & & & & \\
\notext\notext
$\eta\pi^0$       &C2&   8&0.2--4  &$\eta\pi^+$    &C2&  15&1.9--8 \\
$\eta\rho^0$      &C2&  13&0.02--7 &$\eta\rho^+$   &C2&  32& 4--17\\
$\etap\pi^0$      &C2&{\bf11}&0.06--14&$\etap\pi^+$&C2&  31& 1--23\\
$\etap\rho^0$     &C2&  23&0.001--11&$\etap\rho^+$ &C2&  47& 3--24\\
$\eta\eta$        &C2&  18&0.06--2& & & & \\
$\etap\eta$       &C2&  27&0.08--10& & & & \\
$\etap\etap$      &C2&  47&0.02--14& & & & \\
\notext\notext
$\omega\pi^0$     &C2&  14&0.01--12&$\omega\pi^+$  &C2&  23&0.6--8\\
$\omega\rho^0$    &C2&  11&0.005--0.4&$\omega\rho^+$&C2& 61&7--26\\
$\omega\eta$      &C2&  12&0.05--7  & & & & \\
$\omega\etap$     &C2&  60&0.02--19 & & & & \\
$\omega\omega$    &C2&  19& 0.1--3 & & & & \\
\notext\notext
$f_0\pi^0$        &  &    &        &$f_0\pi^+$     &C1& 140&      \\
$f_2\pi^0$        &  &    &        &$f_2\pi^+$     &C1& 240&      \\
\notext\notext
$a_1^\pm\pi^\mp$  &AL& 240&        &$a_1^0\pi^+$   &AR& 900&      \\
$a_1^0\pi^0$      &AR&1100&        &$a_1^+\pi^0$   &AR&1700&      \\
$a_2^\pm\pi^\mp$  &C1& 300&        & & & & \\
\notext\notext
$a_1^0\rho^0$     &AR&2400&        &$a_1^+\rho^0$  &AL& 130&      \\
$a_1^\pm\rho^\mp$ &AR&3400&        &$a_1^0\rho^+$  &  &    &      \\
\notext\notext
$a_1^+a_1^-$      &AL& 570&        &$a_1^+a_1^0$   &AR&13000&     \\
\notext\notext
$p\bar p$         &AL&  18& 0.5--4 & & & & \\
$\Delta^0\bar\Delta^0$&AL&380&      &$\Delta^0\bar p$&AL&   76&     \\
$\Delta^{++}\bar\Delta^{--}$&AL& 47&&$\Delta^{++}\bar p$&AL&{\bf26}&0.1--130\\
\sgline
\end {tabular}
\label{treetab}
\end{center}
\end {table}

Note that all of the decay modes summarized in Table
\ref{sigtab}, some of which will be discussed in the next section, are
decays which are expected to be dominated by penguins.  There is still
no direct evidence for the $b\ra u$ tree modes, though signals with
significance of more than two standard deviations in the \pipi\ and
$\pi^+\pi^0$ channels suggest that such decays may be observed soon.
This dominance of penguins over $b\ra u$ channels suggests
that the problem of penguin pollution in $CP$ violation measurements
for modes such as \Bpipi\ may be particularly severe.

While few conclusions can be reached with the present level of precision in 
the $K\pi$ system, Fleischer and Dighe, Gronau and Rosner \cite{ref:dgr} 
have pointed out the 
promise of such decays for measurements of the weak phases $\alpha$ and 
$\gamma$.  This is of great importance since these are the two phases
that will be hardest to measure with time-dependent asymmetry techniques at 
BaBar\ and Belle.  Subsequently Fleischer and Mannel \cite{flman} pointed out
that if the ratio 

\begin{equation}
R={\calB(B^0\ra K^+\pi^-)+\calB(\bar B^0\ra K^-\pi^+)
\over{\calB(B^+\ra K^0\pi^+)+\calB(B^-\ra\bar K^0\pi^-)}}
\label{eq:flmanR}
\end{equation}
is significantly smaller than 1, useful bounds can be placed on the weak phase 
$\gamma$, namely $\sin^2\gamma<R$.  From Table \ref{sigtab}, we find 
$R=0.65\pm0.38$, indeed less than 1 but certainly not significantly so.  
Improvements upon the original concept have recently been suggested
\cite{flman2,flmangr}.  There 
are potential complications in this analysis due to SU(3) symmetry-breaking
effects, electroweak penguins and especially final-state interactions
as discussed recently by many authors \cite{flmangr,flmancrit}.  The situation
is not yet resolved but it appears promising that future measurements of
these decay rates, with samples 10-100 times larger than existing samples,
may be able to constrain the $CP$ phase $\gamma$.

There is considerable interest in the $CP$ asymmetry for these
and other decay modes.  However, since the observed samples are so small, 
experimentalists cannot yet make meaningful measurements.  The asymmetry, 
$A\equiv {(N_+ - N_-)\over{(N_+ + N_-)}}$, where $N_+$ and $N_-$ are the
number of $K^+$ and $K^-$ events respectively, is a measure of direct
$CP$ violation.  For most modes, predictions are $A\sim0.1$, while experimental
errors are presently greater than 0.3.

\subsubsection{$B\ra\etapr K$, $B\ra\omega K$ and related decays}
\label{sec:etaomega}

Decays with resonances tend to be more difficult to observe because more
particles are involved and the secondary branching fractions to observable 
final states are frequently below 50\%.  The observation of signals 
for $\Betaprk$ \cite{etaprpub} and $\Bomegak$ \cite{omegapub} 
was unexpected, since previous predictions were that the
branching fractions would be smaller than that observed by at least a factor 
of two.

CLEO finds a strong signal for \Betaprkp\ in both the 
\etaprepp\ ($5.2\sigma$) and \etaprrg\ ($4.8\sigma$) channels. 
Combining these with evidence from the chain \etaprepp, \etathreepi\
yields a significance (including systematic errors in the yield) of $7.5\sigma$ 
as shown in Fig.~\ref{fig:etaprk}a.  The combined significance for
the \Betaprkz\ decay is $3.8\sigma$ as shown in Fig.~\ref{fig:etaprkz}a.
Projections onto the $B$ mass axis are also shown. 

\displayscale{htbp}{fig:etaprk}{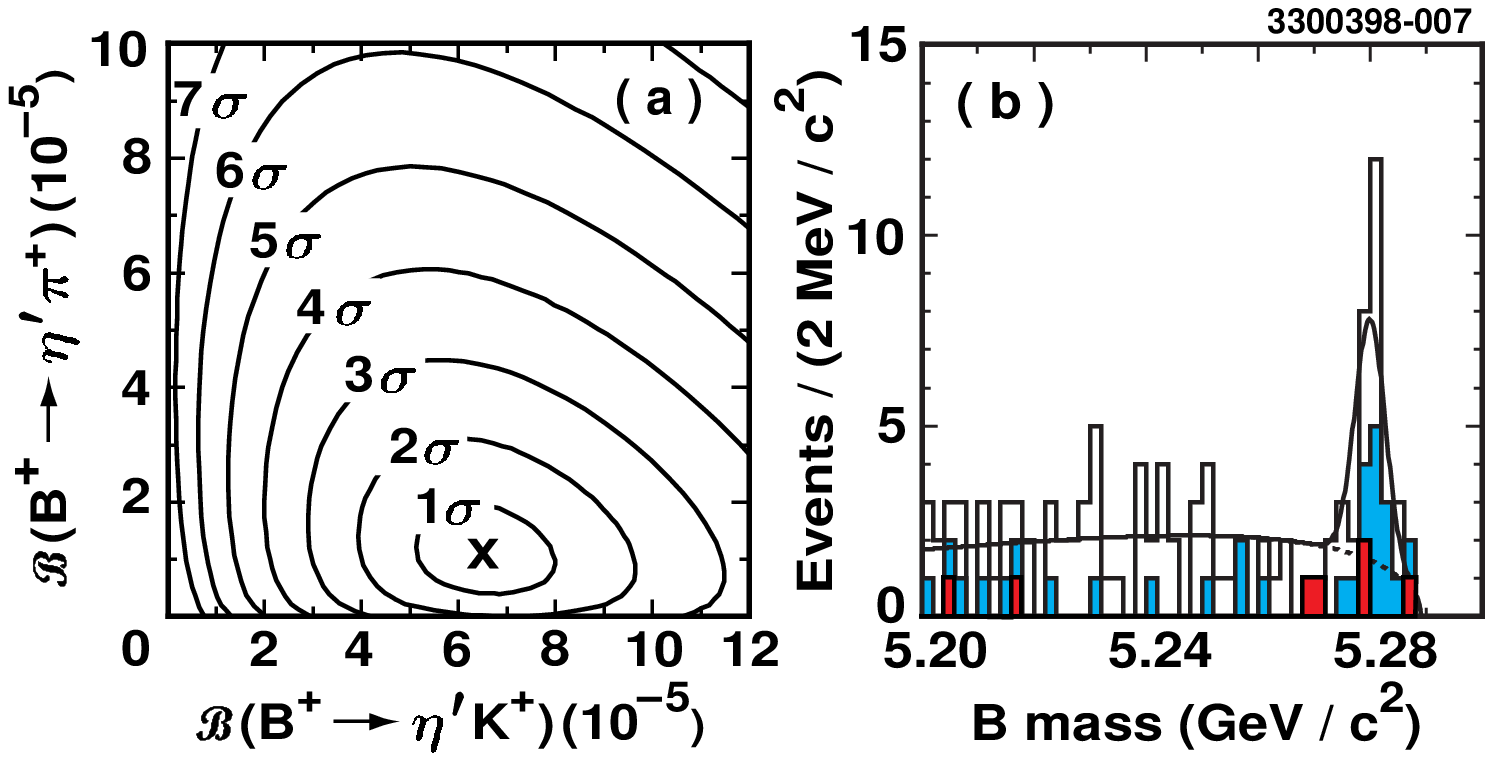}{2.7}
{(a) Likelihood contours and (b) mass projection plot for 
$B^+\ra\eta^\prime h^+$.  For (b), the curves are an overlay of
the best fit function (solid) and background component
(dashed).  The histograms show submodes:
$\eta^\prime\ra\eta\pi\pi\ (\eta\ra\pi^+\pi^-\piz$, dark shaded),
$\eta^\prime\ra\eta\pi\pi\ (\eta\ra\gamma\gamma$, light shaded),
and $\eta^\prime\ra \rho\gamma$ (open).}

\displayscale{htbp}{fig:etaprkz}{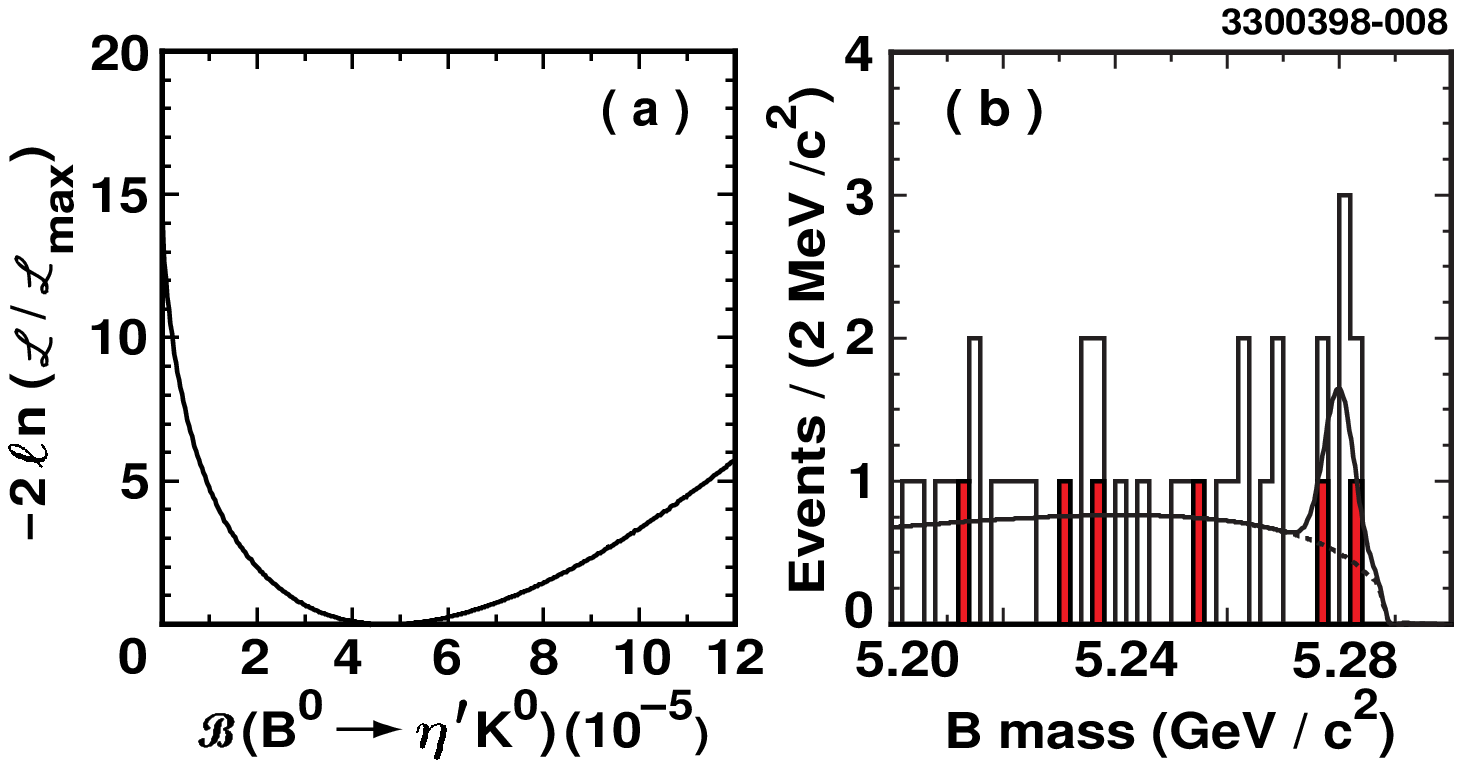}{2.7}
{(a) $-2\ln{\calL/\calL_{\rm max}}$ and (b) mass projection plot for \Betaprkz,
with curves the same as in Fig. \ref{fig:etaprk}.
The histograms show submodes: $\eta^\prime\ra\eta\pi\pi$ with 
$\eta\ra\gamma\gamma$ (shaded) and $\eta^\prime\ra\rho\gamma$ (open).}

Similarly, CLEO finds a signal with $3.9\sigma$
significance for \Bomegakp\ as shown in Fig.~\ref{fig:omegak}.
The mass projection plot is also shown.
The results for all three signals are summarized in Table \ref{sigtab}.
As in the previous section, there are hints of signals for the
corresponding tree decay modes \Betaprpi\ and \Bomegapi, but the
significance of each is only about 2$\sigma$.

\displayscale{htbp}{fig:omegak}{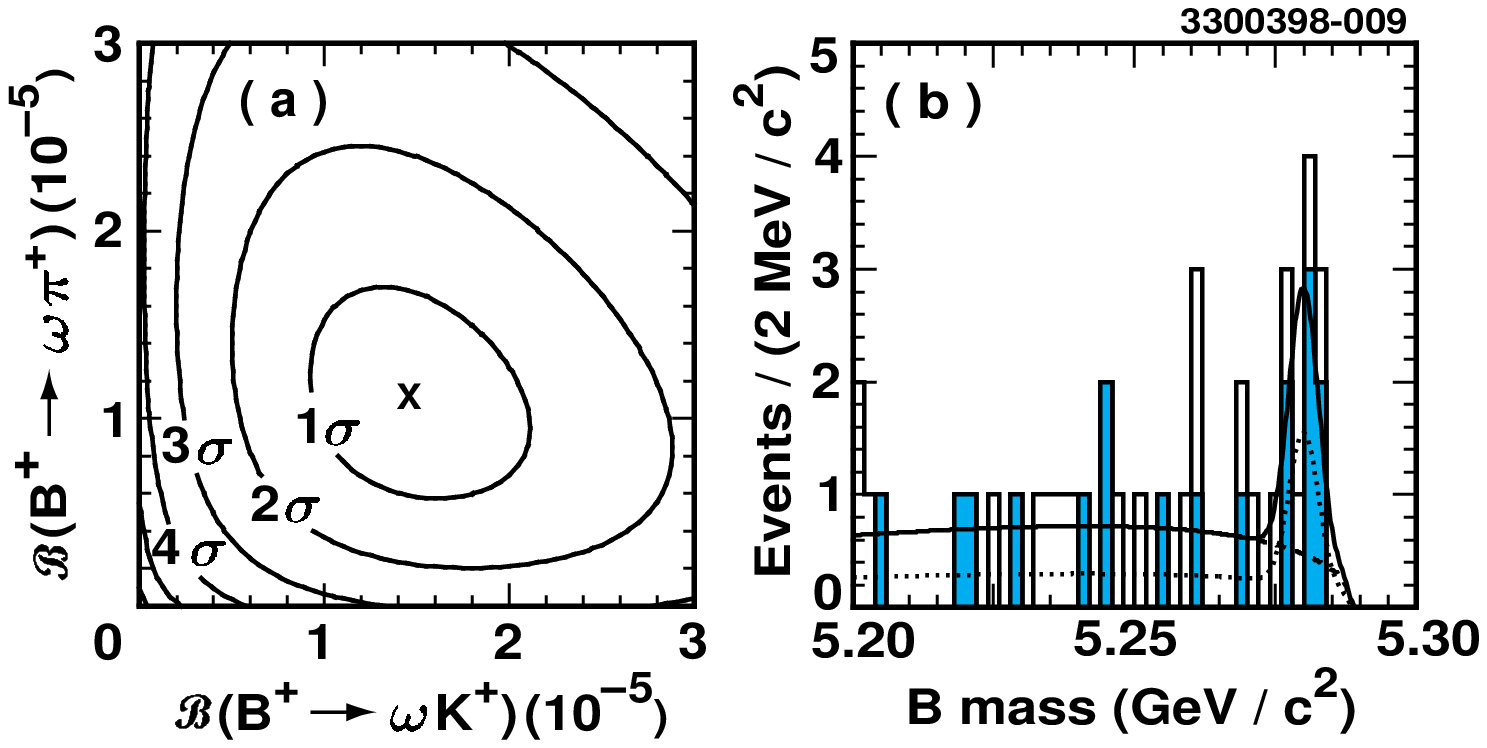}{2.7}
{(a) Likelihood contours and (b) mass projection plot for $B^+\ra\omega h^+$.
In (b), the curves are the same as for Fig. \ref{fig:etaprk}, with the addition 
of a dotted line showing the \Bomegakp\ component of the fit.  The
histograms show \Bomegakp\ (shaded) and \Bomegapi\ (open).}

The recent CLEO work includes limits on many other modes involving
$\eta$, $\etapr$, $\omega$ and $\phi$ mesons.  Tables \ref{pengtab} and 
\ref{treetab} summarize these and other results.  The subset of these results 
from the recent CLEO measurements is also
summarized in graphical form in Figs. \ref{fig:sum_kpi} and \ref{fig:sum_other}.

A number of features of these decays were predicted in advance.  For
instance, Lipkin pointed out \cite{lipkin} that due to interference between the
penguin diagrams shown in Fig. \ref{fig:glufeyn}e and \ref{fig:glufeyn}f,
the rate for \Betak\ is suppressed and the rate for \Betaprk\ is enhanced.
Detailed predictions by Chau \etal\ \cite{oldglurefs} and subsequent 
authors show this pattern, but the predicted enhanced rate for \Betaprk\ 
($\sim$1--2$\times10^{-5}$) was still far below the recently measured value.
Gronau and Rosner suggested \cite{ref:dgr2} the possibility that there could
be significant enhancements to the rate for the decay \Betaprk\ due to
flavor-singlet (hairpin) amplitudes.

It is worthwhile commenting on the evolution of the theoretical predictions
given in Tables \ref{pengtab} and \ref{treetab}.  The basic method of
using effective Hamiltonian theory has remained unchanged since the
original Bauer, Stech and Wirbel \cite{BSWetc} calculations in 1987.
However there have been significant improvements in the detailed
implementation: effective Wilson coefficients \cite{effhamimprov}; 
knowledge of the CKM matrix elements \cite{parodi}; values 
of the relevant quark masses and quark mixing angles (especially important
for $\eta$ and $\etapr$ decay channels); proper treatment of the QCD
anomaly \cite{acgk}; and more complete treatment of non-factorizable
contributions using the empirical color factor $\xi$.  Thus the newest
predictions \cite{newglurefs}, epitomized by the recent comprehensive 
calculations of Ali, Kramer and L\"u, represent a substantial advance 
over the older predictions \cite{oldglurefs}. 
In order to try to capture a sense of this progress, we
have indicated the range of newer calculations with solid lines in Figs.
\ref{fig:sum_kpi} and \ref{fig:sum_other}, while the range of
calculations prior to 1997 is indicated by a dashed line.

Several of the CLEO limits restrict  the range of recent predictions, as
indicated by bold face
in the Tables \ref{pengtab} and \ref{treetab} and overlap of the ``X" and 
the theory line in Figs. \ref{fig:sum_kpi} and \ref{fig:sum_other}.
In other cases they are able to eliminate some 
theoretical hypotheses which have been advanced to account for the surprisingly
large rate for \Betaprk.  For instance the conjecture that the rate is enhanced
by a substantial $c\bar c$ admixture in the $\etapr$, i.e. $\etapr$-$\eta_c$
mixing \cite{zhitnitsky}, appears unlikely.  Such a mechanism would
also yield a large rate for \Betaprkst, in contradiction with the limit
given in Table \ref{pengtab}.  Several authors \cite{etaprccbar} now agree 
that the $c\bar c$ admixture in the $\etapr$ is small and actually leads to 
destructive interference, hence smaller rate, for values of the 
phenomenological parameter $\xi<0.3$ (see Sec. \ref{sec:efftheory}), 
where the rate predictions are largest.  

\displayscale{htbp}{fig:sum_kpi}{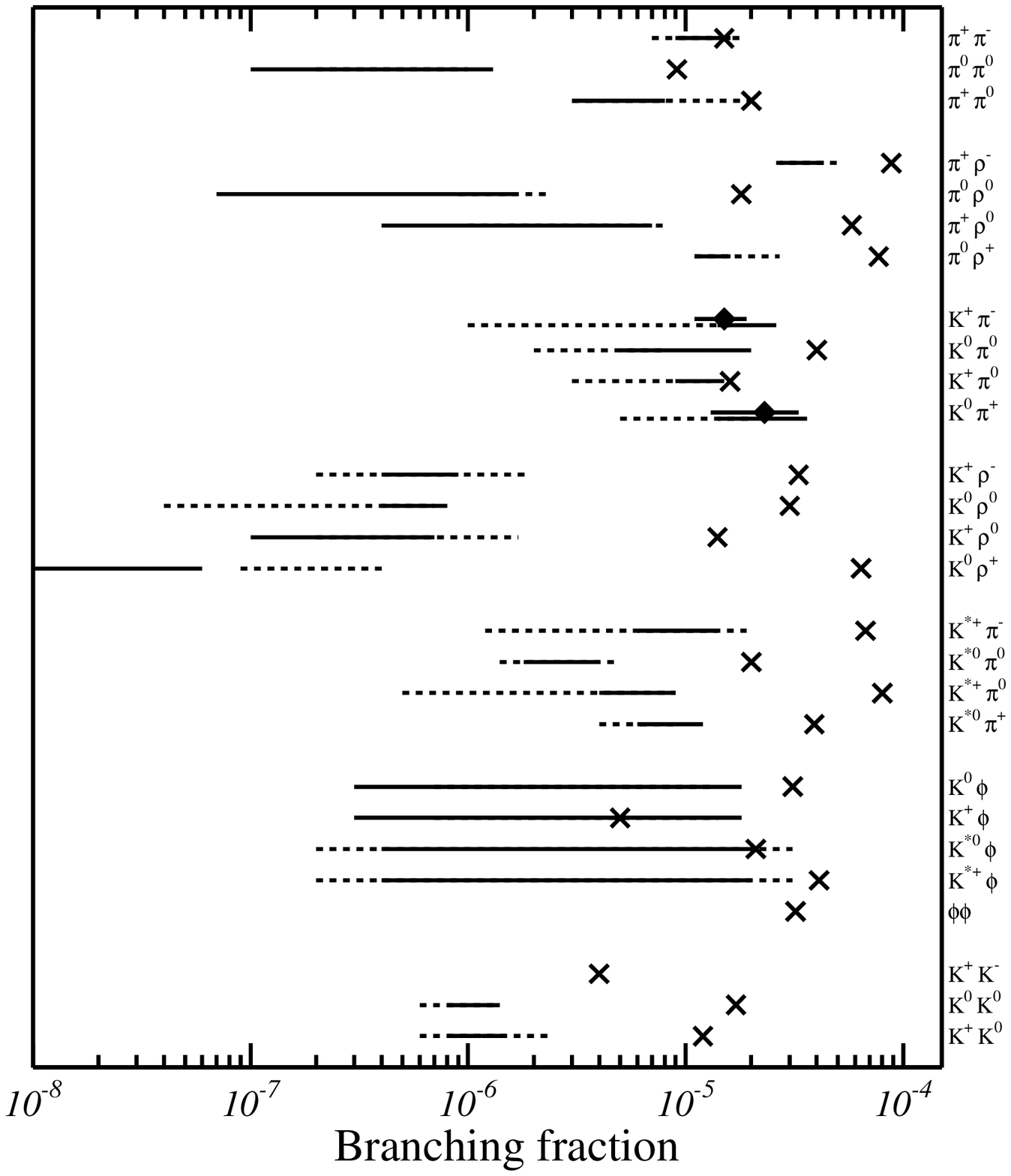}{6.5}
{Graphical summary of various recent CLEO measurements of charmless
hadronic $B$ decays.  Limits are denoted by ``X", significant
measurements by (diamond) points with error bars, and recent (pre-1997) 
theoretical ranges by solid (dashed) lines.}
\displayscale{htbp}{fig:sum_other}{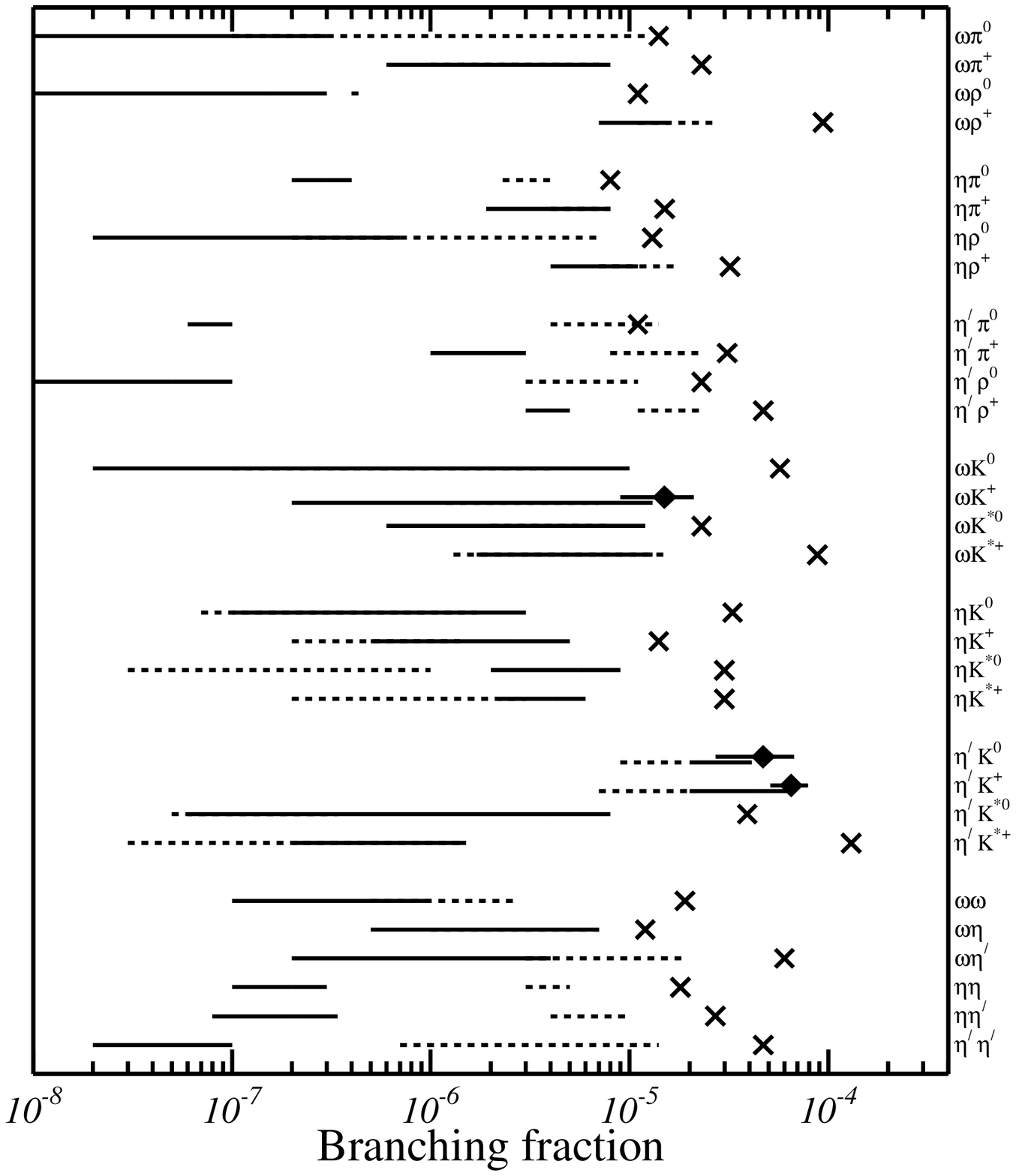}{6.5}
{Same as the previous figure with modes involving $\eta$, $\etapr$ and
$\omega$ mesons.}

A typical range of predictions for the rate for \Betaprk\ is
(2--4)$\times10^{-5}$ \cite{aliGreub,cheng}.  Some theorists believe that the
present experimental value is a fluctuation and will end up in this
range \cite{aliGreub}, while others suggest that the
rate can be enhanced by a variety of mechanisms: use of an even smaller
value of the strange quark mass than is now being used \cite{kagan}; 
use of somewhat larger values of form factors \cite{kagan,datta}; use of 
smaller values of the CKM phases $\gamma$ \cite{oh}; including non-factorizable 
contributions arising from the QCD anomaly \cite{atsoni,ahmady,dky}, or
modification to the Wilson coefficients whereby the value of the
color-factor $\xi$ is different for odd and even penguin coefficients
\cite{cheng}.  In order to account for the \Betaprk\ data, Deshpande \etal\ 
\cite{oh} used constructive 
interference for the $c\bar c$ admixture (now apparently ruled out) and the CKM 
phase $\gamma=35^\circ$; a recent fit \cite{parodi} of all relevant data 
excludes this value: $\gamma=(64\pm12)^\circ$.  Further data 
should help distinguish among the various possibilities.  The rate for
\Betaprkst\ may prove to be crucial in this regard.

The situation for the $\omega$ and $\phi$ decays is quite unclear at present.  
There have been several recent theoretical papers specifically addressing
vector-pseudoscalar final states \cite{cheng,oh,akl,dgr3}, with an
emphasis on the \Bomegakp, \Bomegapi, \Bphik, and \Bphikst\ decays, where
there are measurements or good limits.  The theoretical estimates for
\Bomegakp\ \cite{aliGreub,oh} tend to be $<$10$^{-5}$ except
when $\xi\sim 0$ or $\xi>0.5$.  The decay \Bphik\ is closely related but 
there is as yet no evidence for this decay; the CLEO 90\% CL
limit of $0.5\times10^{-5}$ tends to exclude the large $\xi$ range.
The recent paper of Ali \etal\ \cite{akl} suggests that these decays
belong to a class where factorization may not work well (hence $\xi$ may
be different than it is for other decays), since the largest Wilson 
coefficients are not present in the amplitudes for these processes.
Cheng and Tseng \cite{cheng} come to similar conclusions for quite
different reasons.

There have been a variety of other speculations regarding the rates for
these charmless processes.  Ciuchini \etal\ have suggested that
there could be substantial enhancements in certain decays due to ``charming
penguins", where the charm-quark contribution in the loop becomes large
due to large non-perturbative corrections to the effective-Hamiltonian
calculations \cite{ciuchini}.  They predict that final states such as $\rho K$,
$\phi K^*$, $\omega K$, and $\omega K^*$ would be substantially enhanced,
though even this model is unable to account for the large measured rate for 
\Bomegakp.  Some limits presented here, especially that for $B^+\ra K^0\rho^+$,
tend to indicate that such enhancements are not huge.

\subsection{Non-resonant decays}

The processes discussed so far have been two-body or quasi-two-body decays.
Results also have been reported for a variety of higher multiplicity,
non-resonant decays.  Such studies are important since it is not obvious 
how much of the final state $K\pi\pi$, for example,
is $K^*\pi$, $K\rho$, non-resonant, or some other
possibilities.  Additional incentive for measuring such decays
was provided by a paper \cite{deht} pointing out
that interference effects in the $3\pi$ final state could lead to $CP$
violation, and asymmetry measurements could potentially measure the weak
phase $\gamma$.  However, since this topic is unrelated to penguins, we
do not discuss it further.  The current experimental situation for these
higher multiplicity decays is summarized in Table \ref{nonrestab}.
In a few cases the limits are becoming restrictive but most are far away
from the theoretically expected branching fractions.  For instance the
prediction for $B^+\ra2\pi^+\pi^-$ is 10--50$\times10^{-6}$ \cite{deht}.

\def\notext{\omit&\omit&\omit&\omit&\omit&\omit\\}
\begin {table}[htbp]
\begin{center}
\caption{Results from measurements of non-resonant $B$ decay modes.
We give the 90\% confidence level upper limit on the branching fractions
(UL $\cal B$).  Modes above the line are expected to be dominated by penguins.
Experiment key: AL (ALEPH \cite{alephnonres}), AR (ARGUS \cite{argusnonres}), 
C2 (CLEO II \cite{cleononres}), D (DELPHI \cite{delphirarepub}).}
\vspace{0.3cm}
\begin {tabular}{l c r|l c r}
\dbline
$B^0$ final       &    & UL $\cal B$ &$B^+$ final         &    & UL $\cal B$\\
state             &Expt.&$(10^{-6})$ &state               &Expt.&$(10^{-6})$\\
\sgline
$K^+\pi^+2\pi^-$  &  D &     230     &$K^+\pi^+\pi^-$     & C2 &   28  \\
                  &    &             &$K^-\pi^+\pi^+$     & C2 &   56  \\
                  &    &             &$K^+K^+K^-$         & C2 &   38  \\
                  &    &             &$p\bar pK^+$        & C2 &   89  \\
\sgline
$\bar\Lambda p\pi^-$&AR&     180     &$\bar\Lambda p\pi^+\pi^-$&AR&200 \\
\notext
$\pi^+\pi^-\pi^0$ & AR &     720     &$2\pi^+\pi^-$       & C2 &   41  \\
                  &    &             &$\pi^+2\pi^0$       & AR &  890  \\
                  &    &             &$K^+K^-\pi^+$       & C2 &   75  \\
                  &    &             &$p\bar p\pi^+$      & C2 &   53  \\
\notext
$2\pi^+2\pi^-$    &  D &     230     &2$\pi^+\pi^-\pi^0$  & AR & 4000  \\
$\pi^+\pi^-2\pi^0$& AR &    3100     &                    &    &       \\
$p\bar p\pi^+\pi^-$&AL &     150     &                    &    &       \\
\notext
$2\pi^+2\pi^-\pi^0$&AR &    9000     &$3\pi^+2\pi^-$      & AL &  280  \\
                  &    &             &$p\bar p2\pi^+\pi^-$& AL &  370  \\
\notext
$3\pi^+3\pi^-$    & AL &     660     &$3\pi^+2\pi^-\pi^0$ & AR & 6300  \\
\notext
$3\pi^+3\pi^-\pi^0$&AR &   11000     &                   &  &       \\
\sgline
\end {tabular}
\label{nonrestab}
\end{center}
\end {table}

\subsection{Decays of $B_s^0$ mesons}

\def\notext{\omit&\omit&\omit&\omit\\}
\begin {table}[htbp]
\begin{center}
\caption{Results from measurements of $B_s^0$ decay modes.
We give the 90\% confidence level upper limit on the branching fractions
(UL $\cal B$), and the range of theoretical predictions \cite{dean,xing}.}
\vspace{0.3cm}
\begin {tabular}{l c r c}
\dbline
$B_s^0$ decay mode &Experiment& UL $\calB\ (10^{-6})$ &Theory ($10^{-6}$)\\
\sgline
$K^+K^-$           &DELPHI\cite{delphirarepub}&  46           &   3--21     \\
$K^+\pi^-$         &ALEPH\cite{alephrarepub}  & 210           &  10--18     \\
$\pi^+\pi^-$       &ALEPH\cite{alephrarepub}  & 170           &             \\
$p\bar p$          &ALEPH\cite{alephrarepub}  &  59           &             \\
$\pi^0\pi^0$       &   L3\cite{l3rarepub}     & 210           &             \\
$\eta\pi^0$        &   L3\cite{l3rarepub}     &1000           &0.001--0.02  \\
$\eta\eta$         &   L3\cite{l3rarepub}     &1500           &  0.4--5.6   \\
\sgline
\end {tabular}
\label{bstab}
\end{center}
\end {table}

There have been relatively few studies of the $B^0_s$ meson since the mass,
5370 MeV, is too large for production at the $\Upsilon$(4S) resonance, so 
CLEO and ARGUS have not studied these decays.  Thus all results for $B_s^0$
charmless hadronic decays 
are from the various LEP detectors.  The physics is very similar for 
$B^0_s$ as for $B^0_d$ mesons; imagine the $u$ or $d$ spectator quark in
Fig. \ref{fig:glufeyn} replaced by an $s$ quark, with appropriate changes to
the final state mesons.  Thus the analogue of the penguin decay $B\ra K\pi$ 
is $B^0_s\ra KK$.  In Table \ref{bstab}, we
summarize the present knowledge concerning rare $B^0_s$ decays.
The limit for $B^0_s\ra K^+K^-$ is close to the expectation, but all
others are not.  However,
new information on these decays can be expected in the future from hadronic
colliders.

\subsection{Inclusive decays}
\label{sec:gluinc}

There have been a variety of inclusive searches for gluonic penguins,
which typically involve final states with (hidden or open) strangeness.  
Inclusive decays are intriguing partly because direct $CP$ violation might 
be observed as a difference between the numbers of positively and negatively
charged kaons \cite{browetal}.  We describe several searches for inclusive 
decays, beginning with attempts to measure the inclusive $b\ra sg^*$ rate and 
including one quite surprising observation.

Even though the gluonic penguin $b \ra sg^*$ does not have
a good signature, several experiments have used ingenuity to try
to measure the inclusive process.
The ARGUS collaboration searched for $b \rightarrow sg^*$
in samples where one $B$ of a
\BBbar\ event is fully or partially reconstructed.  They searched for
decays of the other $B$ involving a kaon and multiple pions, a typical
signature of the $b\ra sg^*$ process.  They found two events where
no possible sub-combination is consistent with any charmed particle and 
set a 90\% CL upper limit of 8\% on the branching fraction for $b\ra sg^*$
\cite{ref:argus_incpeng}.

DELPHI \cite{ref:delphi_incpeng} looked for $b \rightarrow sg^*$
by searching for an excess of high $p_t$ kaons, since more 
energy is available for kaons from
$b \ra s$ decays than from $b \ra c \ra s$ decays.
A fit to the kaon $p_t$
distribution provided a preliminary limit of ${\cal B}(b \ra sg^*) < 5\%$
at 95\% confidence level.

There have been two indirect searches for $b\ra sg^*$.  CLEO set an
upper limit of 6.8\% \cite{ref:cleo_dlep} by accounting for all other
types of $B$ decays.  DELPHI \cite{ref:delphi_dd} performs a similar search, 
and finds a limit of 4\% at 90\% confidence level \cite{ref:delphi_sg}.
While none of these searches is yet
sufficiently sensitive to observe the $\sim1$\% signal expected from
Standard Model processes, they appear to exclude models
\cite{ref:gluinos,ref:enhanced1,ref:enhanced2,ref:enhanced3}
where the rate is enhanced by an order of magnitude (see Sec. \ref{sec:NP}).

CLEO has searched \cite{phiinc} for a $\phi$ meson accompanied by an
\xs\ system consisting of a
charged or neutral $K$ meson and zero to four pions, of which at most
one can be a $\piz$.  The \xs\ system was required to have a mass less
than 2 GeV, corresponding to a $\phi$ meson 
momentum of $\sim$2.1 GeV/c; in this region the background is small for
$\phi$ mesons arising from $b\ra c$ processes, potentially allowing
detection of a $b\ra sg^*$ signal.  CLEO observed no signal and, using 
the model of Deshpande \etal\ \cite{deshphi}, they set an upper limit 
$\calB(B\ra\phi\xs)<1.3\times10^{-4}$.   The theoretical expectation
for this process is $(0.6-2.0)\times 10^{-4}$ \cite{deshphi}.

The only positive evidence for inclusive charmless hadronic $B$ decays is 
from the CLEO analysis of the decay \etaprinc\ \cite{etapinc}.  The technique 
is the same as for the $\phi$ inclusive search except that the kaons were
required to be charged.  The momentum of $\etapr$ mesons,
reconstructed with the decay chain \etaprepp, \etatogg, was required to be
in the range $2.0<p_\etapr<2.7$ GeV/c in order to reduce background
from $b\ra c$ processes.  The values of \DE\ and $M$, as defined
above, were required to satisfy $|\DE|<0.1$ GeV and $M>5.275$ GeV.

\displaytwo{htbp}{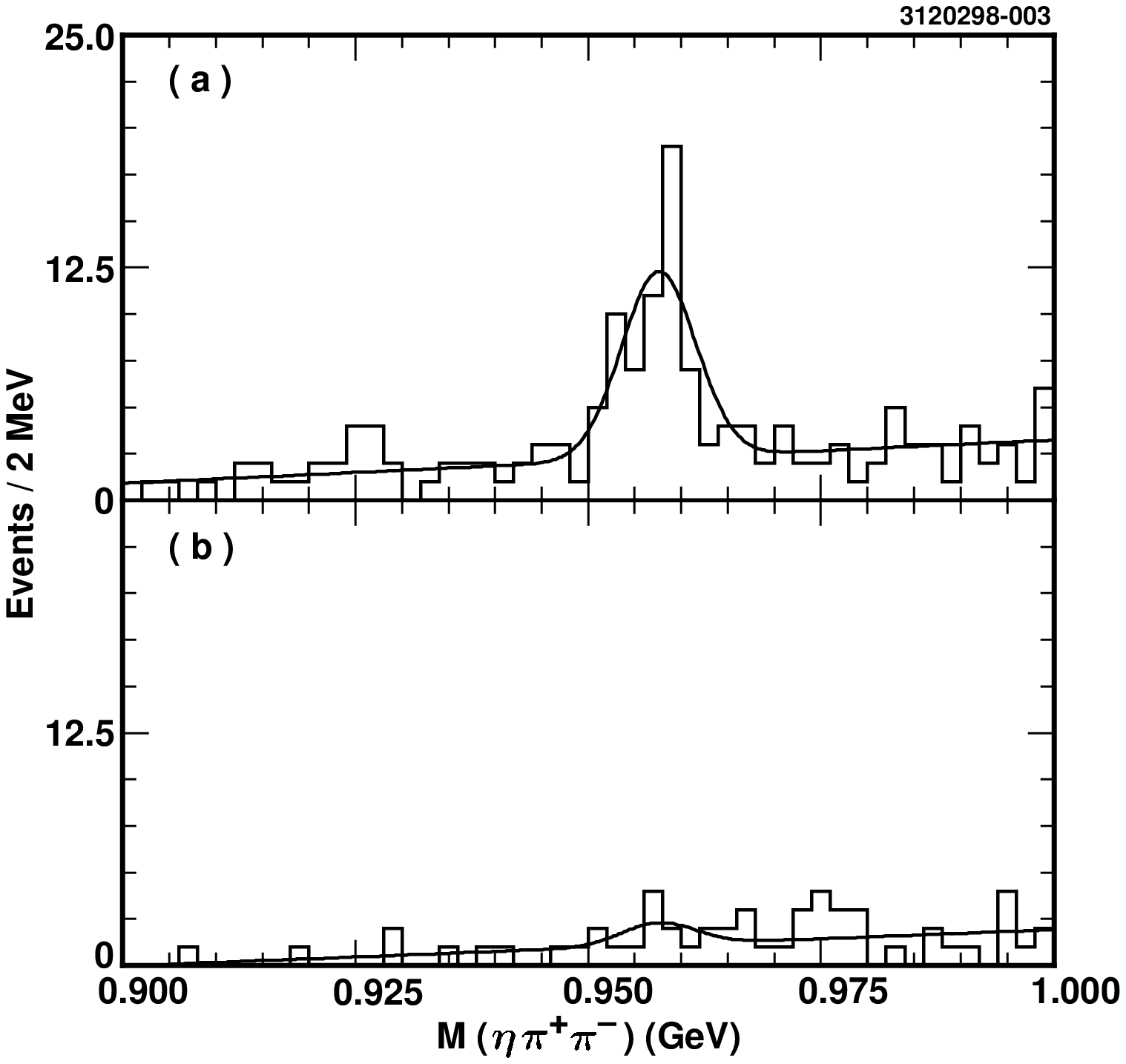}{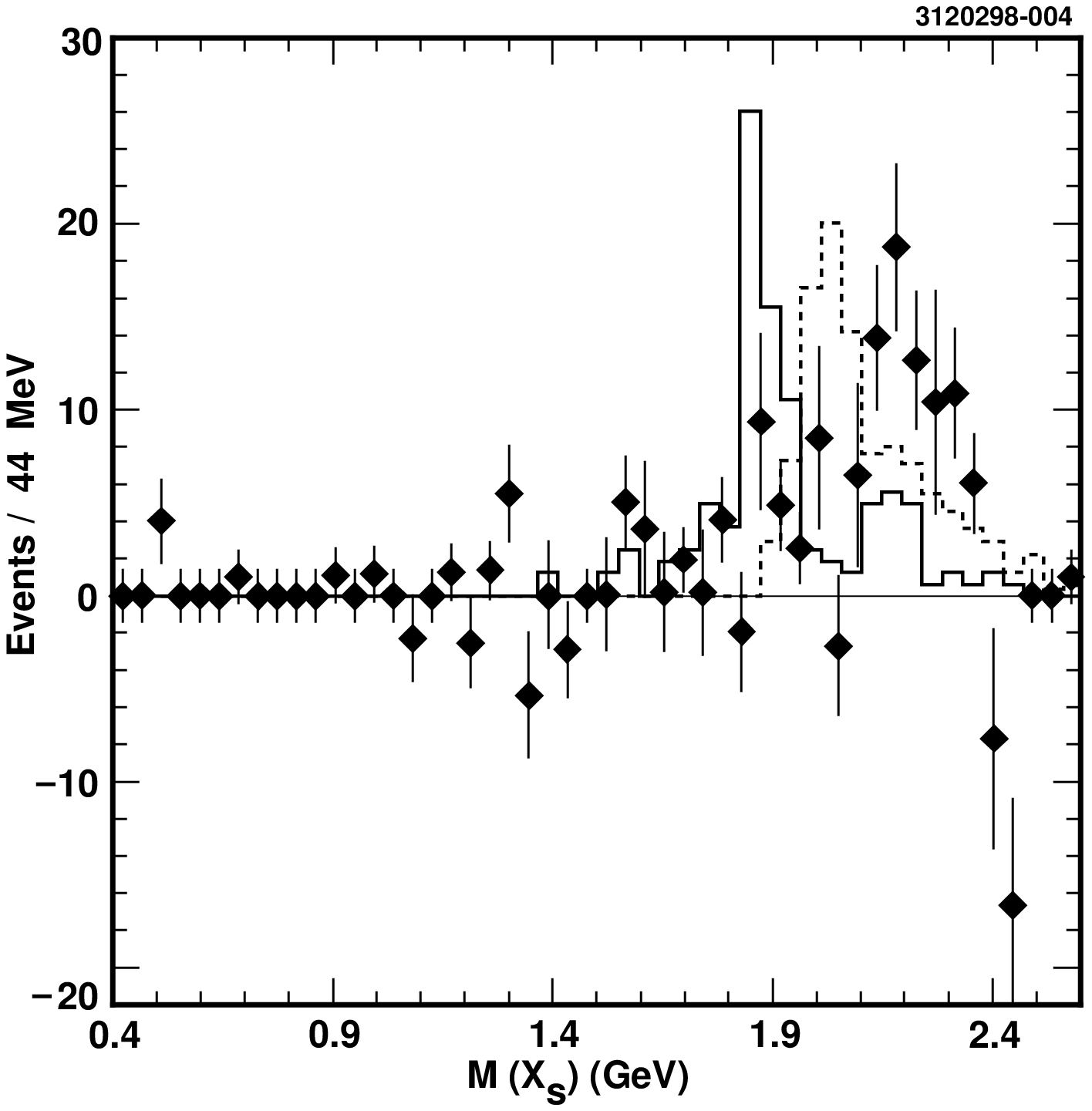}
{The $\etapr$ mass distribution for (a) on-resonance and (b) below-threshold
data.}{Distribution of \xs\ mass for data
(points with error bars) and possible backgrounds: $B\ra D\etapr$ (solid
histogram) and $B\ra D^*\etapr$ (dashed histogram). The normalization of
the backgrounds is arbitrary.}

The $\etapr$ mass distribution is shown in Fig. \ref{etapinc_mass}; a 
clear signal of $39\pm10$ events is seen for on-resonance data and none for 
the below-threshold sample. The signal,
obtained by subtracting the scaled off-resonance data
in bins of \xs\ mass, is plotted in Fig. \ref{xsmass_cbg}.
Note the four events corresponding to \Betaprkp\ and the absence of
events in the $K^*(892)$ mass region, both consistent with the exclusive
results discussed in Sec. \ref{sec:etaomega}.  Also shown in 
Fig. \ref{xsmass_cbg} are distributions for potential background modes such as
$B\ra D\etapr$ and $B\ra D^*\etapr$.  Though these also tend to have large
\xs\ mass, they are more peaked than the data.  These and other
studies suggest that the observed signal does not arise primarily from
color-suppressed $b\ra c$ decays, though it is difficult to rule this
out completely without better models of such processes.  The efficiency
was calculated assuming that the signal arises solely from gluonic penguin
decays, with an equal admixture of \xs\ states from the kaon up to
$K_4^*(2200)$.  The efficiency of ($5.5\pm0.3$)\% leads to
$\calB(\etaprinc)=(6.2\pm1.6\pm1.3)\times10^{-4}$ for $2.0<p_\etapr<2.7$ GeV/c.
The systematic error is dominated by the uncertainty in the \xs\ modeling.

Many theorists suggest that the \etaprinc\ result is the most surprising 
of those included in this review --- the theoretical expectation for the 
high-momentum $\etapr$ branching fraction is $\sim1\times10^{-4}$.  Atwood and 
Soni \cite{atsoni} first suggested that the very large rate could be due to
the QCD anomaly via $b\ra sg^*\ra s \etapr g$.  Later Fritzsch
\cite{fritzsch} suggested a similar anomaly-mediated $b\ra sg$ process.
Both processes have a hard \xs\ mass spectrum in rough agreement with that 
shown in Fig. \ref{xsmass_cbg}, but Fritzsch's 2-body process would be
falling by 2.5 GeV, while the 3-body decay of Atwood and Soni would
still be rising.  There have been several other recent papers discussing
\etaprinc\ \cite{kagan,datta,hou,chao}.  They consider the anomaly and
the possibility of non-SM contributions.  It is still not clear whether
one of these anomaly models can quantitatively account for the \etaprinc\ 
result, though
there don't seem to be any better explanations at present without invoking
new physics.  Since other inclusive processes such as $B\ra\phi\xs$ are not 
enhanced, it seems likely that this result is indeed an ``anomaly"
involving the $\etapr$ meson.

In a similar search for high-momentum $\eta$ mesons, CLEO finds no
evidence for a signal and sets a 90\% CL upper limit for the branching
fraction of $4.4\times10^{-4}$.  The theoretical expectation is that the
rate for $\eta$ mesons would be suppressed relative by about an order of
magnitude since their flavor-singlet component is small.
 
\section{FUTURE PROSPECTS}
\label{sec:future}

\begin{table}[bth]
\caption{Comparison of future $B$ experiments. Parameters which do not
change between different experiments at the same collider are entered only
once.}
\label{tab:bfuture}
\small
\def\1#1{\quad\hskip-0.5cm #1 \hskip-0.5cm\quad}
\vspace{0.3cm}
\begin{tabular}{lllrcrrrrrr}
\dbline
\1{Expt.} &  \1{Collider} &  
   \1{Beams} & \1{$\sqrt{s}$}   &      \1{Year}   &   \1{${\cal L}$ ($10^{33}$} &
       \1{$\sigma(b\bar b)$} &  \1{$b\bar b$ pairs}     & \1{$\beta\gamma c\tau$} 
              & \1{$\sigma(b\bar{b})$} \\ 
         &           & 
         &    &    \1{online}   &   \1{$cm^{-1}$} &
           & \1{($10^7$/yr)} &      &   \1{$/\sigma(q\bar{q})$}    \\   
           &           &
         & \1{\hskip-0.3cm (GeV)}  &             &  \1{$s^{-1}$)}   &  
        \1{($nb$)} &        &   \1{($\mu m$)} &  \\
\hline
\1{CLEO III}   &  \1{CESR}   &  \1{$e^+e^-$}        &    \1{10} &   \1{1999}   &  \1{1.2}      & \1{1}
& \1{1.2}     &  \1{30} & \1{$3\cdot10^{-1}$}  \\
            & \1{CESR-IV} &          &    \1{10} &    \1{?}      &    \1{30}    & \1{1}
& \1{30} & \1{30} & \1{$3\cdot10^{-1}$} \\
\hline
\1{BaBar}      &  \1{PEP-II} &  \1{$e^+e^-{\,^\dag}$}  &    \1{10} &  \1{1999}   &  \1{3-10}   & \1{1}  
& \1{3-10}    & \1{270} & \1{$3\cdot10^{-1}$}  \\
\1{Belle}      &  \1{KEK-B}  &  \1{$e^+e^-{\,^\dag}$}  &    \1{10} &  \1{1999}   &  \1{3-10}   & \1{1}  
& \1{3-10}   &  \1{200} & \1{$3\cdot10^{-1}$}  \\
\hline
\1{HERA-B}     &  \1{HERA}   &  \1{$pN$}              &    \1{40} &  \1{1998}   &  \1{---}      & \1{6-12} 
& \1{50-100} & \1{9000}  & \1{$1\cdot10^{-6}$} \\
\hline
\1{CDF II}     &  \1{Tevatron} & \1{$p\bar p$}      &  \1{1800} &  \1{2000}   &  \1{0.2-1.0}  & \1{100000}   
& \1{20000}   & \1{500}  & \1{$1\cdot10^{-3}$} \\
\1{D0}         &                            &       &         &               &    
& &     &           \\
\1{BTeV${\,^\ddag}$}       &           &       &       &  \1{2004}   &  \1{0.2}      &
&        &  \1{5000} &    \\  
\hline
\1{LHC-B${\,^\ddag}$}      &  \1{LHC}    &  \1{$pp$}            &
                              \1{\hskip-0.3cm 14000} &  \1{2005}   &  \1{0.15}   & \1{500000}   
& \1{75000} &  \1{7000} & \1{$5\cdot10^{-3}$} \\
\1{Atlas}      &         &                  &       &         &       &         &    
& \1{500} &    \\
\1{CMS}        &         &                  &       &         &            &    
& &    \\
\hline
\end{tabular} 
\newline
$^{\dag}$\quad  Asymmetric beam energies.
$^{\ddag}$\quad Forward detector.
\end{table}

Future measurements require
more produced $B$ mesons and good background rejection.
The CLEO II experiment has a substantial amount of new data in the analysis
pipeline (see Table~\ref{tab:benvir}) and continues to accumulate even more
statistics.
There are a variety of new 
experiments poised to do rare-$B$ physics in the near and
longer term as summarized in Table~\ref{tab:bfuture}.

The CESR machine will be upgraded to higher luminosity
and the CLEO detector will get a particle identification
device and a new tracking system (CLEO III detector \cite{CLEOIII}).
Also under construction are two new $\Upsilon(4S)$ colliders,
the Positron-Electron Project-II (PEP-II) 
at the Stanford Linear Accelerator Center (SLAC)
with the BaBar \cite{BaBar} experiment and KEK-B at the 
high energy laboratory in Japan
with the Belle experiment \cite{Belle}.
These new machines have asymmetric beam energies for indirect
$CP$ violation measurements, which will also help background suppression
because of the detached $B$ vertex (see the $\beta\gamma c\tau$
column in Table~\ref{tab:bfuture}).
All ``B Factory" experiments (BaBar, Belle, CLEO III)
are expected to come online in 1999.
They should be able to reach branching fractions
of ${\cal O}(10^{-7})$ at the design luminosity.
Unlike CESR, the new colliders have a double ring structure,
thus in principle, they offer better potential for higher luminosity.
There are some ideas of how to convert CESR into a double ring symmetric
collider (CESR IV) for luminosity upgrades beyond 
the CLEO III phase \cite{CESRIV}.

The clean experimental environment at $e^+e^-$ colliders is ideal
for background suppression when looking for penguin processes
which have very small branching fractions.
Unfortunately, the $e^+e^-\ra b\bar b$ cross-section is only 1 nb,
making it very difficult to reach extremely rare decay modes.
Hadronic colliders, with $b\bar b$ cross-sections approaching 1 mb,
may offer the ultimate experimental avenue to the secrets of $B$ physics.
On the other hand, the total cross-section at hadronic colliders is much
larger, making background rejection the dominant experimental
obstacle. The detector recording rate is limited by technological constraints.
$B$ data can easily be lost by the inability of an experiment to 
reduce background rates to manageable recording rates.
It should be emphasized that hadronic colliders are built with the
primary goal of studying ultra-high-energy processes which
produce particles with large transverse momenta to the beam ($P_t$).
In contrast,
$B$ decays produce particles with transverse momenta not much higher than
ordinary beam interactions. 
So far, high-$P_t$ detectors at the Tevatron (CDF and D0) have been able
to trigger on $B$ physics only in di-muon modes.
Since off-line background suppression has been accomplished
by detached $B$ vertex cuts,
inclusion of a detached vertex requirement in the trigger  
may be the best triggering strategy for non-leptonic rare decays.

There have been some dedicated experiments to study $c$ and $b$ quark
physics with hadronic beams in fixed-target mode in which a hadronic
beam ($\pi^-$ or $p$) collides with a stationary nuclear target ($N$).
While the fixed target charm program has played a complementary role
to $e^+e^-\to c\bar c$ studies, the fixed target experiments barely succeeded 
in observing some inclusive signals from tree-level $b \rightarrow c$ 
decays of $b$ quarks \cite{fixedtarget}.
Since most of the incident beam energy is wasted providing the
motion of the center-of-mass, the effective collision energy was rather low
resulting in a very small $b\bar{b}$ cross-section.
To compound these experimental difficulties, all beam and target
fragments entered the detection apparatus together with $B$ decay
products.
The latest attempt to study $B$ physics in fixed target mode
is represented by the HERA-B experiment \cite{HERAB}
which will collide the proton
beam from the Hadron-Electron Ring Accelerator facility (HERA) at DESY 
with a wire target inserted into the beam pipe.
Use of modern technologies offers hope for success,
though as in the previous fixed target experiments, the HERA-B
experiment will have to contend with cross sections as low as in
$e^+e^-$ colliders and with hadronic backgrounds 
as high as in $p \bar p$ colliders.
The first run is foreseen for 1998, 
with a full capacity run in 1999.

The Tevatron collider is in the midst of an upgrade program
with turn-on planned for 2000.
Luminosity will be substantially increased, with further
luminosity upgrades possible in the long term future (\lq\lq TeV33'').
Improved vertex detectors in CDF-II  \cite{CDFupgrade,cdfd0}
and the D0 upgrade \cite{D0upgrade,cdfd0}
will help background suppression. The D0 experiment will acquire a 
magnetic field, but the tracking system will remain very compact.
CDF-II will implement a $B\to h^+h^-$ trigger. 
Unfortunately, even if this trigger is successful, CDF-II will have
difficulty distinguishing $B^0\to \pi^+\pi^-$ from $B^0\to K^+\pi^-$
and $B_s^0\to K^+K^-$ because of the lack of high momentum particle
identification.

In the farther future (2005), the LEP tunnel will house a new $pp$ collider, 
the Large Hadron Collider (LHC),
with high $P_t$ experiments ATLAS  \cite{ATLAS}
and CMS  \cite{CMS}.
The larger center-of-mass energy will increase the  $b\bar b$ cross-section
and improve the signal-to-background ratio. Since the event rate 
at the LHC design luminosity will be too high to record $b\bar b$ data,
the $B$ physics program in these high $P_t$ experiments may be limited
to initial lower luminosity running.

Both the Tevatron and LHC programs contemplate installation
of dedicated $b$ experiments (BTeV \cite{BTeV} 
and LHC-B \cite{LHCB}).
The machine luminosity would be reduced
for these experiments to keep data rates at manageable
levels. Without the constraints of high $P_t$ physics, the
detectors can be optimized for $b$ physics.
Both detectors would operate in the forward rather than
the central region.
Forward detector geometry offers enough space for
efficient particle identification in the entire momentum range (see below).
Furthermore, the decay length of $B$ mesons is longer than
it is in the central region, allowing for better background suppression
in the off-line analysis and in the trigger.
Last but not least, the entire trigger bandwidth can be saturated with
$b$ physics events.
 
Improvement in particle identification is also crucial at $e^+e^-$
machines.  The separation of gluonic $b \ra s$ penguin decays
from tree-level $b \ra u$ decays is already marginal for CLEO II
(Sec.~\ref{sec:kpi}) and there are more ambitious goals.  We would like to
have measurements of $b \ra d$ penguin decays despite large backgrounds
from the dominant $b \ra s$ decay; measurement of the electroweak
penguin $B^+\ra \phi\pi^+$ with a large $B^0\ra\phi K^+$ background
will be a tremendous challenge.  The BaBar experiment will
use a Detector of Internally Reflected \v Cerenkov (DIRC)
to measure the \v Cerenkov angle of light from
particles traversing quartz bars.  The DIRC gives at least $2.5\sigma$
$\pi$-$K$ separation up to the maximum momentum of 4 GeV. 
The Belle experiment uses an aerogel threshold \v Cerenkov detector
to identify kaons with an 8\% misidentification rate for 
momenta between 1 and 3 GeV. 
The CLEO III detector will use a
Ring Imaging \v Cerenkov (RICH) detector with a solid radiator
to achieve $4 \sigma$ $\pi$-$K$
separation at a momentum of 2.6 GeV/c, the maximum value from $B$ decays.
HERA-B will have a gaseous RICH detector,
which is expected to provide 90\% efficient kaon identification with
less than 2\% misidentification background.  
CDF-II and D0 are limited to kinematic separation of kaons and pions,
with little additional help from $dE/dx$.
The BTeV and LHC-B designs include gaseous RICH detectors.

\section{CONCLUSIONS}

The penguin program already has been a profitable one,
but much penguin physics is still left to be done.
The measured $b \rightarrow s \gamma$ rate has
been found to be consistent with Standard Model predictions and has
provided interesting constraints on new physics models.
Since these measurements are still statistics limited, more data will
provide improved constraints and even allow measurement of $|V_{td}|$
from $b \rightarrow d\gamma$ penguin decays.

Limits on the rates of electroweak penguins $ b\rightarrow s\ell^+\ell^-$
and vertical penguins such as $b \rightarrow \ell^+\ell^-$
are currently orders of magnitude above the predictions
of the Standard Model and give no evidence for non-SM effects.
These processes need to
be measured to confirm Standard Model predictions or to distinguish
between different types of new physics.
Some penguin modes such as $b\to s\nu\bar\nu$ or $B^0\to\gamma\gamma$
will be very difficult to observe, since 
they are not suitable for detection at hadronic colliders and
the $e^+e^-$ machines are likely to
be limited by the $b\bar b$ statistics.

Several suspected gluonic penguins have been observed.
The $B \rightarrow K\pi$ rates are consistent with
theoretical predictions.  However, the $B \rightarrow \omega K$,
$B \rightarrow \eta' K$ and $B \rightarrow \eta' X_s$
rates are all surprisingly large!  These rates have stimulated
many Standard Model and non-SM ideas and more data are needed to resolve
the situation.
More work needs to be done to
separate the gluonic penguin amplitudes from electroweak penguin or
$b \rightarrow u$ amplitudes in hadronic
decays.
The rates of penguin decays such as $B \rightarrow K\pi$,
$B^0 \rightarrow \phi K^0_S$ and $B^0 \rightarrow \eta' K^0_S$
must be measured precisely to give us insight
into $CP$ violation.  The measurement of
many penguin modes is needed to give us a
handle on penguin pollution in indirect $CP$ violation.

The outlook for penguins is very promising.  Within the next five years
we should have good handles on most electromagnetic and gluonic penguin modes,
and perhaps a beginning of an understanding of suppressed modes which
are dominated by electroweak penguins or annihilation diagrams.
Perhaps there can be a reunion in the year 2002 on the 25th birthday of
the penguin and we can look back at the tremendous progress that has
been made!  Darts anyone?

\section*{ACKNOWLEDGMENTS}
The authors would like to thank
A. Ali, H-Y. Cheng, S. Oh, and A. Soni for useful discussions.
We also thank
Ahmed Ali, Bill Ford, and Amarjit Soni
for their careful reading of the manuscript.
We acknowledge our colleagues on CLEO, ARGUS, ALEPH, DELPHI, L3, 
CDF, and D0 for their contributions to the experimental work in this review.
This review was supported by the U.S. Department of Energy.


\begin{thebibliography}{99}
\bibitem{ref:delphi_sg}
Battaglia M, Kluit PM.  Private communication.
Ref.~\cite{ref:delphi_dd} actually quotes a limit on the non-standard model
contributions to $b \rightarrow s g^*$, whereas we quote DELPHI's limit on the
total $b \rightarrow s g^*$ rate
\end{thebibliography}
\end{document}